\documentclass[12pt]{article}
\usepackage{setspace} 
\usepackage{palatino} 
\usepackage[left=1in,right=1in,top=1in,bottom=1in]{geometry} 
\usepackage[semicolon,round,longnamesfirst]{natbib} 
\usepackage[usenames,dvipsnames]{xcolor} 
\usepackage{enumerate}

\usepackage{comment}

\usepackage[backref=page,breaklinks, pdftex]{hyperref}
\hypersetup{
	colorlinks=true, 
	breaklinks=true, 
	urlcolor=Blue, 
	linkcolor=Blue,  
	citecolor=NavyBlue 
	}

\usepackage{amsmath,amsfonts,amssymb,amsthm,mathtools,dsfont}

\newtheorem{theorem}{Theorem}

\newtheorem{corollary}{Corollary}

\newtheorem{proposition}{Proposition}
\newtheorem{lemma}{Lemma}

\newtheorem{definition}{Definition}
\newtheorem{example}{Example}

\usepackage[nameinlink,noabbrev,capitalise]{cleveref}
\crefname{assumption}{Assumption}{Assumptions}
\Crefname{assumption}{Assumption}{Assumptions}

\usepackage{graphics,caption,subcaption}
\usepackage{accessibility}

\usepackage{tikz}
\usetikzlibrary{intersections, quotes, angles}
\usetikzlibrary { decorations.pathmorphing, decorations.pathreplacing, decorations.shapes, }
\usetikzlibrary{calc} 
\usetikzlibrary{patterns}

\usepackage{ifthen}

\newcommand{\showTheo}{0}
\newcommand{\showAlex}{1}

\newcommand{\theo}[1]{%
  \ifthenelse{\equal{\showTheo}{1}}%
    {[Theo] #1}%
    {}%
}

\newcommand{\alex}[1]{%
  \ifthenelse{\equal{\showAlex}{1}}%
    {[Alex] #1}%
    {}%
}

\newtheorem{lemmaalt}{Lemma}[lemma]

\newenvironment{lemmap}[1]{
  \renewcommand\thelemmaalt{#1}
  \lemmaalt
}{\endlemmaalt}

\title{Robust Regulation of Labour Contracts\thanks{We thank Ben Brooks, Gabriel Carroll, Piotr Dworczak, Yingni Guo, Ian Jewitt, Andy Kleiner, Stefan Krasa, Julien Manili, Meg Meyer, Mallesh Pai, Thomas Piketty, François Salanié, Ludvig Sinander, Jean Tirole, Quitzé Valenzuela-Stookey, Thomas Zuber and audiences at Bonn, CRC TR 224 Summer School, EC 2025, Oxford, Princeton, PSE, SAET 2025, Sciences-Po, TSE, UIUC, and Warwick. We also thank Rama Tonpe for excellent research assistance. All errors are our own. Alexis acknowledges support from the European Research Council (ERC) (101040122---IMD).}}
\author{Théo Durandard\thanks{University of Illinois at Urbana-Champaign. Email: \href{mailto:theod@illinois.edu}{theod@illinois.edu}} \quad Alexis Ghersengorin\thanks{University of Bonn. Email: \href{mailto:aghersen@uni-bonn.de}{aghersen@uni-bonn.de}}}
\date{September 08, 2025 \\ \vspace{3mm} \href{https://drive.google.com/file/d/1mDJqSNp9fOGjmhfzbTmLmvGAZE6Zfdnm/view?usp=sharing}{Link to the latest version}
}

\begin{document}

\maketitle
\begin{abstract}
    \begin{singlespace}
    We study the robust regulation of contracts in moral hazard problems. A firm offers a contract to incentivise a worker protected by limited liability. A regulator restricts the set of permissible contracts to (i) improve efficiency and (ii) protect the worker. The regulator faces uncertainty about both the worker’s actions and the firm’s production cost, and evaluates regulations based on their worst-case regret. The regret-minimising regulation mandates a \emph{minimum piece rate compensation} for the worker. This rule simultaneously guarantees a fair share for the worker and preserves enough contractual flexibility to provide incentives.  
    \end{singlespace}
\end{abstract}

\clearpage

\noindent Contracts are the standard instrument to address moral hazard in economic relations. They determine both the surplus generated from the interaction and its distribution among the contracting parties. Because efficiency and distribution are central policy concerns, regulating contracts is a natural intervention in environments with moral hazard. In particular, in labour markets, where firms typically hold greater bargaining power and may distort production to capture a disproportionate share of the surplus \citep{card2022set,naidu2022labor}, policymakers have long regulated labour contracts to mitigate such distortions and ensure a more equitable distribution of the surplus \citep{Berger2025}.\footnote{The U.S. Treasury explicitly mentions the two objectives of \emph{efficiency} and \emph{redistribution} in ``\textit{The State of Labor Market Competition}'' (\citeyear{Treasury2022}).} This raises an important design question: what should be the optimal regulation of contracts in relations affected by moral hazard? We tackle this question by analysing the effect of contract regulations on incentive provision in the canonical principal-agent model with moral hazard. Specifically, we study regulation as a delegation problem: what contracts should the regulator authorise the firm to offer to the worker? 

In practice, regulators lack the knowledge and ability to tailor regulations to individual firm-worker interactions at every point in time. Regulations apply identically across firms with heterogeneous technologies and costs (e.g., within a sector or geographic area). So, regulators face a trade-off between workers' \emph{protection} and firms' \emph{flexibility}. While stricter regulations may benefit employed workers, they also risk inducing novel distortions in firms' production decisions and even deterring some from hiring.\footnote{A similar trade-off exists in the standard monopsonistic model of the labour market: a higher minimum wage protects low-income workers and reduces inefficiencies due to firms' monopsonistic power, but also stops less productive firms from hiring.} Thus, regulations must be designed to perform well in different situations. We capture this by studying optimal robust (non-Bayesian) regulations.

Our model has three players: a firm (she), a worker (he), and a regulator (they). The firm incentivises the worker to take a costly \emph{productive action} by offering a contract. The actions being non-contractible, a contract pays the worker as a function of the output that the actions (stochastically) generate. Hiring entails a fixed \emph{production cost} for the firm, and the pair of this fixed cost and the set of productive actions defines the firm's \emph{technology}. We assume both the firm and the worker are risk-neutral and protected by \emph{limited liability}, and they respectively maximise their profit and surplus. The \emph{regulator} chooses the set of permitted contracts the firm can offer. We assume that the regulator's payoff is a weighted sum of the firm's profit and the worker's surplus, where the latter has a (weakly) greater weight $\alpha \geq 1$. While the firm and the worker know the technology, the regulator does not. The ex post loss that the regulator must tolerate because of this lack of information defines their \emph{regret}. Namely, for a given technology and regulation, the regret is the difference between the regulator's perfect information payoff --- had they known the technology --- and their actual payoff given the equilibria of the contractual relationship. The regulator evaluates regulations by their \emph{worst-case regret}, that is, the maximal regret over all possible technologies and equilibria, and chooses the regulation that minimises it.

Our main result establishes that imposing a minimum piece-rate (MPR) compensation is optimal (\cref{theorem:main}). The optimal MPR regulation requires that a share of the generated output at least equal to $\displaystyle \frac{\alpha - 1}{2\alpha - 1} $ accrues to the worker (where $\alpha$ is the weight the regulator places on the worker's surplus). Formally, this regulation allows every contract pointwise greater than the linear contract $\ell^\star_\alpha \colon y \mapsto \displaystyle \frac{\alpha - 1}{2\alpha - 1} \cdot y$. Two key features underpin the optimality of this regulation. First, the minimum linear contract partially realigns the firm's incentives toward efficient production. It incentivises the worker to choose productive actions, which reduces the firm's incentive to underproduce. It also reduces the firm's incentive to distort production upward, as it forbids overly convex (bonus) contracts that can implement high-output actions cheaply. Hence, the minimum linear contract aligns the firm's incentives with the regulator's preferences for efficiency. Second, the optimal regulation limits the risk of deterring production (and wasting all potential surplus) by granting the firm complete flexibility above the minimum contract. So, the MPR regulation is optimal because it protects the worker's surplus and limits the firm's motives to distort production while preserving the firm's flexibility. 

The central challenge to solving the regulator's problem arises from the size of the delegation possibility set. We are interested in the regulation of \emph{contracts}, not only that of outcomes. The regulator can, therefore, restrict the firm's contractual choice to be in any subset of the infinite-dimensional space of all measurable mappings from outcomes to payments. A key step to tackle the regulator's problem is then to reduce its dimensionality. The proof of \cref{theorem:main} described in \cref{subsec:proof_sketch} starts by doing just that. We show that it is without loss to look for an optimal regulation among MPR ones and that \emph{binary} technologies -- i.e., technologies such that every action induces a binary distribution over outputs $0$ and $\bar{y}$ (where $\bar{y}$ is the upper bound on the expected output of any action) -- maximise the regulator's regret for these regulations. It is without loss because minimum piece-rate regulations and binary technologies constitute \emph{mutual best-response classes} in the zero-sum game between the regulator and the adversarial Nature. That is, for any MPR regulation, there exists a binary regret-maximising technology, and for any binary technology, there exists a MPR regret-minimising regulation. Intuitively, binary distributions can be seen as garblings of non-binary distributions \citep{garrett2023optimal}. So, they minimise the informational content of output, which, by the Informativeness Principle \citep{holmstrom1979moral}), increases incentive costs. This, in turn, raises regret: higher incentive costs either render all actions unprofitable, wasting the entire surplus, or push the firm to incentivise lower production levels to capture more of the surplus, thereby worsening inefficiencies and misallocation. Restricting attention to binary distributions then implies that we can focus on linear contracts to bind the regulator's regret. Finally, we show that allowing all the contracts above an authorised linear contract only ever decreases the regulator's regret. This reduction thus collapses the regulator's infinite-dimensional problem into the one-dimensional problem of setting the slope of the minimum linear contract. Moreover, it guarantees that binary technologies maximise the regulator's regret when they offer an MPR regulation, simplifying the computation of the worst-case regret. Beyond its role in deriving our main result (\cref{theorem:main}), this methodological step highlights a general and promising approach to tackle robustness problems: identifying restricted strategy classes that are mutual best responses provides a powerful way to simplify otherwise intractable minimax robustness problems.

The optimal MPR regulation characterised in \cref{theorem:main} is not uniquely optimal. Our second main result (\cref{theorem:optimalregulationnecessity}) provides necessary conditions for optimality that echo the core features of the optimal MPR regulation. Any optimal regulation must guarantee the worker a minimum payment close to the optimal MPR $\ell^\star_\alpha\cdot y$ for each output $y$. Moreover, it must also offer enough flexibility for the firm to provide incentives. \cref{theorem:main,theorem:optimalregulationnecessity} notably imply that: (i) the optimal MPR regulation is uniquely optimal among the regulations that allow every contract above a minimum one (\cref{corollary:minimaloptimalregulation}); (ii) laissez-faire is optimal if and only if the regulator puts equal weight on the firm's profit and the worker's surplus ($\alpha=1$) (\cref{corollary:comparative_stats}).

In \cref{sec:extension}, we discuss the role of the regulator's knowledge for our main results. We show that the optimal MPR regulation remains optimal when the regulator knows that the effort costs and output distributions are ranked according to the monotone likelihood ratio property or first-order stochastic dominance (\cref{prop:optimalregulationMLRP}) and that the regulator can implement this optimal regulation even when less informed. Furthermore, we also show that the knowledge of quantitative features of the technology does not alter the qualitative feature of optimal regulations. A MPR regulation remains optimal when the regulator knows more precise bounds on the firm's production cost and expected output (\cref{prop:optimalregulationquantitativeknowledge}). Finally, in \cref{subsec:reginpractice}, we discuss the implications of our results for real-world regulations. We argue in particular that our model can help understand the regulation of ``output work'' in the UK and could inform the design of novel legal frameworks to protect the interest of the growing number of workers in the ``platform economy.''

\cref{appendix:proof_theorem_main,app:prooftheorem:optimalregulationnecessity} contain the proofs of \cref{theorem:main,theorem:optimalregulationnecessity}. The Supplemental Appendices in \cite{Durandard25supp} contain supporting results and the omitted proofs.

\paragraph{Related literature.} We view regulation as a delegation problem \citep{holmstrom1980theory} in the canonical principal-agent model with moral hazard \citep{laffont2002theory}. Delegation is a natural framework to study regulations as firms typically know more about the environment, and regulators can constrain firms' choices \citep{laffont1994new}. \cite{amador2022regulating} studied \citeauthor{baron1982regulating}'s \citeyearpar{baron1982regulating} monopoly regulation problem in a setting without transfers. \cite{hitzig23} consider two agents bargaining over pairs of quality and payment (e.g. health insurance coverage and salary if the agents bargain over a labour contract), and they study regulations consisting of restricting the set of authorised qualities and setting the default option. Close to ours are papers that study robust non-Bayesian regulations. In particular, \cite{guo2019robust} derived the regret-minimising monopoly regulation.\footnote{See \cite{naidu2022labor} and the Appendix C of \cite{guo2019robust} for a reinterpretation of \citeauthor{guo2019robust}'s results in terms of labour market regulation.} \cite{malladi2022delegated} studies the robustly optimal approval rules for innovation. \cite{thomas2024regulation} studies robust labour market regulations as a delegation problem. He shows that combining minimum wage, overtime pay, and a cap on hours can help protect the worker and improve efficiency under different bargaining protocols. \cite{guo2019robust} and \cite{malladi2022delegated} consider adverse selection problems and \cite{thomas2024regulation} abstracts from incentives to focus on bargaining. In contrast, we study the effect of regulations on incentive provision in moral hazard problems. 

In the delegated contracting literature that considers moral hazard \citep[e.g.,][]{hiriart2012much,iossa2016corruption} as well as in the papers above, the regulator determines the set of permissible \emph{outcomes} (or allocations). The delegation sets are then subsets of finite-dimensional spaces. Our regulator chooses the set of permissible \emph{contracts}, a subset of the space of all measurable mappings from outputs to payments. Two notable exceptions are \cite{bhaskar2023regulation} and \cite{bhaskar2024regulating}, who, like us, examine regulations as delegation problems in the space of \emph{contracts}. In the former, the authors examine the optimal regulation of contracts in insurance markets. Unlike us, they look at a Bayesian framework, and the regulated interaction suffers from adverse selection. In the latter, they study the regulation of dynamic contracts. They derive the Bayesian optimal regulation when the frictions arise from search inefficiencies and contracts distributes the surplus between the agents but do not incentivise effort to produce. 

We contribute to the growing literature on contract design with a non-Bayesian objective, restarted in economics with \cite{chassang2013calibrated}, \cite{garrett2014robustness}, and \cite{carroll2015robustness} \citep[see][for a survey]{carroll2019robustness}. 
This literature has typically studied the design of robust compensation schemes between a principal and an agent or in a hierarchical organisation \citep{walton2022general}. We depart by looking at the regulation problem and show that the robustness approach yields economically meaningful predictions when the Bayesian problem remains intractable. Notably, linear contracts play a role in our model to realign incentives as in \cite{carroll2015robustness}, \cite{walton2022general}, or \cite{Carroll2023doublemh}. While in the latter, linear contracts ensure that workers' incentives are not too distant from the firms', in our case, they ensure that the firms' incentives are not too distant from the regulator's. 

Finally, labour contract regulations have often been justified to curb firms' monopsony power, and have been primarily assessed on their effects on unemployment \citep{stigler1946economics,card2016myth,naidu2022labor}. Yet, the firms' market --- hence bargaining --- power distorts not only the allocation of workers but also the contractual relationship between firms and workers. Thus, contract regulations also influence the inner workings of firms --- e.g., firms' technology choices and the efforts exerted by workers \citep{obenauer1915effect,coviello2022minimum}. This paper focuses on the intensive margin: the impact of contract regulations on incentive provision.

\section{Model}\label{sec:setting}

There are three players: the regulator (they), the firm (she), and the worker (he). 

\paragraph{Technologies.} An \emph{action} for the agent is a pair $(e,F)$, where $e \geq 0$ is the action's cost (e.g., effort), and $F \in \Delta(\mathbb{R}_+)$ is its distribution over outputs.\footnote{$\Delta(\mathbb{R}_+)$ is the set of all Radon probability measures on $\mathbb{R}_+$. We equip $\Delta\left(\mathbb{R}_+\right)$ with the topology generated by the total variation norm.} A \emph{technology} is a pair $\mathcal{T} = (k, \mathcal{A})$ consisting of a \emph{production cost} $k\geq 0$ for the firm and a nonempty compact set $\mathcal{A}$ of actions for the agent. The only constraint on the set of feasible technologies is that there exists a uniform bound $\bar{y}\geq 0$ on the expected output of any action $(e,F)$: $\mathbb{E}_F[y] \leq \bar{y}$.  We let $\mathbb{T}$ denote the set of all feasible technologies. 

\begin{example}
    In many moral hazard models, the agent chooses an effort level $e$ from a compact set $ [0, \bar{e}] \subset\mathbb{R}$, which costs $c(e)$ and induces the distribution $F(\cdot \vert e)$. In our notations, the set of actions is the image of the mapping from efforts to the associated pairs of cost and distribution: $\mathcal{A} = \left(c(e), F(\cdot \vert e)\right)_{e \in [0,\bar{e}]}$.
\end{example}

\begin{example}[Binary technology]
    \label{example:binary_actions}
    A \emph{binary} action, denoted $(e, B_\mu)$, is an action such that the distribution over outputs $B_{\mu}$ is a binary distribution supported on $\{0,\bar{y}\}$ with mean $\mu$, that is,\footnote{For any pair of output distributions $F$ and $G$ and any real number $a\in[0,1]$, we denote as $aF+(1-a)G$ the output distribution obtained by the usual mixture operation.} $B_{\mu} = \frac{\mu}{\bar{y}}\delta_{\bar{y}}+\left(1-\frac{\mu}{\bar{y}}\right)\delta_{0}$. We refer to a technology $\mathcal{T} = \left( k, \mathcal{A} \right)$ as \emph{binary} if $\mathcal{A}$ only contains binary actions. Binary technologies play a crucial role in our analysis (see \cref{subsec:proof_sketch}).
\end{example}

\paragraph{Contracts and regulations.} The actions being non-contractible, the firm incentivises production by offering a contract that pays the worker as a function of the realised output. We assume that both the worker and the firm are protected by \emph{limited liability} and denote the set of all \emph{contracts} by
\begin{equation*}
    \mathbb{C}_0 = \left\{w\colon\mathbb{R}_+ \to\mathbb{R} \mid w \text{ measurable, and } \forall y, \ 0\leq w(y) \leq y\right\}.
\end{equation*}
A \emph{regulation} is a closed subset of contracts $\mathcal{C} \subset \mathbb{C}_0 $ (for the product topology on $\mathbb{R}_+^{\mathbb{R}_+}$). For any regulation $\mathcal{C} \subset \mathbb{C}_0$, we define the function $\underline{w}^{\mathcal{C}}$ by $\underline{w}^{\mathcal{C}}(y) = \min \left\{ w(y) \mid w\in \mathcal{C} \right\}$ for all $y\geq 0$ (note that $\underline{w}^{\mathcal{C}}$ is not necessarily a contract in $\mathcal{C}$). We refer to the linear contract $y\mapsto \ell\cdot y$ as the contract $\ell$.
\begin{example}[Minimum piece-rate]
    \label{example:MPR_reg}
    The minimum piece-rate (MPR) regulation $\mathcal{C}_\ell$ authorises every contract that exceeds the minimum linear contract $\ell$ pointwise:
    \begin{equation*}
        \mathcal{C}_\ell = \left\{ w \in \mathbb{C}_0  \mid w(y) \geq \ell \cdot y \, \, \text{ for all } y \geq 0 \right\}.
    \end{equation*}
    In this case, $\underline{w}^{\mathcal{C}}$ is a contract in $\mathcal{C}$ and is equal to $\ell$.
\end{example}
As the worker is risk-neutral (see below), limited liability for the worker is the sole agency friction.\footnote{Absent the worker's limited liability, a solution to the contracting problem for the principal is to sell the firm to the agent at a fixed price. Moral hazard plays no role. In this case, we showed in a previous version of the paper that the optimal regulation caps the price the firm can charge.} Limited liability for the firm, on the other hand, prevents the regulator from forcing the firm to pay the worker more than the realised output. However, given that the firm incurs a fixed cost $k$, this constraint does not prevent the regulator from forcing the firm to pay the worker more than her realised revenue $y-k$. It simply puts an exogenous limit on how much the firm can be obliged to pay. An economic justification for this assumption could be that the firm can default or cannot be forced to pay more than the total value of her assets. Still, this assumption also implies that all feasible contracts pay $0$ when the output is $y=0$, a crucial simplification that allows us to solve the regulator's problem of finding an optimal set of contracts.

\paragraph{Timing.} The timing of the game is as follows.
\begin{enumerate}
    \item The regulator chooses the regulation $\mathcal{C} \subset \mathbb{C}_0 $, i.e., the set of authorised contracts.

    \item Nature determines the technology $\mathcal{T} = (k, \mathcal{A})$, the firm and the worker observe it.
    
    \item If the firm does not hire the worker, the game ends, and both players receive their outside option equal to $0$. If the firm wants to hire the worker, she must offer a contract $w \in \mathcal{C}$.
    
    \item If the worker refuses the firm's contract, the game ends, and both players receive their outside option. If the worker accepts the contract, he chooses an action, and the payoffs are realised.  
\end{enumerate}

\paragraph{Payoffs.} Both the firm and the worker are risk-neutral. The firm solves a (constrained) moral-hazard problem. She chooses which authorised contract to offer (if any) to maximise her profit when the worker best responds by choosing the action that maximises his surplus. Formally, given a regulation $\mathcal{C} \subset \mathbb{C}_0 $ and a technology $\mathcal{T}=(k, \mathcal{A}) \in \mathbb{T}$, the firm's payoff is\footnote{\cref{lemma:existenceoptimalcontract} in \cref{app:existence} guarantees the existence of an optimal contract and associated optimal action under mild regularity conditions that are satisfied by all the regulations we consider. Note, however, that we do not impose these conditions as part of the definition of a regulation, and thus writing the maximum for an arbitrary $\mathcal{C}$ is a small abuse of notation.}
\begin{align}
    \Pi(\mathcal{C,T}) = \max_{w \in \mathcal{C}, (e,F) \in \mathcal{A}} & \big\{0, \mathbb{E}_F[y-w(y)]-k\big\} \tag{$\Pi$} \label{eq:profit_program}\\
    \text{s.t.} \quad & \mathbb{E}_F[w(y)] - e \geq 0, \tag{$IR_W$} \label{eq:IR_W} \\
    \text{ and } \quad & \mathbb{E}_F[w(y)] - e \geq \mathbb{E}_{\Tilde{F}}[w(y)] - \Tilde{e} \quad \text{ for all } (\tilde{e},\tilde{F}) \in \mathcal{A}. \tag{$IC_W$} \label{eq:IC_W}
\end{align}
Since the firm anticipates the worker's response to any contract, the maximisation program \eqref{eq:profit_program} equivalently views the firm as directly choosing the worker's action, taking his optimising behaviour as a constraint.\footnote{As is standard in principal-agent models, we implicitly assume that the firm can select her preferred action when the worker is indifferent between several.} Namely, the worker accepts a contract $w \in \mathcal{C}$ and takes action $(e,F)$ if and only if this gives a payoff greater than his outside option (constraint \eqref{eq:IR_W}) and maximises his surplus (constraint \eqref{eq:IC_W}).

The worker's surplus when the regulation is $\mathcal{C}$, the technology is $\mathcal{T}$, the firm and the worker best-reply to each other with the strategy profile $\sigma = (w^{\star}, (e,F))$, is
\begin{align*}
    WS(\mathcal{C,T}, \sigma) =  \mathbb{E}_F[w^\star(y)] - e.
\end{align*}
If the firm optimally decides not to offer any contract, we denote the equilibrium strategy as $\sigma = (\emptyset, \emptyset)$ and $WS(\mathcal{C, T}, \sigma) = 0$. We denote as $\Sigma(\mathcal{C, T})$ the set of subgame perfect equilibria of the contracting game. 

The regulator's payoff for a regulation $\mathcal{C}$, a technology $\mathcal{T}$, and an equilibrium $\sigma$, is 
\begin{equation*}
    \Pi(\mathcal{C,T}) + \alpha WS(\mathcal{C,T}, \sigma),
\end{equation*}
where $\alpha \geq 1$ is the weight the regulator places on the worker's surplus.
 
\paragraph{Regulator's regret.} At the time of the regulation choice, the regulator does not know the technology. If they were perfectly informed of the technology $\mathcal{T} \in \mathbb{T}$, they would choose the contract and the action that maximise their payoff, solving the following program:\footnote{It is easily seen that the regulator does not gain from allowing multiple contracts. A solution exists by \cref{lemma:existenceoptimalcontract} in \cref{app:existence}.}
\begin{align}
    V(\mathcal{T}) = \underset{w \in \mathbb{C}_0 , (e,F) \in \mathcal{A}}{\max} \, &  \mathbb{E}_F[y-w(y)]-k + \alpha\big(\mathbb{E}_F[w(y)] - e\big) \tag{$R^{\star}$} \label{eq:FB_program} \\
    \text{s.t.} \quad & \mathbb{E}_F[y-w(y)]-k \geq 0, \tag{$IR_F$} \label{eq:IR_F} \\
    & \mathbb{E}_F[w(y)] - e \geq 0, \tag{\ref{eq:IR_W}} \\
    & \mathbb{E}_F[w(y)] - e \geq \mathbb{E}_{\Tilde{F}}[w(y)] - \Tilde{e} \text{ for all } (\tilde{e},\tilde{F}) \in \mathcal{A}. \tag{\ref{eq:IC_W}}
\end{align}
Compared to the firm's program \eqref{eq:profit_program}, the regulator's choice of contract and action in \eqref{eq:FB_program} must additionally satisfy the firm's participation constraint \eqref{eq:IR_F}.\footnote{As for the firm's program, we assume that, in case of indifference, the regulator can break ties in their favour in the perfect information problem.} We refer to any action (contract) that jointly solves \labelcref{eq:FB_program} with a contract (an action) as being \labelcref{eq:FB_program}-optimal.

The \emph{regret} captures the loss the regulator suffers from their ignorance about the technology. For a given regulation $\mathcal{C} \subset \mathbb{C}_0 $ and technology $\mathcal{T} \in \mathbb{T}$, the regulator's regret is
\begin{align*}
    R(\mathcal{C, T}) = \underset{\sigma \in \Sigma(\mathcal{C, T})}{\sup} \left\{ V(\mathcal{T}) - \left[\Pi\left(\mathcal{C},\mathcal{T}\right) + \alpha WS\left(\mathcal{C},\mathcal{T}, \sigma\right)\right] \right\}.
\end{align*}
The difference between the perfect information payoff and the realised one is maximised over all possible equilibria of the contracting game between the firm and the worker.\footnote{Being robust against worst-case equilibria means that when there are multiple profit-maximising contracts and actions, the firm chooses the worst one for the regulator.}

\paragraph{Regulator's objective.} The regulator evaluates a regulation $\mathcal{C} \subset \mathbb{C}_0 $ according to the greatest regret it generates across all technologies $\mathcal{T} \in \mathbb{T}$, what we name its \emph{worst-case regret}:
\begin{align*}
    WCR(\mathcal{C}) = \underset{\mathcal{T} \in \mathbb{T}}{\sup} \, R(\mathcal{C},\mathcal{T}).
\end{align*}
This is our sole departure from the Bayesian approach, which would assign a prior belief over the set of possible technologies $\mathbb{T}$. The regulator would then minimise the \emph{expected} regret.\footnote{Minimising the regulator’s expected regret is equivalent to maximising their expected payoff because the regulator's complete information payoff does not depend on the regulation.} Instead, the regulator is cautious and wishes to find the regulation that performs the best is the worst-case scenario, namely to find $\mathcal{C} \subset \mathbb{C}_0 $ that minimises the worst-case regret:\footnote{Note that the regulator could not be better off by offering a menu of regulations, as this would be equivalent to a single regulation equal to the union of all these regulations.}
\begin{align}
    \underset{\mathcal{C}}{\inf } \, WCR(\mathcal{C}). \tag{P} \label{eq:minmax_regret}
\end{align}
A regulation $\mathcal{C}$ is \emph{optimal} if $WCR(\mathcal{C}) \leq WCR(\mathcal{C'})$ for every other regulation $\mathcal{C}'$. One interpretation of the regulator's problem consists of the following zero-sum game. The regulator selects a regulation $\mathcal{C}$, and an adversarial Nature then chooses the technology $\mathcal{T} \in \mathbb{T}$ and the equilibrium $\sigma \in \Sigma$ of the contracting game.

\section{Optimal Regulation}\label{sec:OptimalRegulation}

\subsection{Main Result}\label{subsec:MainResult}

While regulations can be complicated --- the regulator must consider every closed set of contracts --- our main result derives a simple optimal regulation. This regulation imposes a minimum piece-rate compensation for the worker that only depends on the regulator's weight $\alpha$ on the worker's surplus. 

\begin{theorem}\label{theorem:main}
    Let
    \begin{align}
        \bar{R} & = \min_{\ell \in \left[0,\frac{1}{2}\right]}  \ \max \left\{\alpha e^{\frac{2\ell-1}{1-\ell}}(1-\ell)\bar{y}, \, \alpha e^{-\frac{1}{\alpha}} (1-\ell)\bar{y}\right\}; \label{equation:wcr} \\
        \text{and} \quad \mathcal{C}^\star_\alpha & = \left\{w \in \mathbb{C}_0  \, \colon w(y) \geq \ell^\star_\alpha\cdot y \coloneqq  \frac{\alpha-1}{2\alpha -1} \cdot y \ \text{ for all } y \geq 0 \right\} \subset \mathbb{C}_0 .\label{equation:minimum_linear_contract_regulation}
    \end{align}
    \begin{enumerate}[(i)]
        \item The worst-case regret of any regulation $\mathcal{C} \subset \mathbb{C}_0 $ is at least $\bar{R}$.
        \item The worst-case regret of the regulation $\mathcal{C}^\star_\alpha$ is $\bar{R}$, hence $\mathcal{C}^\star_\alpha$ is optimal.
    \end{enumerate}
\end{theorem}

\Cref{theorem:main} states that $\bar{R}$ is a lower bound on the worst-case regret of any regulation, as well as an upper bound on the worst-case regret of the regulation $\mathcal{C}^\star_\alpha$. Therefore, $\mathcal{C}^\star_\alpha$ is optimal and $\bar{R}$ is the minmax regret of the regulator. The optimal regulation $\mathcal{C}^\star_\alpha$ imposes a minimum piece-rate, that is, any authorised contract must be pointwise above the minimum linear contract $\ell^\star_\alpha = \displaystyle \frac{\alpha-1}{2\alpha-1}$. It is worth noting that $\ell_\alpha^\star$ only depends on $\alpha$; in particular, it does not depend on $\bar{y}$ (see \cref{subsec:less_knowledge}).

The regulator has to trade off the firm's flexibility to incentivise production and her ability to distort production and extract surplus from the worker. Either the regulation is \emph{too restrictive} and deters the firm from hiring the worker, thus generating regret from wasted production opportunities. Or the regulation is \emph{too permissive} and the firm produces but regret results from inefficient production levels and the lack of worker protection. 

The two terms in the maximand of the lower bound $\bar{R}$ reflect this trade-off for the minimum piece-rate (MPR) regulation\footnote{Only $\ell\in [0,\frac{1}{2}]$ are considered as we show that higher MPR can never be optimal (see \cref{lemma:lowerbound_no_production_1,lemma:binding_cases}).} $\mathcal{C}_\ell$. The first term is the regulator's greatest regret from scenarios where the regulation prevents the firm from hiring the worker. The second term is the regulator's greatest regret for allowing inefficient distortions and insufficiently protecting the worker. The $\min$ and the $\max$ operators come from, respectively, the regulator's ability to choose a regulation (in this case, the slope of the minimum linear contract) and the regulator's concern about the worst-case scenario. The proof sketch in the next section (\labelcref{subsec:proof_sketch}) explains why MPR regulations are a right class of regulations to consider for optimality. The optimal minimum linear contract in $\mathcal{C}^\star_\alpha$ is then obtained by solving for the optimal $\ell$ in \cref{equation:wcr}. Hence, $\mathcal{C}^\star_\alpha$ equalises the regulator's maximal regrets from surplus destruction and rent-extraction distortions. We give further economic interpretations of the optimality of $\mathcal{C}_\alpha^\star$ in \cref{subsec:taking_stocks}.

\subsection{Proof Sketch}\label{subsec:proof_sketch}

The proof of \cref{theorem:main} proceeds in two steps. We first construct, for each regulation, binary technologies (see \cref{example:binary_actions} for the definition) that generate a regret of at least $\bar{R}$, with equality for $\mathcal{C}^{\star}_{\alpha}$ (\cref{appendix:proof_lowerbound}). We then show that these constructed technologies maximise regret for $\mathcal{C}^{\star}_{\alpha}$ (\cref{app:prooftheorem:optimalregulation}). Although sufficient, this direct approach conceals an insightful result on which it relies. Namely, binary technologies and MPR regulations are the right classes of worst-case scenarios and optimal regulations to consider. In the proof sketch, we add a preliminary step that shows this result (\cref{app:proof_step1} contains the supporting lemmas) and clarifies the economic forces driving the optimality of $\mathcal{C}^{\star}_{\alpha}$. It also highlights a method that can be used in similar settings.

\paragraph{Step 1: Simplifying the search for optimal regulations and worst-case technologies.} 

We show that it is without loss to look for an optimal regulation among MPR ones and that binary technologies generate the greatest regret for these regulations. The key step to simplifying the regulator's and Nature's problems is to show that the classes of MPR regulations and binary technologies constitute \emph{mutual best-response classes} in the zero-sum game between the regulator and Nature: that is, for any strategy in one class, a best-response lies in the other. We hope this approach can prove useful in other similar robustness problems. Formally, our preliminary result follows from the following sequence of inequalities, where "linear $\mathcal{C}$" refers to regulations that only contain linear contracts:
\begin{align}
        \inf_{\text{\textbf{all} } \mathcal{C}} \quad \sup_{\text{\textbf{all} } \mathcal{T}} R(\mathcal{C,T}) 
        \geq & \inf_{\text{\textbf{all} } \mathcal{C}}\quad \sup_{\text{\textbf{binary} } \mathcal{T}} R(\mathcal{C,T}) \label{ineq:binary} \\
        \geq & \inf_{\text{\textbf{linear} } \mathcal{C}} \quad \sup_{\text{\textbf{binary} } \mathcal{T}} R(\mathcal{C,T}) \label{ineq:linear} \\
        \geq  & \inf_{\text{\textbf{MPR} } \mathcal{C}} \quad \sup_{\text{\textbf{binary} } \mathcal{T}} R(\mathcal{C,T}) \label{ineq:MPR}\\
        \geq & \inf_{\text{\textbf{MPR} } \mathcal{C}} \quad \sup_{\text{\textbf{all} } \mathcal{T}} R(\mathcal{C,T}) \label{ineq:all_tech}\\
        \geq & \inf_{\text{\textbf{all} } \mathcal{C}} \quad \sup_{\text{\textbf{all} } \mathcal{T}} R(\mathcal{C,T}). \label{ineq:all_reg}
\end{align}
First, the regulator is weakly better off when Nature is restricted to a narrower class of technologies (yielding inequality \eqref{ineq:binary}) or when they can choose from a broader set of regulations (proving inequality \eqref{ineq:all_reg}). Second, under binary technologies, only the payments at output levels $0$ and $\bar{y}$ are payoff-relevant. Hence, restricting attention to linear regulations entails no loss of generality, establishing inequality \eqref{ineq:linear}. The core of the argument then lies in the remaining two inequalities, \eqref{ineq:MPR} and \eqref{ineq:all_tech}, which we now explain in detail.

\Cref{ineq:MPR} states that there exists an optimal MPR regulation when Nature maximises regret among binary technologies. This result follows from \cref{lemma:ineq:MPR} (in \cref{app:proof_step1}), which shows that for any linear regulation $\mathcal{L}$ with smallest contract $\ell$, the corresponding MPR regulation $\mathcal{C}_\ell$, with minimum linear contract $\ell$, yields a weakly lower worst-case regret against binary technologies. To understand this, remember that the regulator's worst-case regret results from either (i) the firm staying out of business or (ii) the firm distorting production and capturing too much of the generated surplus. Consider case (i). If the firm does not hire under $\mathcal{C}_{\ell}$, she certainly does not hire under the more restrictive regulation $\mathcal{L} \subset \mathcal{C}_{\ell}$, so the regulator's regret is unchanged. Now consider case (ii). Then, the regulator wishes their regulation were more constraining to limit the firm's distortion and protect a larger share of the surplus for the worker. Thus, one would expect some constraints in $\mathcal{C}_\ell$ to be binding in the worst-case scenario. The minimum contract $\ell$ being the only constraint in $\mathcal{C}_\ell$, the worst-case technology involves the firm offering contract\footnote{See \cref{lemma:support_simplification_production} in \cref{app:supporting_results}.} $\ell$. Given that $\ell \in \mathcal{L}$ and $\mathcal{L} \subset \mathcal{C}_\ell$, the firm also offers $\ell$ (for the same technology) under regulation $\mathcal{L}$, leading again to the same regret. In short, $\mathcal{C}_\ell$ imposes the same minimum contract as $\mathcal{L}$ but allows full flexibility otherwise. Hence, when it is too constraining, so is $\mathcal{L}$; when it is not sufficiently constraining, the worst-case scenarios are driven by the minimum contract $\ell$ being too low, implying that $\mathcal{L}$ is not sufficiently constraining either. Consequently, by allowing more flexibility, $\mathcal{C}_\ell$ weakly reduces the worst-case regret compared to $\mathcal{L}$.

The final step is to establish \cref{ineq:all_tech}, which states that a binary worst-case technology exists against MPR regulations. We now sketch the argument, which is contained in \cref{lemma:ineq:all_tech} (in \cref{app:proof_step1}). Fix an MPR regulation $\mathcal{C}_\ell$. Let $\mathcal{T}$ be any technology, for which we write the regret as $R(\mathcal{C_\ell,T}) = \Pi^\star + \alpha WS^\star - \Pi -\alpha WS$, where the presence of the superscript $\star$ refers to the payoffs under the \eqref{eq:FB_program}-optimal action and contract and its absence refers to the payoffs under the profit-maximising action and contract. Reorganising, we can write the regret as 
\begin{align}
    R(\mathcal{C_\ell,T}) & = \alpha \big(WS^\star-WS\big) + \Pi^\star - \Pi\notag\\
     & = \alpha \big(S^\star-WS\big) -(\alpha-1) \Pi^\star - \Pi, \label{eq:proofsketch_simpleregret_step}
\end{align}
where $S^\star = \Pi^\star + WS^\star$ is the total surplus resuting from \eqref{eq:FB_program}. We can make a helpful simplification in our search for a worst-case technology by considering technologies for which (i) the firm earns zero profit under the regulator’s optimal contract (i.e., $\Pi^\star=0$), and (ii) the worker receives zero surplus under the firm’s preferred contract (i.e., $WS=0$).\footnote{See \cref{lemma:support_simplification_production,lemma:ineq:all_tech} for the complete arguments.} Hence, the worst-case regret takes the form 
\begin{align}
    WCR(\mathcal{C}_\ell) = \alpha S^\star - \Pi. \label{eq:proofsketch_simpleregret}
\end{align}

\cref{eq:proofsketch_simpleregret} implies that, for any \emph{fixed} profit level $\Pi$, maximising regret amounts to maximising the surplus $S^\star$ generated by the \eqref{eq:FB_program}-optimal action. Maximising $S^\star$ while holding the firm’s profit fixed requires maximising the incentive rents associated with the \eqref{eq:FB_program}-optimal action, so the firm does not deviate to the \eqref{eq:FB_program}-optimal action and increases her profit. This is where binary technologies play a crucial role as they maximise incentive rents.

We explain this key point in more detail. Recall that output plays a dual role in moral-hazard problems. It is both the firm’s revenue and the signal about the worker’s action that the firm uses to provide incentives. Replacing each action in a given technology with an ``equivalent'' binary action (i.e., one with the same cost and mean output) then reduces the informativeness of the output realisations on which the firm relies to incentivise the worker. According to the logic of the Informativeness Principle \citep{holmstrom1979moral}, this should increase the implementation costs, as \cite{garrett2023optimal} highlight. Indeed, consider a candidate regret-maximising technology $\mathcal{T} = (k, \mathcal{A})$ for the MPR regulation $\mathcal{C}_\ell$. Construct the ``equivalent'' binary technology $\mathcal{T}_b = \left(k, \mathcal{A}_b = \left\{(e, B_{\mu_F}) \right\}_{(e,F) \in \mathcal{A}}\right)$, where $\mu_F = \mathbb{E}_F[y]$. Under the binary technology $\mathcal{T}_b$, any contract $w \in \mathcal{C}_\ell$ is outcome-equivalent to the linear contract $\ell^w = \frac{w(\bar{y})}{\bar{y}}$. But, for any $w\in \mathcal{C}_\ell$, by definition of $\mathcal{C}_\ell$, it is the case that $\ell^w \in \mathcal{C}_\ell$. So, if the contract $w \in \mathcal{C}_\ell$ incentivises an action in $\mathcal{A}_b$, then the linear contract $\ell^w \in \mathcal{C}_\ell$ incentivises the ``equivalent'' action in $\mathcal{A}$, since, for all $(e,F) \in \mathcal{A}$,
\begin{align*}
    \mathbb{E}_{B_{\mu_F}}[w(y)] -e = \ell^w \cdot \mu_F -e = \mathbb{E}_F [\ell^w \cdot y] - e.
\end{align*}
In other words, transforming $\mathcal{T}$ into $\mathcal{T}_b$ effectively reduces the set of contracts available, and therefore increases implementation costs. \cref{lemma:ineq:all_tech} builds on this argument to construct, for any technology, a binary technology that increases the regulator's regret. For the sake of conciseness, we avoid the details of this construction and refer the interested readers to \cref{app:proof_step1}. We simply highlight the key property which makes binary technologies increase implementation costs for MPR regulations: that, for any contract $w\in \mathcal{C}$, the linear contract $\ell^w$ is also in $\mathcal{C}$. This property does not generally hold for an arbitrary regulation. As a result, binary technologies may fail to increase implementation costs for non-MPR regulations, as the following simple example illustrates. 

\begin{example}
    \label{example:worst_non_binary}
    Let $\mathcal{\hat{C}}$ be the regulation that only imposes the (pointwise) minimum contract $w_\varepsilon$, for a small $\varepsilon>0$, where $w_\varepsilon(y) = (y-\varepsilon)\mathds{1}_{\{y \in [y_1, y_2]\}}$, for some $0<y_1 <y_2<\bar{y}$. Then replacing the technology $\hat{\mathcal{T}} =(0, \{(0,\delta_y)\})$ for some $y\in [y_1, y_2]$ with its equivalent binary technology $\hat{\mathcal{T}}_b=(0, \{(0, B_y)\})$ decreases the firm's implementation cost from $y-\varepsilon$ to $0$.
\end{example}

This concludes Step 1, which reduces the regulator's infinite-dimensional delegation problem to the one-dimensional problem of finding the optimal MPR regulation. Furthermore, it also shows that it suffices to consider binary technologies to analyse worst-case scenarios. Still, Nature's problem for a given MPR regulation $\mathcal{C}_\ell$,
\begin{align}\label{eq:Nature'sproblem}
    WCR(\mathcal{C}_\ell) = \sup_{\text{binary } \mathcal{T}} R\left(\mathcal{C}_{\ell}, \mathcal{T}\right),
\end{align}
remains complicated. Nature can choose any production cost $k$ and any (compact) subset of the product space of efforts and binary distributions. The remainder of the proof proceeds in two steps. First, fixing an arbitrary MPR regulation, we construct candidate worst-case binary technologies and obtain a lower bound on its worst-case regret. The minimum of this family of lower bounds is thus a lower bound for the minmax regret problem \eqref{eq:minmax_regret} and is equal to $\bar{R}$. This proves \cref{theorem:main} (i). Second, we show that this lower bound is tight (i.e., it is also an upper bound) for the regulation $\mathcal{C}_\alpha^\star$, which establishes \cref{theorem:main} (ii).

\paragraph{Step 2: Regret-maximising binary technologies.} We construct two candidate worst-case binary technologies, differentiating the case where the firm does not hire the worker (\cref{lemma:lowerbound_no_production_2}) from the case where she does (\cref{lemma:lowerbound_surplus_extraction_2}). We thus obtain a lower bound on Nature's value \eqref{eq:Nature'sproblem}, for any $\mathcal{C}_\ell$ with\footnote{When $\ell \geq \frac{1}{2}$, \cref{lemma:lowerbound_no_production_1,lemma:binding_cases} show that the regulator's regret for the MPR regulation $\mathcal{C}_{\ell}$ exceeds the regret for $\mathcal{C}_{\frac{1}{2}}$. So, we can focus on MPR regulations $\mathcal{C}_{\ell}$ with $\ell\leq \frac{1}{2}$.} $\ell \leq \frac{1}{2}$, equal to the maximum of the regret generated by those two scenarios. Precisely, this lower bound is
\begin{align}\label{eq:lowerbound_MPR}
    \sup_{\mathcal{T}} R\left(\mathcal{C}_{\ell}, \mathcal{T}\right) \geq \max \left\{
    \alpha e^{\frac{2\ell -1}{1-\ell}} (1-\ell) \bar{y} , \
    \alpha e^{-\frac{1}{\alpha}} (1- \ell) \bar{y}
    \right\}.
\end{align}
The minimum of the lower bound in \eqref{eq:lowerbound_MPR} across $\ell \in [0,1/2]$ is by definition equal to $\bar{R}$. Hence, proving the inequality \eqref{eq:lowerbound_MPR} proves \cref{theorem:main} (i). 

We sketch the constructions that lead to the lower bound in \eqref{eq:lowerbound_MPR}. The details are in \cref{appendix:proof_lowerbound}. Since the ideas in both scenarios (production and no production) are similar, we focus on the production case. We fix an MPR regulation $\mathcal{C}_\ell$ with $\ell \leq 1/2$. \cref{lemma:support_simplification_production} (in \cref{app:supporting_results}), which we already used in Step 1, simplifies the search for worst-case scenarios: we can focus on binary technologies for which the firm extracts the entire surplus (i.e., $WS =0$), the production cost $k$ is $0$, and the firm implements the least productive action by offering the minimum contract $\ell$. We first fix the expected output of the least productive action to $a\bar{y}$ for some $a \in [0,1]$, which also fixes the firm's profit to $(1-\ell)a\bar{y}$. Remember from \cref{eq:proofsketch_simpleregret} that, holding profit fixed, the regulator's regret increases with the expected surplus $S^\star$ of the \eqref{eq:FB_program}-optimal action. To maximise the surplus $S^\star$ while making sure that the firm does not implement the \eqref{eq:FB_program}-optimal action, we must maximise the implementation costs of this action. To do so, we need to create potential deviations for the worker (i.e., actions) such that incentivising the \eqref{eq:FB_program}-optimal action is too costly for the firm. At the same time, each of these (deviating) actions must be less profitable than the least productive action for the firm (otherwise she would deviate); that is, their implementation costs must also be sufficiently high. Finally, all the incentive compatibility constraints must bind in order to maximise the expected surplus of the \eqref{eq:FB_program}-optimal action. As a result, we construct a binary technology with the following key features (see \cref{lemma:lowerbound_surplus_extraction_2} for its precise definition):
\begin{enumerate}[(i)]
    \item there is a continuum of binary actions indexed by $i \in [a\bar{y}, \bar{y}]$ such that $i$ is the expected output of action $i$ and $e_i$ its cost;
    \item the implementation costs are such that all the actions $i\in [a\bar{y}, \bar{y})$ can only be incentivised by a contract that yields exactly a profit $(1-\ell)a\bar{y}$ for the firm (while all contracts that implement action $\bar{y}$ yield at most a profit $(1-\ell)a\bar{y}$).
\end{enumerate}
\begin{figure}
    \centering
    \begin{tikzpicture}[scale=0.45]
	\draw[->] (-1,0) -- (11,0) node[right] {output};
	\draw[->] (0,-1) -- (0,11) node[above] {payment};
		
	\draw (0,0) node[below=3mm, left=1mm] {$0$};
	\draw (10,-0.1) -- (10,0.1) node[below=2mm] {$\bar{y}$};
	\draw (-0.1,10) -- (0.1,10) node[left=2mm] {$\bar{y}$};

        \coordinate (A) at (0, 0);
        \coordinate (TR) at (10,10);
        \coordinate (M1) at (5, 0); 
        \coordinate (L1) at (5, 2); 
        \coordinate (U1) at (5, 5); 
        
	\draw[dotted] (10,0) -- (10,10);
        \draw[dotted] (0,10) -- (10,10);

	\draw[black, thick] (0,0) -- (10,10); 

        \draw[thick, densely dashed] (0, 0) -- (10, 4);
        \draw[dotted] (0,4) -- (10,4);
        \draw (8,3.2) node[right, below] {$\ell$};

        \draw ($(M1) + (0 , -0.1)$) -- ($(M1) + (0 ,0.1)$) node[below=2mm] {$a \overline{y}$};
        \draw[decorate,decoration={brace,amplitude=5pt},xshift=4pt,yshift=0pt] (L1) -- (U1) node[midway, left=1mm] {$(1-\ell)a\bar{y}$}; 
	\draw ($(L1) + (-0.1, 0)$) -- ($(L1) + (0.1, 0)$); 
	\draw ($(U1) + (-0.1, 0)$) -- ($(U1) + (0.1, 0)$);
        \draw[decorate,decoration={brace,amplitude=5pt},xshift=4pt,yshift=0pt] (M1) -- (L1) node[black,midway,xshift=-5mm] {$e_{a\bar{y}}$}; 

        \draw[domain=5:10, smooth, variable=\x, thick] 
        plot ({\x}, {\x-2*ln(\x/5)-3}); 
        \draw (8,5.3) node[right] {$e_i$};

        \draw[decorate,decoration={brace,amplitude=5pt, mirror},xshift=4pt,yshift=0pt] (10, 5.6138) -- (10,10) node [black,midway,xshift=12mm] {$\bar{y}-e_{\bar{y}}$};
        
    \end{tikzpicture}
    \caption{Regret is equal to $\alpha (\bar{y}-e_{\bar{y}}) - (1-\ell)a\bar{y}$.}
    \label{fig:production}
\end{figure}
\cref{fig:production} depicts this technology. Point (ii) implies that the cost graph $\{(i, e_i)\colon i\in[a\bar{y},\bar{y}]\}$ satisfies the differential equation
\begin{equation}
    \frac{\partial e_i}{\partial i} =\frac{i-(1-\ell)a\bar{y}}{i}. \tag{D}\label{eq:DE_production}
\end{equation}
To understand this equation, note first that $\frac{i-(1-\ell)a\bar{y}}{i}$ is the slope of the linear contract such that the firm's profit is exactly $(1-\ell)a\bar{y}$ if the worker takes action $i$. Hence, point (ii) implies that this must be the only contract that incentivises action $i <\bar{y}$.But, the worker prefers action $i$ to $j < i$ (resp. $j>i$) under any contract $\tilde{\ell}$ if and only if $\tilde{\ell} > \frac{e_i-e_j}{i-j}$ (resp. $\tilde{\ell} < \frac{e_i-e_j}{i-j}$) --- that is, the marginal benefit of choosing $i$ instead of $j$ must exceed the marginal cost. Graphically, this means that the slope of the contract must be greater (resp. lower) than the slope of the segment between the points $(i,e_i)$ and $(j, e_j)$ if $j<i$ (resp. $j>i$). Thus, for the contract $\frac{i-(1-\ell)a\bar{y}}{i}$ to be the only one implementing action $i \in [a\bar{y}, \bar{y})$, this property must hold for every $j\neq i$, which graphically implies that the line with slope $\frac{i-(1-\ell)a\bar{y}}{i}$ passing by $(i, e_i)$ must be everywhere below the cost curve. Given that this has to be true for every $i \in [a\bar{y}, \bar{y})$, we obtain the differential equation \eqref{eq:DE_production}. Furthermore, as the worker's surplus is zero under the offered contract $\ell$ and the implemented action $a\bar{y}$, the cost curve must satisfy the initial condition $e_{a\bar{y}}=\ell\cdot a\bar{y}$, leading to a unique solution to this differential equation.

Since the cost $e_i$ increases less than one to one with the expected output $i$, action $(e_{\bar{y}}, \delta_{\bar{y}})$ is \eqref{eq:FB_program}-optimal. The regulator implements it by offering the contract $y\mapsto y$, giving the entire surplus to the worker. Since the firm implements action $a\bar{y}$ in the worst-case equilibrium, the regulator's regret equals $\alpha (\bar{y}-e_{\bar{y}}) - (1-\ell)a\bar{y}$. Maximising the regret over profit levels (i.e., the parameter $a$) gives a regret of $\alpha e^{-\frac{1}{\alpha}} (1-\ell)\bar{y}$ (see \cref{lemma:lowerbound_surplus_extraction_2}). 

Following a similar logic, we construct a technology for which the firm does not produce under the MPR regulation $\mathcal{C}_{\ell}$ that generates a regret of $\alpha e^{\frac{2\ell -1}{1-\ell}} (1-\ell) \bar{y}$ (see \cref{lemma:lowerbound_no_production_2}). Together, these two regrets lead to the lower bound in \cref{eq:lowerbound_MPR}.

\paragraph{Step 3: Optimal MPR regulation.} 

Direct computations show that the lower bound in \eqref{eq:lowerbound_MPR} is minimised when $\alpha e^{\frac{2\ell -1}{1-\ell}} (1-\ell) \bar{y} = \alpha e^{-\frac{1}{\alpha}} (1- \ell) \bar{y}$, that is, for $\ell_\alpha^\star = \frac{\alpha-1}{2\alpha-1}$. This minimum is by definition equal to $\bar{R}$. Thus, the final step shows that this lower bound is tight for $\ell_\alpha^\star$ (i.e., it is also an upper bound), which proves the optimality of $\mathcal{C}_\alpha^\star$ and completes the proof of \cref{theorem:main}. We again prove it by differentiating the no-production (\cref{lemma:boundonregretnoproduction}) and the production (\cref{lemma:boundonregretproduction}) cases.\footnote{We note that the proofs of \cref{lemma:boundonregretnoproduction,lemma:boundonregretproduction} work verbatim for any MPR regulation $\mathcal{C}_\ell$ with $\ell \leq 1/2$. This means that the lower bound in \eqref{eq:lowerbound_MPR} is the worst-case regret of $\mathcal{C}_\ell$.} We only sketch the argument for the production case, as the no-production case works similarly. We refer the readers to \cref{app:prooftheorem:optimalregulation} for the details. 

Following again \cref{lemma:support_simplification_production}, we can write the worst-case regret as in \eqref{eq:proofsketch_simpleregret}: $\alpha S^\star -\Pi$. To bound the regret, we thus bound $S^\star$, holding fixed the firm's profit $\Pi$. Then we maximise over profit levels and obtain the desired upper bound $\alpha e^{-\frac{1}{\alpha}}(1-\ell_\alpha^\star)\bar{y} = \bar{R}$. 

Fix an arbitrary technology $\mathcal{T}$, let $\Pi$ be the firm's profit and $(e_R, F_R)$ the \eqref{eq:FB_program}-optimal action whose expected output we denote $\mu_R$. Since the firm maximises her profit, the linear contract with slope $\frac{\mu_R - \Pi}{\mu_R} - \epsilon$ (if allowed) cannot implement $(e_R, F_R)$ for any $\epsilon>0$ (otherwise, the firm would implement $(e_R, F_R)$ and increase her profit). So, there must exist a less productive action $(e_1, F_1)$ with mean $\mu_1 < \mu_R$ such that the worker (weakly) prefers $(e_1, F_1)$ to $(e_R, F_R)$ when offered the linear contract with slope $\frac{\mu_R - \Pi}{\mu_R}$, which implies that:
\begin{align*}
    e_R - e_1 \geq \left(\frac{\mu_R - \Pi}{\mu_R} - \epsilon\right) \left(\mu_R - \mu_1\right).
\end{align*}
If the linear contract with slope $\frac{\mu_1 - \Pi}{\mu_1} - \epsilon$ is authorised, offering it cannot increase the firm's profit. So, there must again exist another less productive action $(e_{2},F_{2})$ with mean $\mu_2 < \mu_1$ that the worker (weakly) prefers over $(e_1,F_1)$ when offered the contract $\frac{\mu_1 - \Pi}{\mu_1}$. Proceeding inductively, we construct a chain of actions $\left(e_R, F_R\right) \to \left(e_1, F_1\right) \to \dots $ such that, for all $k$,
\begin{align}\label{eq:inequalitiessketch_2}
    e_{k-1} - e_k \geq \left(\frac{\mu_{k-1} - \Pi}{\mu_{k-1}} -\epsilon\right) \left(\mu_{k-1} - \mu_k\right),
\end{align}
until we reach an action $(e_N, F_N)$ such that the linear contract with slope $\frac{\mu_N - \Pi}{\mu_N}$ falls below the minimum contract $\ell^\star_\alpha$. In this case, we can take $(e_N, F_N)$ to be the action implemented by the firm (see \cref{lemma:boundonregretnoproduction} for details). The chain of inequalities in \cref{eq:inequalitiessketch_2} then yields a lower bound on the effort cost of the \eqref{eq:FB_program}-optimal action,
\begin{align}\label{eq:lowerboundeffort_claim2}
    e_R \geq \ell^\star_\alpha\cdot \mu_N +  \sum_{k\leq N} \frac{\mu_k - (1-\ell_\alpha^\star)\mu_N}{\mu_k} \left(\mu_{k-1} - \mu_k\right),
\end{align}
where $\Pi = (1-\ell_\alpha^\star)\mu_N$ and $e_N =\ell^\star_\alpha\cdot \mu_N$ because the firm offers the contract $\ell_\alpha^\star$ and the worker's surplus is zero (\cref{lemma:support_simplification_production}). This gives an upper bound on $S^\star$ for each $\mu_N$ and $\mu_R$. Maximising over $\mu_N$ and $\mu_R$ leads to the desired upper bound $\alpha e^{-\frac{1}{\alpha}} (1- \ell) \bar{y}$. 

\cref{lemma:boundonregretnoproduction} proceeds similarly to show that the regulator's regret is bounded above by $\alpha e^{\frac{2\ell^{\star}_\alpha -1}{1-\ell^{\star}_\alpha}} (1-\ell^{\star}_\alpha) \bar{y}$ for all technologies $\mathcal{T}$ such that firm produces under regulation $\mathcal{C}_\alpha^\star$.

\subsection{Characteristics of optimal regulations}\label{subsection:characteristics}

The optimal regulation $\mathcal{C}^\star_\alpha$ characterised in \cref{theorem:main} is not the only regulation that minimises the worst-case regret. For instance, the proof of \cref{theorem:main} shows that the linear regulation allowing all and only the linear contracts above $\ell^\star_\alpha$ is also optimal. Still, we argue that all optimal regulations share the key features of $\mathcal{C}^{\star}_{\alpha}$. Remember that we define $\underline{w}^\mathcal{C}$ as $\underline{w}^\mathcal{C}(y)=\min\{w(y) \mid w\in\mathcal{C}\}$ for any $y$, what we now refer to as the regulation's \emph{minimum guarantee}.
\begin{theorem}\label{theorem:optimalregulationnecessity}
    Any optimal regulation $\mathcal{C} \subset \mathbb{C}_0$ satisfies the following properties.\footnote{$\underline{conv}(w)(\cdot) = \sup \left\{g \colon g \text{ convex and } g\leq w\right\}$ stands for the convex envelope of $w$ and $\mathrm{Im}(\mathcal{C})(y) = \left\{ v \geq 0 \colon \exists w \in \mathcal{C} \text{ such that } w(y)=v \right\} $ is the image of the regulation.}
    \begin{enumerate}[(i)]
        \item (Protection vs. flexibility) \ For all $y\in [0, \bar{y}]$,
        \begin{equation*}
            \max\left\{0, y-(1-\ell^\star_\alpha)\bar{y}) \right\} \leq \underline{w}^{\mathcal{C}}(y) \leq \min\{y,\ell^\star_\alpha\cdot\bar{y}\};
        \end{equation*}
        and for all $y\geq \bar{y}$,
        \begin{equation*}
            \underline{w}^{\mathcal{C}}(y)=\ell^\star_\alpha\cdot y.
        \end{equation*}

        \item (Robust to gaming) \ For any $w\in \mathcal{C}$, if $w(y)=\ell_\alpha^\star\cdot y$ for some $y\geq \bar{y}$, then $\underline{conv}(w)(y) \geq \ell^\star_\alpha \cdot y$ for all $y\geq \bar{y}$.

        \item (Minimal flexibility) \ For all $y \geq \bar{y}$, $[\ell^\star_\alpha \cdot y, (1 - \rho^*)\cdot y] \subset \mathrm{Im}(\mathcal{C})(y)$, where $\rho^* = e^{-\frac{1}{\alpha}} (1-\ell^{\star}_{\alpha})$.
    \end{enumerate}
\end{theorem}

\begin{figure}
    \centering
    \begin{tikzpicture}[scale=0.5]
	\draw[->] (-1,0) -- (11,0) node[right] {output};
	\draw[->] (0,-1) -- (0,11) node[above] {payment};
		
	\draw (0,0) node[below=3mm, left=1mm] {$0$};
	\draw (7,-0.1) -- (7,0.1) node[below=2mm] {$\bar{y}$};
	\draw (-0.1,7) -- (0.1,7) node[left=2mm] {$\bar{y}$};

	\draw[dotted] (7,0) -- (7,7);
        \draw[dotted] (0,7) -- (7,7);

	\draw[] (0,0) -- (11,11); 
        \draw[decorate,decoration={crosses}] (0,0) -- (2.8, 2.8);
        \draw[line width=0.05pt] (0,0) -- (11, 4.4) node[right] {$\ell^\star_\alpha(y)$}; 
        \draw[decorate,decoration={crosses}] (2.8, 2.8) -- (7, 2.8);
        \draw[dash dot dot, line width = 2pt] (4.2,0) -- (7, 2.8);
        \draw[dash dot dot, line width = 2pt] (0,0) -- (4.2, 0);


        \begin{scope}
            \clip (7, 2.8) -- (4.2, 0) -- (0, 0) -- (2.8, 2.8);
            \fill[pattern=north west lines, pattern color=black] (7, 2.8) -- (4.2, 0) -- (0, 0) -- (2.8, 2.8) -- cycle;
        \end{scope}
        \begin{scope}
            \clip (7, 2.8) -- (11, 4.4) -- (11, 4/7*11) -- (7, 4) -- cycle;
            \fill[pattern=crosshatch dots, pattern color=black] (7, 2.8) -- (11, 4.4) -- (11, 4/7*11) -- (7, 4) -- cycle;
        \end{scope}
    \end{tikzpicture}
    \caption{Region of $\underline{w}^{\mathcal{C}}$ and minimal flexibility of optimal regulations.}
    \label{fig:optimal_regulation_necessity}
\end{figure}
Although the details of optimal regulations may vary, \cref{theorem:optimalregulationnecessity} shows that the fundamental elements that ensure optimality are invariant. Point (i) says that any optimal regulation must include a minimum guarantee for the worker, $\underline{w}^{\mathcal{C}}$, which cannot be ``too far'' from the optimal minimum linear contract $\ell^\star_\alpha$. Specifically, $\underline{w}^{\mathcal{C}}$ must lie in the hatched region of \cref{fig:optimal_regulation_necessity} for outputs below $\bar{y}$ and be equal to $\ell^\star_\alpha$ for outputs above $\bar{y}$. In addition, point (ii) states that optimal regulations do not permit Nature to circumvent this requirement by giving the firm a chance to offer less \textit{in expectation} than the required minimum guarantee $\ell^\star_\alpha$ for outputs above $\bar{y}$. Specifically, although an optimal regulation may allow contracts that pay the worker less than $\ell^\star_\alpha$ for outputs lower than $\bar{y}$, the lower convex envelope of such contracts must be at least $\ell^\star_\alpha \cdot y$ for all\footnote{The statement requires that $w(y)=\ell_\alpha^\star\cdot y$ for some $y\geq \bar{y}$ because a contract such that $conv(w) < \ell_\alpha^\star$ but that is `high' everywhere beyond $\bar{y}$, although it allows the firm to pay little the worker in expectation, it is hard to implement as it provides incentives to the agent to choose more productive actions.} $y\geq \bar{y}$. Point (iii) says that optimal regulations must afford the firm enough flexibility above the minimum guarantee by having contracts whose images cover the dotted region in \cref{fig:optimal_regulation_necessity}. Therefore, \cref{theorem:optimalregulationnecessity} shows that the key requirements of the optimal regulation $\mathcal{C}_\alpha^\star$ must be met by any optimal regulation: a minimum guarantee which is close to $\ell_\alpha^\star$ and enough flexibility for the firm to incentivise the worker.

\paragraph{Proof intuition.} To prove \cref{theorem:optimalregulationnecessity}, for each requirement, we construct technologies that generate a regret greater than $\bar{R}$ when the regulation does not satisfy it (see \cref{app:prooftheorem:optimalregulationnecessity}). If $\underline{w}^\mathcal{C}$ does not satisfy point (i) for some $y'$, then we can replicate the construction in \cref{lemma:lowerbound_no_production_2,lemma:lowerbound_surplus_extraction_2} to increase the regulator's regret beyond $\bar{R}$. Instead of constructing binary technologies where actions are supported on $\{0, \bar{y}\}$, as in Step 2 of the proof sketch in \cref{subsec:proof_sketch}, we construct a technology where each action is supported on $\{0,y'\}$. Depending on whether $\underline{w}^\mathcal{C}(y')$ is too high (i.e., above $\mathds{1}_{\{y'<\bar{y}\}} \ell_\alpha^\star\cdot \bar{y} + \mathds{1}_{\{y'\geq\bar{y}\}}\ell_\alpha^\star\cdot y'$) or too low (i.e., below $\mathds{1}_{\{y'<\bar{y}\}} (y'-(1-\ell_\alpha^\star) \bar{y}) + \mathds{1}_{\{y'\geq\bar{y}\}}\ell_\alpha^\star\cdot y'$), we respectively construct a technology such that firm does not produce and wastes production opportunity or such that the firm excessively distorts production and extracts surplus from the worker. 

The proof of point (ii) follows a similar logic. Suppose the lower convex envelope of some authorised contract $w$ is less than $\ell^\star_\alpha$ at some $y'\geq\bar{y}$. Thus, there exist $y_1$ and $y_2$ such that $y_1<y'< y_2$, $py_1+(1-p)y_2=y'$ for some $p\in (0,1)$, but $pw(y_1)+(1-p)w(y_2)<\ell^\star_\alpha\cdot y'$. We show that Nature can then augment the worst-case technology constructed in \cref{lemma:lowerbound_surplus_extraction_2} by adding a \emph{non-binary action} supported on $\{0, y_1, y_2\}$, allowing the firm to extract additional surplus from the worker and pushing the regulator’s regret beyond $\bar{R}$.

Finally, suppose that a regulation $\mathcal{C}$ does not satisfy point (iii). This means that there exists $y' \geq \bar{y}$ for which a non-empty open interval $(w_1, w_2)$ from $[\ell^\star_\alpha\cdot y', (1-\rho^\star)y']$ is not covered at all by any contract in $\mathcal{C}$ evaluated at\footnote{The fact that this is an open interval comes from the regulation $\mathcal{C}$ being closed.} $y'$. We exploit this additional constraint from the regulation to modify the construction in \cref{lemma:lowerbound_no_production_2} of a technology that prevents the firm from hiring the worker under $\mathcal{C}$. As we previously explained in \cref{subsec:proof_sketch}, the proof of \cref{lemma:lowerbound_no_production_2} constructs a technology with a continuum of actions supported on $\{0, \bar{y}\}$, such that each action can generate a profit of exactly zero, in the case where a regulation allows every linear contract above a minimum. As a consequence, the curve of the cost of the actions as a function of the average output was increasing smoothly, such that every small increase in the slope of a contract increases the average output of the incentivised action by a proportionally small amount. Here, our actions are supported on $\{0,y'\}$, and we exploit the fact that no marginal increase of slope is available for the slopes within $\left(\frac{w_1}{y'}, \frac{w_2}{y'}\right)$ to make the cost curve increase less. This results in a greater wasted surplus and thus a greater regret.

\subsection{Taking Stocks on Optimal Regulations}
\label{subsec:taking_stocks}

\subsubsection{The role of linear contracts}

\cref{theorem:main}, and in particular the crucial role played by binary technologies in its proof, raises a natural concern about the role of other output levels in our model. Indeed, if all actions were supported on $\{0,\bar{y}\}$ only, a similar regulation would be optimal, suggesting that the linearity of the minimum contract in our optimal regulation $\mathcal{C}_\alpha^\star$ might be artificial. \cref{theorem:optimalregulationnecessity} clears this concern and clarifies the benefits of linear contracts. 

First, it shows that output levels other than $\bar{y}$ cannot be ignored. As the proof of point (i) demonstrates, if the minimum guarantee is either too restrictive or too permissive at some output level $y'$, Nature can exploit it with actions supported on $\{0, y'\}$ to increase the regulator's regret. Similarly, the proof of point (iii) shows that insufficient flexibility at output levels different than $\bar{y}$ allows Nature to construct technologies with the corresponding binary support to generate a greater regret. Finally, point (ii) highlights the advantages of a linear minimum guarantee. Non-linear minimum guarantees leave scope for Nature to design \emph{non-binary} technologies whose implemented contracts pay the worker less than the minimum guarantee \emph{in expectation}. A linear minimum payment eliminates Nature's opportunity to circumvent the worker's protection with such riskier actions. This is why binary technologies are sufficient to maximise the regulator's regret under MPR regulations, but not under more general ones (see \cref{example:worst_non_binary} above). 

\cref{theorem:optimalregulationnecessity} deepens our economic understanding of the role of the two key features of $\mathcal{C}^{\star}_{\alpha}$: the minimum linear contract and complete flexibility above it. The minimum linear contract $\ell^\star_\alpha$ not only protects the worker by guaranteeing him a minimal share of the output, it also limits the firm's incentives to distort production to extract surplus and Nature's ability to lower the worker's expected payment. First, $\ell^\star_\alpha$ already provides the worker with incentives to choose more productive actions, which limits the firm's ability to distort production downward. Second, $\ell^\star_\alpha$ also restricts upward distortions: because the firm has to pay the worker a minimum share $\ell^\star_\alpha$, she cannot offer very convex contracts (e.g., bonus contracts) that cheaply incentivise over-production. So, the minimum linear contract curtails rent extraction by lowering the firm's potential gain from under- and overproduction. The firm implements an action closer to the \labelcref{eq:FB_program}-optimal one, which reduces the regulator's regret. Third, $\ell^{\star}_{\alpha}$ also prevents Nature from gaming the regulation by offering riskier technologies. In addition, the more flexible a regulation is (i.e., the more contracts it includes), the more actions the firm can implement, thus limiting the risk of wasting the entire surplus by impeding production. Since the minimum linear contract already realigns the firm's and regulator's motives, the regulator does not gain from further restricting the firm's autonomy. 

In line with this last economic interpretation, \cref{theorem:optimalregulationnecessity} implies that $\mathcal{C}^{\star}_{\alpha}$ is uniquely optimal among the subclass of \emph{minimal contract regulations}, that is, regulations that authorise all contracts above a minimum one.

\begin{definition}\label{def:minclosedregulation}
    A \emph{minimum contract regulation} $\mathcal{C} \subset \mathbb{C}_0 $ is equal to $\mathcal{C} = \left\{ w \in \mathbb{C}_0 \colon w \geq \underline{w}^\mathcal{C} \right\}$.
\end{definition}
\begin{corollary}\label{corollary:minimaloptimalregulation}
    $\mathcal{C}^{\star}_{\alpha}$ is the unique optimal minimum contract regulation.
\end{corollary}

\subsubsection{When is regulation beneficial?}

\cref{theorem:main,corollary:minimaloptimalregulation} also provides insights into the evolution of the optimal minimum contract regulation as a function of the regulator's redistributive weight $\alpha$.

\begin{corollary}\label{corollary:comparative_stats}
    \
    \begin{enumerate}[(i)]
        \item The optimal regulation $\mathcal{C}^\star_\alpha$ is more constraining as $\alpha$ increases: $\alpha \geq \alpha' \implies \mathcal{C}^\star_\alpha \subseteq \mathcal{C}^\star_{\alpha'}$.
        \item Laissez-faire (i.e., regulation $\mathbb{C}_0$) is optimal if and only if $\alpha =1$.
    \end{enumerate}
\end{corollary}
Point (i) indicates that the more the regulator values the worker surplus, the more willing they are to risk preventing production (when the surplus is small) in exchange for guaranteeing the worker a larger share of the surplus when production happens. This is reminiscent of the equity-efficiency trade-off in the minimum wage literature. While a higher minimum wage may distort the market and prevent firms from hiring, it ensures a greater salary for employed workers.

In the extreme case, when $\alpha=1$, increasing the firm's profit at the expense of the worker does not increase the regret. However, the firm's profit-maximising behaviour may. Distortions to extract the agent's incentive rent are costly for the regulator: they waste potential surplus. \cref{corollary:comparative_stats} (ii) reveals that the risk of pushing the firm out of business outweighs any potential gain from limiting these distortions. Indeed, any regulation reduces the firm's potential profit and, hence, jeopardises the firm's viability. When $\alpha =1$, mitigating this risk becomes the regulator's primary objective. The optimal policy, therefore, grants full flexibility to the firm. In other words, the information asymmetry between the regulator and the firm is too severe for the regulator to correct the inefficient distortions introduced by moral hazard. For any regulation, there is a scenario where the regulation prevents the firm from producing, and that scenario becomes the main concern of a regulator whose objective is to maximise total surplus. So, the regulator does not gain from regulating when they only care about efficiency ($\alpha=1$).

\section{Incorporating Additional Knowledge}
\label{sec:extension}

Our baseline model assumes that the regulator has no information on the firm’s technology beyond the uniform upper bound on expected output $\bar{y}$. This minimally informed regulator is a natural starting point. It allows us to isolate the regulator's fundamental trade-off between the worker's protection and the firm's flexibility and to derive a simple solution: a minimal piece-rate compensation for the worker. We now explore whether our insights are robust to the knowledge the regulator has. 

\subsection{MLRP-ranked actions}\label{subsec:MLRP}

The regulator may reasonably assumes that more costly actions are ``more productive.'' Moral-hazard models often capture this by ranking actions according to the Monotone Likelihood Ratio Property (MLRP) \citep[see, e.g.,][]{jewitt1988justifying,innes1990limited,jewitt2008moral}. Incorporating such knowledge restricts the set of feasible technologies to be\footnote{To allow for general distributions (e.g., mixtures of absolutely continuous and singular distributions), we work with the generalised definition of MLRP by \cite{athey2002monotone} (Definition A.1).} 
\begin{align*}
    \mathbb{T}^{MLRP} = \left\{ \mathcal{T} = \left(k, \mathcal{A} = \left\{(e_i,F_i) \right\}_{i \in I \subset \mathbb{R}_+}\right) \in \mathbb{T} \colon  i \geq j \Rightarrow e_i \geq e_j \text{ and } F_i \geq_{MLRP} F_j \right\}.
\end{align*}
The regulator then solves
\begin{align*}
    \underset{\mathcal{C}\subset \mathbb{C}_0 }{\inf} \, \underset{\mathcal{T}\in \mathbb{T}^{MLRP}}{\sup}\, R\left(\mathcal{C,T}\right).
\end{align*}
\begin{proposition}\label{prop:optimalregulationMLRP}
    The MPR regulation $\mathcal{C}^{\star}_{\alpha}$ minimises the regulator's regret when they know that $\mathcal{T} \in \mathbb{T}^{MLRP}$. 
\end{proposition}
\cref{prop:optimalregulationMLRP} (proved in \cref{app:proofprop:optimalregulationMLRP}) demonstrates that the minimum linear contract regulation $\mathcal{C}^{\star}_{\alpha}$ is still optimal when the regulator knows that the worker's actions are always ranked according to MLRP. It also implies that $\mathcal{C}^{\star}_{\alpha}$ is optimal for any intermediate levels of additional knowledge. In particular, $\mathcal{C}^{\star}_{\alpha}$ remains optimal if the regulator knows that actions are ranked according to average output or first-order stochastic dominance \citep[e.g.,][]{Georgiadis2024}. 
\begin{corollary}\label{corollary:optimalregulationadditionalknowledge}
    For any set of technologies $\mathbb{T}'$ such that $\mathbb{T}^{MLRP} \subset \mathbb{T}' \subset \mathbb{T}$, the regulation $\mathcal{C}^{\star}_{\alpha}$ minimises the regulator's regret when the regulator knows that $\mathcal{T} \in \mathbb{T}'$.
\end{corollary}

\subsection{Quantitative Knowledge}\label{subsec:QuantKnowledge}

\cref{prop:optimalregulationMLRP,corollary:optimalregulationadditionalknowledge} only incorporate the knowledge of ``qualitative'' features. They leave open the question of what happens when the regulator additionally knows ``quantitative'' features of the possible technologies. We show that further knowledge on actions' expected profitability and the firm's production cost does not alter the qualitative feature of optimal regulations. Formally, let $0 \leq \underline{k} \leq \bar{k}$, $\bar{k} > 0$, and $0\leq \underline{y} < \bar{y}$ and consider the subset of technologies\footnote{Note that when $\bar{k} = 0$, no optimal regulation exists under our tie-breaking rule. The regulator would like to force the firm to pay the worker $w(y)=y$. However, the firm is then indifferent between participating or not, and, in the worst-equilibrium chooses not to participate. Still, the regulator can approach the optimum by offering MPR regulations with minimum contract $\ell = 1-\epsilon$ for $\epsilon>0$ small.}
\begin{align*}
    \mathbb{T}^c = \left\{ \mathcal{T} = (k, \mathcal{A}) \in \mathbb{T} \ : \ k \in [\underline{k}, \bar{k}] \ \text{ and } \ \mu_F \in [\underline{y}, \bar{y}] \ \forall (e,F) \in \mathcal{A} \right\} \subset \mathbb{T}.
\end{align*}

\begin{proposition}\label{prop:optimalregulationquantitativeknowledge}
    An MPR regulation $\mathcal{C}_{\ell^{\star}}$ is optimal when the regulator knows that\footnote{In \cref{app:proofprop:optimalregulationquantitativeknowledge}, we obtain an implicit characterization of the optimal $\ell^{\star}$.} $\mathcal{T} \in \mathbb{T}^c$. 
\end{proposition}

\cref{prop:optimalregulationquantitativeknowledge} (proved in \cref{app:proofprop:optimalregulationquantitativeknowledge}) shows that the regulator's quantitative knowledge of the technologies does not alter the qualitative features of the optimal regulation. An MPR regulation remains optimal, although the slope of the minimum linear contract generally differs from $\ell^{\star}_{\alpha}$. In our baseline model, ${k}$ represents the total cost of employing a worker. Thus, a natural interpretation of the minimum production cost $\underline{k}$ is that of a statutory minimum wage: a flat minimum wage does not provide incentives to worker while it imposes to the firm a minimum fixed labour cost regardless of performance. \cref{prop:optimalregulationquantitativeknowledge} then implies that our regulation could complement flat minimum wage policies. For instance, the MPR regulation could govern the pay-for-performance clauses of contracts to ensure that performance-based pay aligns with the regulator's objectives, on top of existing minimum wage regulations.

Finally, the lower bounds on the regulator's regret when the regulator knows that $\mathcal{T} \in \mathbb{T}^c$, derived in \cref{lemma:lowerbound_no_production_2constrained} and \cref{lemma:lowerbound_surplus_extraction_2constrained} in \cref{app:proofprop:optimalregulationquantitativeknowledge}, are attained by the same families of technologies as in \cref{lemma:lowerbound_no_production_1,lemma:lowerbound_no_production_2,lemma:lowerbound_surplus_extraction_2}. In particular, the actions in the production set are ranked according to MLRP, ensuring that the optimal MPR regulation continues to be optimal when the regulator possesses \emph{both qualitative and quantitative} knowledge.

\begin{corollary}\label{corollary:optimalregulationadditionalqualandquantknowledge}
    The MPR regulation $\mathcal{C}_{\ell^\star}$ that is optimal for the set of technologies $\mathbb{T}^c$ remains optimal for any set of technologies $\mathbb{T}'$ such that $\mathbb{T}^{MLRP} \cap \mathbb{T}^c \subset \mathbb{T}' \subset \mathbb{T}^c$.
\end{corollary}

Taken together, \cref{prop:optimalregulationMLRP,prop:optimalregulationquantitativeknowledge,corollary:optimalregulationadditionalknowledge,corollary:optimalregulationadditionalqualandquantknowledge} reassure us that our key insight stems from economically plausible scenarii. MPR regulations are optimal because they solve the regulator's fundamental trade-off between flexibility and protection. 

\subsection{Robustness to Less Knowledge}\label{subsec:less_knowledge}

One may symmetrically worry that the regulator knows \emph{less} than our baseline model assumes. The question is then: how sensitive is our optimal regulation to the regulator's \emph{exact} knowledge of the maximal expected output $\bar{y}$? In other words, we would like to assess the ``robustness'' of our robust regulation. Following the approach proposed by \cite{Ball2024}, we examine how the regret guarantee and optimal regulations vary when we perturb the set of possible technologies around the knowledge $\bar{y}$.

\cref{theorem:main} provides a direct answer. The regulator's regret is continuous in $\bar{y}$. So, the effect of perturbations becomes arbitrarily small as those perturbations vanish. Our regulation passes the ``robust robustness test.'' It turns out that the optimal MPR regulation $\mathcal{C}^{\star}_{\alpha}$ identified in \cref{theorem:main} does even better than passing the test, as it is independent of $\bar{y}$. It remains \emph{exactly} optimal when we vary $\bar{y} \in \mathbb{R}_+$ and is robust to any possible misspecifications. Moreover, \cref{theorem:optimalregulationnecessity} establishes that $\mathcal{C}^{\star}_{\alpha}$ is the most flexible optimal regulation that is robust to any misspecification. Indeed, points (i) and (ii) imply that any regulation $\mathcal{C}$ optimal for all possible values of $\bar{y} \in \mathbb{R}_+$ must be more restrictive than $\mathcal{C}^{\star}_{\alpha}$, i.e., $\mathcal{C} \subset \mathcal{C}^{\star}_{\alpha}$.

\section{Regulations in practice}\label{subsec:reginpractice}

Our model speaks to economic relations where the involved parties attempt to mitigate moral hazard by using pay-for-performance compensation schemes and the party incentivising effort holds significant bargaining power. 

These characteristics are present in many labour interactions. Although employees typically receive a fixed salary, their contracts often also contain a variable component \citep[see][]{lucifora2015IZA,fuhrmans2024WSJ}. Variable pay is common among insurance and retailing salespersons \citep{coviello2022minimum}, real estate agents, or agriculture workers \citep{hill2018minimum,ku2022does}. Moreover, many people work for firms not as employees but as independent workers. In this case, their compensation often depends on performance indicators. Such pay-for-performance contracts have existed for a long time for freelancers (i.e., self-employed workers), and have become increasingly more prevalent with the growth of the platform economy (or gig economy).\footnote{As of August 2021, $16\%$ of Americans had earned money through a platform at least once \citep{anderson2021state_gigwork}. According to the European Parliament, 28.3 million people were working for digital labour platforms in the EU in 2022, and they expect this figure to rise to 43 million by the end of 2025 \citep{europarl2024gig}.} Indeed, the increasing number of digital platform workers lead the International Labour Organization to put  ``decent work in the platform economy'' on the agenda of the 113th International Labour Conference in 2025, and to recommend in their conclusions that ``each member should take measures to ensure that the remuneration, including when calculated on a piece-rate basis, which is due to digital platform workers under national laws, regulations, collective agreements or contractual obligations, is adequate'' \citep{ILO2025_DecentWorkPlatformEconomy}.

Regulators have thus debated about the appropriate regulatory framework for labour contracts based on performance. While some have advocated for reclassifying independent workers, particularly platform workers, as employees to better protect their economic interests, new legal frameworks have also been explicitly designed.\footnote{For instance, while U.S. politicians have voiced their concerns about misclassification of gig workers \citep{rosenberg2021WP}, Uber and Lyft won a legal battle in California to classify their workers as independent contractors \citep{nyt2024gigworker_decision}. See also \cref{footnote:UK_regulation} about the regulation in the UK and \cite{europarl2024gig} for the European law.} Our results can inform these debates. Specifically, if the regulators hold strictly redistributive preferences towards workers (that is, $\alpha>1$ in our model) and if minmax regret is an acceptable criterion, then \cref{theorem:main} justifies setting a simple minimum piece-rate compensation. This prescription aligns with several real-world regulations. 

For instance, the UK regulates ``output work'' --- defined as ``work that is paid only according to the number of things that a worker makes or tasks they perform,'' which includes, e.g., freelancers, pieceworkers, some production, manufacturing, and agricultural workers, data entry operators, sales worker, etc --- by mandating a \emph{fair} (linear) pay rate for output.\footnote{\label{footnote:UK_regulation}This fair rate is determined as follows \citep[see][]{UK2021}: ``Test some or all of the workers. The group you test must be typical of the whole workforce -- not just the most efficient or fastest ones. Work out how many pieces of work have been completed in a normal working hour. Divide this by the number of workers to work out the average rate. If the work changes significantly, do another test to work out the new average rate. It's not necessary to do another test if the same work is being done in a different environment, for example work previously done in a factory being done at home. Divide the average rate of work by 1.2 (this means new workers will not be disadvantaged if they're not as fast as the others yet). Divide the hourly minimum wage rate by that number to work out the fair rate for each piece of work completed.''} Notably, this regulation has applied to platform work. In North America, comparable regulations have been implemented to protect ``gig workers.'' New York City imposed a minimum pay rate per minute of trip time for app-based restaurant delivery workers \citep{NYC2023}. The Government of British Columbia set minimum per-engaged-minute and per-kilometre compensations \citep{BC2025}. Washington State instituted a minimal payment per course, mile and engaged minute for Uber and Lyft drivers \citep{shepardson2022}. In all these cases, firms voiced concerns that the regulations constrained their ability to tailor incentive schemes (e.g., use surge pricing or performance bonuses for completing more trips), potentially leading to reduced hiring and efficiency losses. These complaints are consistent with our results: the implemented regulations limit firms' flexibility, running the risk of reducing employment, to protect a fair share of the surplus for the employed workers.\footnote{To make the illustration complete, let us translate this application into the language of our paper. Platforms act as firms, and drivers as workers. Drivers generate revenue per trip through the prices paid by final consumers. The prices usually depend on the distance, the time, the demand and a flat fee. Platforms set compensation contracts based on this output. Therefore, the regulations that implement a minimum payment per mile or per minute are indeed imposing a minimum sharing rule of the final output. Drivers decide when to work, where to go, what ride to accept, and how long to drive, which affects the output. Moreover, drivers may have different cars and live in different locations, the prices of gas may vary across time, etc. So, the mapping from effort (i.e., drivers' decisions) to output is individual-specific and uncertain.}

Our results can also explain the collective agreements and sectoral regulations that govern specific industries. For instance, in the music industry, the US Copyright Royalty Board has negotiated a minimum share of streaming revenues for artists and a minimum per-subscriber payment, a settlement known as Phonorecords IV. In France, the ``Code de la propriété intellectuelle, Article L212-3,'' guarantees remuneration that is ``appropriate and \emph{proportionate} to the actual or potential economic value of the rights assigned, considering their contribution to the work.''\footnote{The authors' translation.} Again, our analysis provides a theoretical justification for implementing such a percentage-of-revenue floor on redistributive grounds.

Finally, one can view our optimal MPR regulation (in particular \cref{prop:optimalregulationquantitativeknowledge}) as a rationale for a minimum profit-sharing rule (or bonus) for employees \citep{atkinson72capital,Weitzman1986}. This is, for instance, a rule implemented in France since 1967 for firms with more than 100 employees,\footnote{The eligibility threshold was reduced to 50 employees in 1990, and to 11 employees in 2025.} and intended to align shareholders' and employees' goals to improve productivity.\footnote{\cite{Nimier-David2023} find evidence that the mandated profit-sharing increased the labour share at the firm level and primarily benefited low-skill workers.} Sometimes, profit-sharing rules do not come from regulation but instead from an agreement negotiated by unions (see, e.g., the agreements that the United Auto Workers union in the U.S. obtains for its workers).

Beyond labour contracts, our model could also apply to business-to-business relationships. Outsourcing and subcontracting often exhibit moral-hazard. For instance, evidence shows that outsourcing public service is frequently accompanied by quality deterioration \citep{Andersson19outsourcing}. In franchising arrangements, franchisors hold bargaining power while franchisees may shirk operational or quality standards \citep{Fan17financial}. In many of these cases, the incentive provider hold substantial bargaining power and the regulator may be willing to protect the party exerting effort. Our results suggests that imposing a minimum piece-rate could be appropriate.


\section{Concluding Remarks}

The robust approach taken in this paper prevents optimal regulations from depending too finely on the regulator's speculative probabilistic knowledge. Akin to \cite{carroll2015robustness}'s \citeyear{carroll2015robustness} solution to the moral hazard problem, our robust criterion exhibits the effectiveness of a simple regulatory instrument: a minimum piece-rate. This optimal MPR highlights the regulator's essential trade-off: it protects the worker in a way that limits the firm's benefits from distorting production, while offering the firm sufficient flexibility to respond to different technologies, reducing the risk of exit. 

Our analysis illustrates how contract regulation can serve fairness objectives. This instrument pertains to the class of \emph{predistributive} policies, that ``encourage a more equal distribution of economic power and rewards" \emph{ex-ante} \citep{Hacker2011}. However, regulators also use extensively \emph{redistributive} instruments --- notably taxation ---  that aim to equalise production benefits \emph{ex-post}. One reason we focus on contract regulation and shut down taxation is the lesser attention predistribution has received, despite evidence suggesting its central role in explaining Europe–US inequality differences \citep{Blanchet2022, Bozio2024}.\footnote{\cite{Kuziemko2023} provide evidence that predistribution is also politically important. They argue that the Democrats' shift from promoting predistribution to favouring redistribution in the 1970s explains why less-educated voters started turning their backs on their historical party in favour of the Republicans.} This paper thus contributes to the normative investigation of predistribution by examining a prevalent instrument: the regulation of contracts. In practice, pre- and re-distributive instruments interact, and understanding their relative benefits is a promising direction for future research.

\bibliographystyle{apalike}
\bibliography{Bibliography}

\appendix

\section{Proof of \cref{theorem:main}}\label{appendix:proof_theorem_main}

This section contains the main results to obtain \cref{theorem:main}. We commonly refer to supplementary results that are in \cref{app:supporting_results}. The specific supporting results for Step 1 of the proof sketch in \cref{subsec:proof_sketch} are in \cref{app:proof_step1}.

\subsection{Proof of \cref{theorem:main} (i)}\label{appendix:proof_lowerbound}

\cref{lemma:lowerbound_surplus_extraction_2,lemma:lowerbound_no_production_1,lemma:lowerbound_no_production_2} prove slightly more general results than the ones we refer to in the proof sketch (\cref{subsec:proof_sketch}), as those results are also used in the proof of \cref{theorem:optimalregulationnecessity}. Remember that for any regulation $\mathcal{C}$, we write $\underline{w}^{\mathcal{C}}(y') = \inf \left\{ w(y') \mid w \in \mathbb{C}_0 \right\}$.

\begin{lemma}\label{lemma:lowerbound_no_production_1}
    For any regulation $\mathcal{C} \subset \mathbb{C}_0 $,
    \begin{align*}
        WCR(\mathcal{C}) \geq \underset{0 \leq y \leq \bar{y}}{ \sup } \, \alpha \underline{w}^{\mathcal{C}}(y).
    \end{align*}
\end{lemma}
\begin{proof}
    Let $\mathcal{C} \subset \mathbb{C}_0 $. Suppose that the production set $\mathcal{A}$ contains a single action producing the output level $0\leq y' \leq \bar{y}$ with probability 1 at zero cost, $\mathcal{A} = \left\{\left(0, \delta_{\{ y' \}}\right)\right\}$, and that the production cost $k$ is such that $0\leq k < y'$. Given the technology $\mathcal{T}=\left(k, \mathcal{A} \right)$ and the regulation $\mathcal{C}$, the maximal \emph{profit} the firm can obtain is
    \begin{align*}
        y'- \underline{w}^{\mathcal{C}}(y') - k.
    \end{align*}
    If $k > y'- \underline{w}^{\mathcal{C}}(y')$, no contract is signed between the firm and the worker, as the firm strictly prefers her outside option to the best available contract in $\mathcal{C}$. The regret for the regulator is then
    \begin{align*}
        R(\mathcal{C,T}) = \alpha(y' -k),
    \end{align*}
    since the contract $y \mapsto (y - k)\mathds{1}_{\{y'=y\}}$ implements action $\left(0, \delta_{\{ y' \}}\right)$ and allocates the entire surplus to the worker. Taking the supremum over $k >y'-\underline{w}^{\mathcal{C}}(y')$, we obtain
    \begin{align*}
        WCR(\mathcal{C}) \geq \alpha \underline{w}^{\mathcal{C}}(y').
    \end{align*}
    Finally, we obtain the desired bound by taking the supremum over $0\leq y' \leq \bar{y}$.
\end{proof}

\begin{lemma}\label{lemma:lowerbound_no_production_2}
    For any regulation $\mathcal{C} \subset \mathbb{C}_0 $,
    \begin{align}\label{eq:lowerbound_no_production_2}
        WCR(\mathcal{C}) \geq \alpha \,  \underset{0\leq y \leq \bar{y}}{\sup } \,  e^{\frac{2\underline{w}^{\mathcal{C}}(y)- y}{y- \underline{w}^{\mathcal{C}}(y)}}(y-\underline{w}^{\mathcal{C}}(y)) \mathds{1}_{\left\{ \underline{w}^{\mathcal{C}}(y) \leq \frac{1}{2}y \right\}}.
    \end{align}
\end{lemma}
\begin{proof}
    Let $\mathcal{C} \subset \mathbb{C}_0$ be a regulation. We can assume that there exists $y'$ such that $\underline{w}^{\mathcal{C}}(y') \leq \frac{1}{2}y'$, as otherwise the result is vacuous as the regret is non negative by definition. 

To show that the right-hand side of \cref{eq:lowerbound_no_production_2} is a lower bound on the regulator's worst-case regret, it is enough to exhibit a technology that yields a regret at least equal to it. So, in the remainder of the proof, we construct such a technology. 

Let $k$ be (strictly) smaller than $ y' -\underline{w}^{\mathcal{C}}(y')$. Denote by $B_i$ the binary output distribution with support $\left\{0, y'\right\}$ and mean $i$. Define $\underline{i}=\displaystyle y' \frac{k}{y' - \underline{w}^{\mathcal{C}}(y')}$ and the compact action set $\mathcal{A}^k = \left\{ (e_i, {B}_{i}) \mid i\in I = \left[\underline{i} , \ y' \right] \right\}$,
where, for all $i\in I$,
\begin{align*}
    e_i = i - \frac{k y'}{y' - \underline{w}^{\mathcal{C}}(y')} + k \left( \ln \left(\frac{k y'}{y'-\underline{w}^{\mathcal{C}}(y')} \right) - \ln (i) \right).
\end{align*}
We consider the family of technologies $\mathcal{T}^k= (k, \mathcal{A}^k)$, $k < y' -\underline{w}^{\mathcal{C}}(y')$.

By definition of $e_i$, for all $w \in \mathcal{C}$,
\begin{align*}
    \frac{\partial}{ \partial i} \left( \ell^w \cdot i - e_i \right) = \ell^w - \frac{i - k}{i} \geq 0 \Leftrightarrow \ell^w \geq \frac{i - k}{i},
\end{align*}
where $\ell^w = \frac{w(y')}{y'}$. So, for $\underline{i}<i<y'$, $(e_i, B_{i})$ is implementable by $w \in \mathcal{C}$ if and only if
\begin{align*}
    w(y') = \frac{i - k}{i} y',
\end{align*}
by a simple first-order condition. Direct computations then shows that action $\underline{i}$ is only implementable by $w$ if $w(y') = \underline{w}^{\mathcal{C}}(y')$, and action $i = y'$ is implementable by $w \in \mathcal{C}$ if and only if
\begin{align*}
    w(y') \geq \frac{y' - k}{y'} y' = y'-k.
\end{align*}
Since $\frac{\partial}{\partial i} e_i \leq 1$, action $y'$ maximises total surplus. Moreover, any contract $w$ such that $w(y') = y'-k$ implements this action and gives the entire surplus to the worker and zero profit to the firm. Thus, the regulator offers such a contract to solve \eqref{eq:FB_program}, obtaining
\begin{align*}
    V(\mathcal{T}^k) = \alpha \left(y' - k - e_{y'}\right).
\end{align*}
On the other hand, observe that, for all $i\in [\underline{i}, y']$, any contract that implements $(e_i, B_i)$ gives a profit of at most zero to the principal. So, in the worst-case scenario for the regulator, the firm chooses her outside option. The regulator's regret is thus
\begin{align*}
    R\left(\mathcal{C,T}^k\right) = V(\mathcal{T}^k) = \alpha \left(y' - k - e_{y'}\right). 
\end{align*}
Plugging in the effort cost, the regret is equal to
\begin{align}
     \alpha \left(y' - k - e_{y'}\right) & =  \alpha \left(y' - k - y' + \frac{k y'}{y' - \underline{w}^{\mathcal{C}}(y')} - k \left( \ln \left(\frac{k y'}{y'-\underline{w}^{\mathcal{C}}(y')} \right) - \ln (y') \right) \right) \notag \\
     & = \alpha k \left(\frac{y'}{y'-\underline{w}^{\mathcal{C}}(y')} - \ln \left(\frac{k}{y' - \underline{w}^{\mathcal{C}}(y')} \right) -1 \right). \label{eq:lemma:lowerbound_surplus_extration_2}
\end{align}
Since the worst-case regret is obtained by maximising over all possible technologies,  taking the supremum of the above expression with respect to $k \in[0, y' - \underline{w}^{C}(y'))$ yields a lower bound on the worst-case regret. Taking derivatives with respect to $k$, we obtain
\begin{align*}
    & \frac{\partial}{\partial k} \left( \alpha k \left(\frac{y'}{y'-\underline{w}^{\mathcal{C}}(y')} - \ln \left(\frac{k}{y' - \underline{w}^{\mathcal{C}}(y')} \right) -1 \right) \right) \geq 0 \\
    \Leftrightarrow \, & \frac{y'}{y'-\underline{w}^{\mathcal{C}}(y')} - \ln \left(\frac{k}{y' - \underline{w}^{\mathcal{C}}(y')} \right) -2 \geq 0 \\
    \Leftrightarrow \, & k \leq e^{\frac{2 \underline{w}^{\mathcal{C}}(y') - y'}{y' - \underline{w}^{\mathcal{C}}(y')}} \left( y' - \underline{w}^{\mathcal{C}}(y') \right).
\end{align*}
So, $\alpha \left(y' - k - e_{y'}\right)$ is maximised over $k\in [0, y' - \underline{w}^{\mathcal{C}}(y')]$ at
\begin{align*}
    k^{\star} = e^{\frac{2 \underline{w}^{\mathcal{C}}(y') - y'}{y' - \underline{w}^{\mathcal{C}}(y')}} \left( y' - \underline{w}^{\mathcal{C}}(y') \right).
\end{align*}
It follows that the regulator's worst-case regret is bounded below by
\begin{align*}
    WCR(\mathcal{C}) = \underset{\mathcal{T}}{\sup} \, R\left(\mathcal{C,T}\right) \geq \underset{k \in [0, y' - \underline{w}^{\mathcal{C}}(y'))}{\sup} \,  R\left(\mathcal{C}, \mathcal{T}^k \right) = \alpha e^{\frac{2\underline{w}^{C}(y')-y'}{y'- \underline{w}^{\mathcal{C}}(y')}}(y'-\underline{w}^{C}(y')).
\end{align*}
Taking the supremum over $y'$ leads to the desired lower bound.
\end{proof}

\begin{lemma}\label{lemma:lowerbound_surplus_extraction_2}
    For any regulation $\mathcal{C} \subset \mathbb{C}_0 $,
    \begin{align}\label{eq:lowerbound_surplus_extraction_2}
        WCR(\mathcal{C}) \geq \alpha e^{-\frac{1}{\alpha}} \,  \underset{ 0\leq y \leq \bar{y}}{\sup } \, \left(y-\underline{w}^{\mathcal{C}}(y)\right) 
    \end{align}
\end{lemma}
\begin{proof}
    Let $\mathcal{C} \subset \mathbb{C}_0 $ be a regulation. As in the proof of \cref{lemma:lowerbound_no_production_2}, to show that the right-hand side of \cref{eq:lowerbound_surplus_extraction_2} is a lower bound on the regulator's worst-case regret, it is enough to exhibit a technology that yields a regret at least equal to it. So, in the remainder of the proof, we construct such a technology. 
    
    Let $\mu_F, y'$ such that $0 \leq \mu_F \leq y' \leq \bar{y}$, and $\Pi^{\mu_F} = \left(1-\frac{\underline{w}^{\mathcal{C}}(y')}{y'}\right) \mu_F$. We denote by $B_i$ the binary output distribution with support $\left\{0, y'\right\}$ and mean $i$ and we define the compact action set $\mathcal{A}^{\mu_F} = \left\{ (e_i, {B}_{i}) \, :\, i\in I = \left[\mu_F, y' \right] \right\}$, where, for all $i\in I$,
    \begin{align*}
        e_i = i - \mu_F + \Pi^{\mu_F} \left( \ln \left(\mu_F \right) - \ln i \right) + \frac{\underline{w}^{\mathcal{C}}(y')}{y'} \mu_F.
    \end{align*}
    We consider the family of technologies $\mathcal{T}^{\mu_F} = \left(0, \mathcal{A}^{\mu_F}\right)$, $\mu_F \leq y'$.

    By definition of $e_i$, for all $w \in \mathcal{C}$,
    \begin{align*}
        \frac{\partial}{ \partial i} \left( \ell^w i - e_i \right) = \ell^w - \frac{i - \Pi^{\mu_F}}{i} \geq 0 \Leftrightarrow \ell^w \geq \frac{i - \Pi^{\mu_F}}{i},
    \end{align*}
    where $\ell^w = \frac{w(y')}{y'}$. So, $(e_i, B_{i})$ with $i<y'$ is implementable by $w \in \mathcal{C}$ if and only if
    \begin{align*}
        w(y') = \frac{i - \Pi^{\mu_F}}{i} y',
    \end{align*}
    while action $i = y'$ is implementable by $w \in \mathcal{C}$ if and only if
    \begin{align*}
        w(y') \geq \frac{y' - \Pi^{\mu_F}}{y'} y'.
    \end{align*}
    
    The contract $w(y) =y$ implements the surplus maximising action $(e_{y'}, B_{y'})$ (since $\frac{\partial}{\partial i} e_i \leq 1$), and leaves no profit to the firm, hence allocating the entire surplus to the worker. So, it is the regulator's preferred contract and
    \begin{align*}
        V(\mathcal{T^{\mu_F}}) = \alpha \left(y' - e_{y'}\right).
    \end{align*}
    
    Next, observe that, for all $i$, any contract that implements $(e_i, B_i)$ gives a profit of at most $\Pi^{\mu_F}$ to the principal, since
    \begin{align*}
        \left(1- \frac{i-\Pi^{\mu_F}}{i} \right) \mathbb{E}_{B_i}\left[y\right] = \Pi^{\mu_F}.
    \end{align*}
    So, in the worst-case scenario, the firm implements the least productive action, generating the average output $\mu_F$. After simple calculations, the regulator's regret rewrites
    \begin{align*}
        R\left(\mathcal{C,T}^{\mu_F}\right) = \alpha \left(y' - e_{y'} - \left( \mu_F - e_{\mu_F}\right)\right) + (\alpha-1) \Pi^{\mu_F}. 
    \end{align*}
    Plugging in the effort cost, the regret is
    \begin{align*}
        \alpha \left(y' - e_{y'} - \left( \mu_F - e_{\mu_F}\right)\right) +(\alpha-1) \Pi^{\mu_F} & =  - \alpha \Pi^{\mu_F}\left( \ln \left(\mu_F \right) - \ln y' \right) + (\alpha-1) \Pi^{\mu_F} \\
        & = \alpha \int_{\mu_F}^{y'} \frac{1- \frac{\underline{w}^{\mathcal{C}}(y')}{y'}}{\mu}\mu_F d\mu + (\alpha -1) \left(1- \frac{\underline{w}^{\mathcal{C}}(y')}{y'}\right) \mu_F.
    \end{align*}
    Since the worst-case regret is obtained by maximising over all possible technologies, we can obtain a lower bound by taking the supremum of the above expression with respect to $\mu_F \in [0, y']$, which is achieved for $\mu_F = y' e^{-\frac{1}{\alpha}}$. It follows that the regulator's worst-case regret is bounded below by
    \begin{align*}
        WCR(\mathcal{C}) = \underset{\mathcal{T}}{\sup} \, R\left(\mathcal{C,T}\right) & \geq \underset{\mu_F \in [0, y']}{\sup} \,  R\left(\mathcal{C},\left(0, \mathcal{A}^{\mu_F} \right) \right) \\
        & =  \alpha e^{-\frac{1}{\alpha}} \left( y' - \underline{w}^{\mathcal{C}}(y') \right).
    \end{align*}
    Since $y'$ was arbitrary, the lower bound in \cref{lemma:lowerbound_surplus_extraction_2} follows.
\end{proof}

\begin{proof}[Proof of \cref{theorem:main} (i).] By \cref{lemma:lowerbound_surplus_extraction_2,lemma:lowerbound_no_production_1,lemma:lowerbound_no_production_2}, for all $\mathcal{C} \subset \mathbb{C}_0 $, $WCR(\mathcal{C})$ is bounded below by
\begin{align*}
     \underset{0\leq y \leq \bar{y}}{ \sup } \max \left\{ 
        \alpha \underline{w}^{\mathcal{C}}(y),  
        \alpha e^{\frac{2\underline{w}^{\mathcal{C}}(y)- y}{y- \underline{w}^{\mathcal{C}}(y)}}(y-\underline{w}^{\mathcal{C}}(y)) \mathds{1}_{\left\{ \underline{w}^{\mathcal{C}}(y) \leq \frac{1}{2}y \right\}}, 
        \alpha e^{-\frac{1}{\alpha}} \left(y-\underline{w}^{\mathcal{C}}(y)\right) \right\},
\end{align*}
which is bounded below by the function whose supremum is taken, evaluated at $\bar{y}$,
\begin{align*}
    \max \left\{ 
        \alpha \underline{w}^{\mathcal{C}}(\bar{y}),
        \alpha e^{\frac{2\underline{w}^{\mathcal{C}}(\bar{y})- \bar{y}}{\bar{y}- \underline{w}^{\mathcal{C}}(\bar{y})}}(\bar{y}-\underline{w}^{\mathcal{C}}(\bar{y})) \mathds{1}_{\left\{ \underline{w}^{\mathcal{C}}(\bar{y}) \leq \frac{1}{2}\bar{y} \right\}}, 
        \alpha e^{-\frac{1}{\alpha}} \left(\bar{y}-\underline{w}^{\mathcal{C}}(\bar{y})\right)\right\},
\end{align*}
which itself is greater than
\begin{align*}
    \min_{0\leq \underline{w} \leq \bar{y}} \max \left\{ 
        \alpha \underline{w},
        \alpha e^{\frac{2\underline{w}- \bar{y}}{\bar{y}- \underline{w}}}(\bar{y}-\underline{w}) \mathds{1}_{\left\{ \underline{w} \leq \frac{1}{2}\bar{y} \right\}}, 
        \alpha e^{-\frac{1}{\alpha}} \left(\bar{y}-\underline{w}\right) \right\}.
\end{align*}
Direct computations show that $ e^{\frac{2\underline{w}- \bar{y}}{\bar{y}- \underline{w}}}(\bar{y}-\underline{w}) = \underline{w}$ when $\underline{w}=\frac{1}{2}\bar{y}$. Then, applying \cref{lemma:binding_cases} in \cref{app:supporting_results}, we obtain that this equals $\bar{R}$, hence showing \cref{theorem:main} (i). 
\end{proof}

\subsection{Proof of \cref{theorem:main} (ii)}\label{app:prooftheorem:optimalregulation}

Since $\bar{R}$ is a lower bound on the regulator's regret for any regulation $\mathcal{C} \subset \mathbb{C}_0 $, we show the optimality of $\mathcal{C}^{\star}_{\alpha} = \left\{w \in \mathbb{C}_0  \mid w(y) \geq  \frac{\alpha-1}{2\alpha -1} \cdot y \text{ for all } y \geq 0 \right\} \subset \mathbb{C}_0 $ by proving that
\begin{align}\label{eq:WCRproofoptimalregulation}
    WCR\left(\mathcal{C}^{\star}_{\alpha}\right) \leq \bar{R}.
\end{align}
\cref{eq:WCRproofoptimalregulation} follows from:
\begin{align*}
    WCR\left(\mathcal{C}^{\star}_{\alpha}\right) & = \underset{\mathcal{T} \in \mathbb{T}}{\sup} \, R\left(\mathcal{C}^{\star}_{\alpha}, \mathcal{T}\right) \\
    & = \max\,\left\{ \underset{\mathcal{T} \in \mathbb{T} \colon \Pi\left(\mathcal{C}^{\star}_{\alpha}, \mathcal{T}\right) =0 }{\sup} \, R\left(\mathcal{C}^{\star}_{\alpha}, \mathcal{T}\right), \, \, \underset{\mathcal{T} \in \mathbb{T} \colon \Pi\left(\mathcal{C}^{\star}_{\alpha}, \mathcal{T}\right) >0 }{\sup}\, R\left(\mathcal{C}^{\star}_{\alpha}, \mathcal{T}\right) \right\} \\
    & \leq \max \left\{ e^{ \frac{2\ell^{\star}_\alpha\cdot\bar{y}-\bar{y}}{\bar{y} - \ell^{\star}_\alpha\cdot\bar{y}}}\left(\bar{y} - \ell^{\star}_\alpha\cdot\bar{y}\right), \, \alpha e^{-\frac{1}{\alpha}} \left(\bar{y} - \ell^{\star}_\alpha\cdot\bar{y}\right) \right\} \\
    & =\bar{R},
\end{align*}
where the inequality follows from \cref{lemma:boundonregretnoproduction,lemma:boundonregretproduction} below and the last equality follows from the definition of $\ell_\alpha^\star$. \hfill \qed

\begin{lemma}\label{lemma:boundonregretnoproduction}
    \begin{align}\label{eq:regret_claim_1}
       \underset{\mathcal{T} \in \mathbb{T} \colon \Pi\left(\mathcal{C}^{\star}_{\alpha}, \mathcal{T}\right) =0 }{\sup} \, R\left(\mathcal{C}^{\star}_{\alpha}, \mathcal{T}\right) \leq  \alpha e^{ \frac{2\ell^{\star}_\alpha\cdot\bar{y}-\bar{y}}{\bar{y} - \ell^{\star}_\alpha\cdot\bar{y}}}\left(\bar{y} - \ell^{\star}_\alpha\cdot\bar{y}\right).
    \end{align}
\end{lemma} 
\begin{proof}
Let $\mathcal{T} = \left(k , \mathcal{A} = \left\{ \left(e_i, F_i\right) \right\}_{i\in I} \right)$ be a technology such that $\Pi\left(\mathcal{C}^{\star}_{\alpha}, \mathcal{T}\right) =0$. For all $i\in I$, we write $\mu_i$ for $\mathbb{E}_{F_i}\left[y\right]$. Denote by $\left(e_R, F_R\right)$ the \eqref{eq:FB_program}-optimal action. Since $\Pi\left(\mathcal{C}^{\star}_{\alpha}, \mathcal{T}\right) =0$, we can assume that the firm takes her outside option, as it maximises regret among all possible contracting equilibria. We distinguish two cases:
\begin{enumerate}
    \item[a)] if $\mathbb{E}_{F_R} \left[ y \right] =\mu_R \leq \frac{k}{1-\ell^{\star}_\alpha}$, we show that
    \begin{align*}
        R\left(\mathcal{C}^{\star}_{\alpha}, \mathcal{T}\right) \leq \alpha \ell^{\star}_\alpha\cdot\bar{y};
    \end{align*}

    \item[b)] if $ \mu_R > \frac{k}{1-\ell^{\star}_\alpha}$, we show that
    \begin{align*}
        R\left(\mathcal{C}^{\star}_{\alpha}, \mathcal{T}\right) \leq \alpha e^{ \frac{2\ell^{\star}_\alpha\cdot\bar{y} - \bar{y}}{\bar{y}- \ell^{\star}_\alpha\cdot\bar{y}}}\left(\bar{y} - \ell^{\star}_\alpha\cdot\bar{y}\right).
    \end{align*}
\end{enumerate}
The claim then follows \cref{lemma:binding_cases} point 2., in \cref{app:supporting_results}, that shows
\begin{align*}
    \alpha \ell^{\star}_\alpha\cdot\bar{y} \leq \alpha e^{ \frac{2\ell^{\star}_\alpha\cdot\bar{y} - \bar{y}}{\bar{y}- \ell^{\star}_\alpha\cdot\bar{y}}}\left(\bar{y} - \ell^{\star}_\alpha\cdot\bar{y}\right),
\end{align*}

\noindent a) If $\mu_R \leq \frac{k}{1-\ell^{\star}_\alpha}$, then $\mu_R - k \leq \ell^{\star}_\alpha \mu_R$. Therefore, the regulator's regret is bounded above by
\begin{align*}
    R\left( \mathcal{C}^{\star}_{\alpha}, \mathcal{T} \right) & \leq \alpha \left( \mu_R -e_R -k\right) \\
    & \leq \alpha \left( \ell^{\star}_\alpha \mu_R - e_R\right) \\
    & \leq \alpha \ell^{\star}_\alpha\cdot\bar{y}.
\end{align*}

\noindent b) If $\mu_R > \frac{k}{1-\ell^{\star}_\alpha}$, then $\frac{\mu_R - k}{\mu_R} > \ell^{\star}_\alpha$. So, the linear contract $y\to 1- \frac{ k + \epsilon_1}{\mu_R} \cdot y$ belongs to the regulation $\mathcal{C}^{\star}_{\alpha}$ for $\epsilon_1>0$ small. Since $\left(\frac{\epsilon_1+k}{\mu_R}\mu_R\right) -k = \epsilon_1 >0$, but $\Pi\left(\mathcal{C}^{\star}_{\alpha}, \mathcal{T}\right) =0$, there exists an action $i_1 \in I$ that the worker prefers to action $R$, that is,
\begin{align*}
    & \left(1-\frac{\epsilon_1 + k}{\mu_R}\right) \mu_{i_1} - e_{i_1} \geq \left(1-\frac{\epsilon_1 + k}{\mu_R}\right) \mu_R -e_R, \\
    \text{which can be written} \quad &  \left(e_R -e_{i_1}\right) \geq  \left(1-\frac{\epsilon_1 + k}{\mu_R}\right) \left( \mu_R - \mu_{i_1} \right);
\end{align*}
that gives the firm a negative profit,
\begin{align*}
    & \left( 1-\frac{\mu_R -\epsilon_1 - k}{\mu_R} \right) \mu_{i_1} - k \leq 0, \quad \text{which implies that} \quad \mu_{i_1}<\mu_R.
\end{align*}
If there exist multiple such actions in $\mathcal{A}$, take $(e_{i_1}, F_{i_1})$ to be the action with the smallest mean among them.\footnote{We can always find such action as the set of ``deviating'' actions is a closed subset of the compact $\mathcal{A}$.} If $\mu_{i_1} \leq \frac{k}{1-\ell_\alpha^\star}$, we stop here. Otherwise, we proceed inductively, always selecting the smallest such action, to construct a sequence $\left\{i_j\right\}_{j\in \mathbb{N}}$ such that, for all $j \in \mathbb{N}$, $\mu_{i_{j+1}} < \mu_{i_{j}}$, $0<\epsilon_{{j+1}} \leq \frac{1}{2}\epsilon_{j}$, and 
\begin{align}\label{eq:inequalities_proof_claim_1}
    (e_{i_j} -e_{i_{j+1}}) \geq  \left(1- \frac{k + \epsilon_{i_j}}{\mu_{i_j}} \right) (\mu_{i_j} - \mu_{i_{j+1}}),
\end{align}
until we reach $N \in \mathbb{N}$ such that $\mu_N \leq \frac{k}{1-\ell_\alpha^\star}$. 

If no such $N$ exists, we obtain a sequence of actions, $(e_{i_n}, F_{i_{n}})_n$. By construction, $(e_{i_n}, \mu_{i_n}, \epsilon_{i_n})_n$ is decreasing. Since it is bounded below and $\mathcal{A}$ is compact, $(e_{i_n}, F_{i_{n}})_n$ converges to some action $(e_{i_{\infty}}, F_{i_\infty}) \in \mathcal{A}$ such that $\mu_{i_{\infty}} < \mu_{i_j}$ for all $j \in \mathbb{N}$. We claim that 
\begin{align*}
    \mu_{i_{\infty}}\leq \frac{k}{1-\ell^{\star}_\alpha}.
\end{align*}
Suppose by contradiction that $\mu_{i_{\infty}} > \frac{k}{1-\ell^{\star}_\alpha}$. Thus, the linear contract $y \to  \frac{\mu_{i_{\infty}} - \epsilon_{\infty}- k}{\mu_{i_{\infty}}} \mu_{i_{\infty}} y$ belongs to $\mathcal{C}^{\star}_{\alpha}$ for $\epsilon_{\infty} > 0$ small. Since $\Pi\left(\mathcal{C}^{\star}_{\alpha}, \mathcal{T}\right) =0$, there exists $i \in I$ such that 
\begin{align*}
    & \left( 1- \frac{\mu_{i_{\infty}} - \epsilon_{\infty}- k}{\mu_{i_{\infty}}} \right) \mu_{i} - k \leq 0 < \left( 1-\frac{\mu_{i_{\infty}} - \epsilon_{\infty}- k}{\mu_{i_{\infty}}} \right) \mu_{i_{\infty}} - k
    \quad \Rightarrow \quad  \mu_{i} < \mu_\infty,
\end{align*}
and
\begin{align*}
    & \frac{\mu_{i_{\infty}} - \epsilon_{\infty}- k}{\mu_{i_{\infty}}} \mu_{i} - e_{i} > \frac{\mu_{i_{\infty}} - \epsilon_{\infty}- k}{\mu_{i_{\infty}}} \mu_{i_{\infty}} -e_{i_{\infty}} \iff   e_{i_{\infty}} -e_{i} > \frac{\mu_{i_{\infty}} - \epsilon_{\infty}- k}{\mu_{i_{\infty}}} \left(  \mu_{i} - \mu_{i_{\infty}} \right).
\end{align*}
But, this contradicts the convergence of the sequence $(e_{i_j}, F_{i_j})_{j\in \mathbb{N}}$ to $(e_{i_{\infty}}, F_{i_\infty})$. 

Summing the inequalities in \cref{eq:inequalities_proof_claim_1} over the $j$'s, we obtain
\begin{align*}
    -e_R \leq -e_R + e_{i_{\infty}} & = - \sum_{j=0}^{\infty} (e_{i_j} - e_{i_{j+1}}) \\
    & \leq \epsilon_1 - \sum_{j=0}^{\infty} \left(1- \frac{k}{\mu_{i_j}} \right) (\mu_{i_j} - \mu_{i_{j+1}}) \\
    & \leq \epsilon_1 - \int_{\frac{k}{1-\ell^{\star}_\alpha}}^{\mu_R} \left(1-\frac{k}{\mu}\right)d\mu.
\end{align*}
Since $\epsilon_1>0$ was arbitrary, it follows that
\begin{align}
    R\left(\mathcal{C}^{\star}_{\alpha}, \mathcal{T}\right) & = \alpha \left( \mu_R - e_R -k \right) \notag \\
    & \leq \alpha \left( k\frac{\ell^{\star}_\alpha}{1-\ell^{\star}_\alpha} - k \left( \ln \left( \frac{k}{1- \ell^{\star}_\alpha} \right) - \ln \left( \mu_R\right) \right) \right). \label{eq:proof_claim_1_upperbound}
\end{align}
The supremum of the upper bound in \cref{eq:proof_claim_1_upperbound} over $\mu_R\in [0, \bar{y}]$ and $k \in \left[0, \mu_R \left(1 - \ell^{\star}_\alpha\right) \right]$ is obtained for $\mu_R^* = \bar{y}$ and
\begin{align}\label{eq:k^*}
    k^* = e^{ \frac{2\ell^{\star}_\alpha\cdot\bar{y}-\bar{y}}{\bar{y} - \ell^{\star}_\alpha\cdot\bar{y}}}\left(\bar{y} - \ell^{\star}_\alpha\cdot\bar{y}\right).
\end{align}
It follows that
\begin{align*}
    R(\mathcal{C}^{\star}_{\alpha}, \mathcal{T}) & \leq \alpha e^{ \frac{2\ell^{\star}_\alpha\cdot\bar{y}-\bar{y}}{\bar{y} - \ell^{\star}_\alpha\cdot\bar{y}}}\left(\bar{y} - \ell^{\star}_\alpha\cdot\bar{y}\right) \left( \frac{\ell^{\star}_\alpha}{1-\ell^{\star}_\alpha} - \left( \ln \left( \frac{e^{ \frac{\ell^{\star}_\alpha\cdot\bar{y}}{\bar{y}-\ell^{\star}_\alpha\cdot\bar{y}} - 1}\left(\bar{y} - \ell^{\star}_\alpha\cdot\bar{y}\right)}{1- \ell^{\star}_\alpha} \right) - \ln \left( \bar{y}\right) \right) \right) \\
    & = \alpha e^{ \frac{2\ell^{\star}_\alpha\cdot\bar{y}-\bar{y}}{\bar{y} - \ell^{\star}_\alpha\cdot\bar{y}}}\left(\bar{y} - \ell^{\star}_\alpha\cdot\bar{y}\right).
\end{align*}
This concludes the proof of \cref{lemma:boundonregretnoproduction}.
\end{proof}

\begin{lemma}\label{lemma:boundonregretproduction}
    \begin{align}\label{eq:regret_claim_2}
        \underset{\mathcal{T} \in \mathbb{T} \colon \Pi\left(\mathcal{C}^{\star}_{\alpha}, \mathcal{T}\right) >0 }{\sup}\, R\left(\mathcal{C}^{\star}_{\alpha}, \mathcal{T}\right) \leq e^{- \frac{1}{\alpha}}\left(\bar{y} - \ell^{\star}_\alpha\cdot\bar{y}\right) = \bar{R}.
    \end{align}
\end{lemma}
\begin{proof}
Let $\mathcal{T} = \left(k , \mathcal{A} = \left\{ \left(e_i, F_i\right) \right\}_{i\in I} \right)$ be a technology such that $\Pi\left(\mathcal{C}^{\star}_{\alpha}, \mathcal{T}\right) >0$. For all $i\in I$, we write $\mu_i$ for $\mathbb{E}_{F_i}\left[y\right]$. Denote by $\left(e_R, F_R\right)$ the \eqref{eq:FB_program}-optimal action, and by $\left(e_F, F_F\right)$ the action optimally implemented by the firm in the regret-maximising equilibrium. By \cref{lemma:support_simplification_production} (in \cref{app:supporting_results}), we can assume that (i) $k=0$, (ii) the firm optimally implements $\left(e_F, F_F\right)$ by offering the minimum linear contract $\ell^{\star}_{\alpha}$, and (iii) the worker's surplus $\mathbb{E}_{F_F}\left[ \ell^{\star}_{\alpha} \cdot y \right] - e_F$ is zero. We show that
\begin{align}\label{eq:proofclaim2_upperbound}
    R\left(\mathcal{C}^{\star}_{\alpha}, \mathcal{T}\right) \leq \max \, \left\{ (\alpha-1)
    \left(\bar{y} - \ell^{\star}_\alpha\cdot\bar{y}\right), \, \alpha e^{-\frac{1}{\alpha}} \left(\bar{y} - \ell^{\star}_\alpha\cdot\bar{y}\right) \right\}.
\end{align}
\cref{lemma:claim2inequality} (in \cref{app:supporting_results}) shows that $(\alpha-1)
\left(\bar{y} - \ell^{\star}_\alpha\cdot\bar{y}\right) \leq \alpha e^{-\frac{1}{\alpha}} \left(\bar{y} - \ell^{\star}_\alpha\cdot\bar{y}\right)$. Therefore, \cref{lemma:boundonregretproduction} follows from \cref{eq:proofclaim2_upperbound}.

To prove \eqref{eq:proofclaim2_upperbound}, first observe that the worker's incentive compatibility constraint \eqref{eq:IC_W} implies that, for all $(e, F) \in \mathcal{A}$, 
\begin{align*}
    \mathbb{E}_F\left[ \ell^{\star}_{\alpha} \cdot y \right] - e \leq \mathbb{E}_{F_F}\left[ \ell^{\star}_{\alpha} \cdot y \right] - e_F = 0.
\end{align*}
In particular, 
\begin{align}\label{eq:ICproofpropsublinearregulations}
    e_R \geq \mathbb{E}_{F_R} \left[ \ell^{\star}_{\alpha}\cdot y\right] = \ell^{\star}_{\alpha} \cdot \mu_R.
\end{align}
Since $k=0$, the regulator offers the contract $y\mapsto y$ in \eqref{eq:FB_program}, hence $(e_R,F_R)$ maximises total surplus. So, $\mu_F - e_F \leq \mu_R -e_R$. 

\noindent a) Suppose first that $\mu_R -e_R = \mu_F -e_F$. \cref{eq:ICproofpropsublinearregulations} then implies that $\mu_R -e_R \leq \mu_R - \mathbb{E}_{F_R} \left[ \ell^{\star}_{\alpha} \cdot y\right] = (1-\ell^\star_\alpha)\mu_R$. The regulator's regret is then bounded above by
\begin{align*}
    R\left(\mathcal{C}^{\star}_{\alpha}, \mathcal{T}\right) = (\alpha -1) \left(\mu_R - e_R \right) & \leq (\alpha-1) (1-\ell^\star_\alpha)\mu_R \\
    & \leq \left(\alpha -1\right) (1-\ell^\star_\alpha)\bar{y},
\end{align*}
where the last inequality follows since $\ell^{\star}_{\alpha}\leq 1$ and $\mu_r\leq \bar{y}$.

\noindent b) Suppose now that the firm does not implement an efficient action: $\mu_R -e_R > \mu_F -e_F$. Then $\mu_F < \mu_R$, since $e_R \geq\ell^{\star}_{\alpha} \cdot \mu_R$ and $e_F = \ell^{\star}_{\alpha} \cdot \mu_F$. The regulator's regret is
\begin{align}\label{eq:proofclaim2_regret}
    R\left(\mathcal{C}^{\star}_{\alpha}, \mathcal{T}\right) = \alpha \left( \mu_R - e_R \right) - \left(1- \ell^{\star}_\alpha\right) \mu_F.
\end{align}
\textbf{Claim:} 
\begin{align*}
    e_F - e_R \leq - \int_{\mu_R}^{\mu_F} \left(1 - \frac{1-\ell^\star_\alpha}{\mu}\mu_F \right) d\mu.
\end{align*}
\begin{proof}[Proof of the Claim]
    Given that $\ell^\star_\alpha <1$ (as otherwise the firm would not produce), for all $\mu > \mu_F$, $1- \frac{1-\ell^{\star}_\alpha}{\mu} \mu_F > \ell^{\star}_\alpha$. In particular, given that $\mu_R > \mu_F$, this implies that the linear contract 
\begin{align*}
    y\mapsto \left(1- \epsilon_R - \frac{1 -\ell^{\star}_\alpha}{\mu_R} \mu_F \right) \cdot y
\end{align*}
belongs to the regulation $\mathcal{C}^{\star}_{\alpha}$ for $\epsilon_R>0$ small. Firm's optimality implies that offering this linear contract cannot increase the firm's profit. So, it cannot implement $(e_R, F_R)$, and there exists an action $(e_1, F_1) \in \mathcal{A}$ with $\mu_{1} < \mu_R$ such that 
\begin{align*}
   & \left( 1 - \epsilon_R - \frac{1-\ell^{\star}_\alpha}{\mu_R} \mu_F\right) \mu_1 - e_1 > \left( 1- \epsilon_R - \frac{1-\ell^{\star}_\alpha}{\mu_R} \mu_F\right) \mu_{R} - e_{R} \\
   \Leftrightarrow \, \, &  e_{R} - e_1 > \left( 1- \epsilon_R - \frac{1-\ell^{\star}_\alpha}{\mu_R} \mu_F\right) \left(\mu_R - \mu_{1}\right).
\end{align*}
If $\mu_1 \leq \mu_F$, then the above also holds for the profit-maximising action $(e_F, \mu_F)$ (as $(e_F, F_F)$ is incentivised with the minimum contract $\ell_\alpha^\star$). In that case, choose $(e_1, F_1) = (e_F, F_F)$ and stop here. If, instead, $\mu_1 > \mu_F$, we proceed inductively. Since $\mu_1 > \mu_F$, the linear contract 
\begin{align*}
    y\mapsto \left(1- \epsilon_1 - \frac{1 -\ell^{\star}_\alpha}{\mu_1} \mu_F \right) \cdot y
\end{align*}
belongs to the regulation $\mathcal{C}^{\star}_{\alpha}$ for $\frac{1}{2} \epsilon_R \geq \epsilon_1>0$ small. Firm's optimality implies that offering this linear contract cannot increase the firm's profit. So, there exists an action $(e_2, F_2) \in \mathcal{A}$ with $\mu_{2} < \mu_1$ such that 
\begin{align*}
   & \left( 1 - \epsilon_1 - \frac{1-\ell^{\star}_\alpha}{\mu_1} \mu_F\right) \mu_2 - e_2 > \left( 1- \epsilon_1 - \frac{1-\ell^{\star}_\alpha}{\mu_1} \mu_F\right) \mu_{1} - e_{1} \\
   \Leftrightarrow \, \, &  e_{1} - e_2 > \left( 1- \epsilon_1 - \frac{1-\ell^{\star}_\alpha}{\mu_1} \mu_F\right) \left(\mu_1 - \mu_{2}\right).
\end{align*}
If $\mu_1 \leq \mu_F$, then the above also holds for the profit maximising action $(e_F, \mu_F)$. In that case, choose $(e_1, F_1) = (e_F, F_F)$ and stop here. If, instead, $\mu_2 > \mu_F$, we continue our construction until we can choose $(e_N, F_N) = (e_F, F_F)$. So, we obtain a sequence $\left\{\left(e_j, F_j\right)\right\}_{0\leq j \leq N \leq \infty}$, where $(e_0, F_0) = (e_R,F_R)$,  such that, for all $1 \leq j\leq N-1$, $\mu_{j+1} < \mu_j$ and
\begin{align}\label{eq:inequalities_proof_claim_2}
    e_{j} - e_{j+1} > \left( 1- \epsilon_1 - \frac{1-\ell^{\star}_\alpha}{\mu_j} \mu_F\right) \left(\mu_j - \mu_{j+1}\right),
\end{align}
for some $0 < \epsilon_j < \frac{1}{2}\epsilon_{j-1}$. 

If $N = \infty$, we obtain a sequence of actions, $(e_{i_n}, F_{i_{n}})_n$. By construction, $(e_{i_n}, \mu_{i_n})_n$ is decreasing. Since it is bounded below and $\mathcal{A}$ is compact, $(e_{i_n}, F_{i_{n}})_n$ converges to some action $(e_{i_{\infty}}, F_{i_\infty}) \in \mathcal{A}$ such that $\mu_F \leq \mu_{\infty} \leq \mu_{j}$ for all $j \geq 0$. 

We show that $\mu_{\infty} = \mu_F$. Suppose by contradiction that $\mu_{\infty} > \mu_F$. Then the linear contract 
\begin{align*}
    y \mapsto \left( 1- \epsilon_{\infty} - \frac{1-\ell^{\star}_\alpha}{\mu_{\infty}} \mu_F\right) \cdot y
\end{align*}
belongs to $\mathcal{C}^{\star}_{\alpha}$ for $\epsilon_{\infty}>0$ small. So, there exists $m \in I$ such that $\mu_m < \mu_{\infty}$ and
\begin{align*}
    \left( 1- \epsilon_{\infty} -  \frac{1-\ell^{\star}_\alpha}{\mu_{\infty}} \mu_F\right) \mu_{m} - e_{m} > \left( 1- \epsilon_{\infty} - \frac{1-\ell^{\star}_\alpha}{\mu_i} \mu_P\right) \mu_{\infty} - e_{\infty}.
\end{align*}
But, then, the sequence $\left\{\left(e_j, F_j\right)\right\}_{0\leq j \leq \infty}$ does not converge to $(e_{\infty}, F_{\infty})$ : a contradiction. So, $\mu_F = \mu_{\infty}$. Given that the worker chooses action $(e_F, F_F)$ with contract $\ell_\alpha^\star$, this implies that  $e_{F} \leq e_{\infty}$. So, $e_F - e_R \leq e_{\infty} - e_R$. 

Multiplying the inequalities in \cref{eq:inequalities_proof_claim_2} by $-1$ and summing over the $j$'s, we have
\begin{align*}
    e_{F} - e_R \leq e_\infty-e_R & = \sum_{j=0}^{N} \left( e_{j+1} - e_{j}\right) \\
    & \leq \epsilon_R (\mu_R - \mu_F) - \sum_{j=0}^{\infty} \left( 1- \frac{1-\ell^{\star}_\alpha}{\mu_{j}} \mu_F\right) \left(\mu_{j} - \mu_{j+1}\right) \\
    & = \epsilon_R (\mu_R - \mu_F)  - \sum_{j=0}^{N} \int_{\mu_{j+1}}^{\mu_{j}} \left(1- \frac{1-\ell^{\star}_\alpha}{\mu_{j}} \mu_F\right) d\mu \\
    & \leq \epsilon_R (\mu_R - \mu_F)  - \sum_{j=0}^{N} \int_{\mu_{j+1}}^{\mu_{j}} \left(1- \frac{1-\ell^{\star}_\alpha}{\mu} \mu_F\right) d\mu \\
    & = \epsilon_R (\mu_R - \mu_F) - \int_{\mu_F}^{\mu_{R}} \left(1- \frac{1-\ell^{\star}_\alpha}{\mu} \mu_F\right) d\mu.
\end{align*}
Since $\epsilon_R$ was arbitrary, this proves the claim. 
\end{proof} 

As a result, we can bound the regret in \cref{eq:proofclaim2_regret} from above:
\begin{align*}
    R\left(\mathcal{C}^{\star}_{\alpha}, \mathcal{T}\right) & \leq \alpha \left( \mu_R - \mu_F - \int_{\mu_F}^{\mu_{R}} \left(1- \frac{1-\ell^{\star}_\alpha}{\mu} \mu_F\right) d\mu\right) + (\alpha -1) \left( 1-\ell^{\star}_\alpha \right) \mu_F \\
    & \leq \alpha \int_{\mu_F}^{\mu_{R}} \frac{1-\ell^{\star}_\alpha}{\mu} \mu_F d\mu + (\alpha -1) \left(1-\ell^{\star}_\alpha \right) \mu_F.
\end{align*}
Maximising the right-hand side with respect to $\mu_F\leq \mu_R$, we obtain
\begin{align*}
    R\left(\mathcal{C}^{\star}_{\alpha}, \mathcal{T}\right) \leq \alpha e^{-\frac{1}{\alpha}} (\bar{y} - \ell^{\star}_\alpha\cdot\bar{y}).
\end{align*}

Combining cases a) and b) gives \cref{eq:proofclaim2_upperbound}, which concludes the proof.
\end{proof}

\section{Proof of \cref{theorem:optimalregulationnecessity}}\label{app:prooftheorem:optimalregulationnecessity}

\begin{proof}[Proof of point (i)]
    We prove the contrapositive. Let $\mathcal{C} \subset \mathbb{C}_0 $. We show that, if
    \begin{equation*}
        \max\left\{0, y-(\bar{y}-\ell^\star_\alpha\cdot\bar{y}) \right\} > \underline{w}^{\mathcal{C}}(y)  \text {  or  } \underline{w}^{\mathcal{C}}(y) > \min\{y,\ell^\star_\alpha\cdot\bar{y}\}
    \end{equation*}
    for some $y$, then $\mathcal{C}$ is not optimal.
    
    First, \cref{lemma:lowerbound_surplus_extraction_2,lemma:lowerbound_no_production_1,lemma:lowerbound_no_production_2} imply that, if $\underline{w}^{\mathcal{C}}(y) < (y-(\bar{y}-\ell^{\star}_\alpha\cdot\bar{y}))^+$ or $\underline{w}^{\mathcal{C}}(y) > \min\{y,\ell^{\star}_\alpha\cdot\bar{y}\}$ for some $0 \leq y < \bar{y}$, then 
    \begin{align*}
        WCR\left(\mathcal{C}\right) > \bar{R},
    \end{align*}
    Hence, $\mathcal{C}$ is not optimal. 
    
    Second, \cref{lemma:maximalminimalwageabovebary} deals with the case where $y > \bar{y}$.
\begin{lemma}\label{lemma:maximalminimalwageabovebary}
    If $\underline{w}^{\mathcal{C}}(y) \neq \frac{\alpha-1}{2\alpha-1} y$ for some $y>\bar{y}$, then $\mathcal{C}$ is not optimal.
\end{lemma}

\begin{proof}
    Let $\mathcal{C} \subset \mathbb{C}_0 $ be a regulation such that $\underline{w}^{\mathcal{C}}(y') > \frac{\alpha-1}{2\alpha-1} y'$ for some $y'>\bar{y}$. We show that $WCR(\mathcal{C}) > \bar{R}$. The proof adapts the argument made in the proof of \cref{lemma:lowerbound_no_production_2}. Assume that the production cost $k$ is such that $ k< \bar{y} -\frac{\underline{w}^{\mathcal{C}}(y')}{y'} \bar{y}$. Denote by $B_i$ the binary output distribution with support $\left\{0, y'\right\}$ and mean $i$. Define the production set $\mathcal{A}^k$ as
    \begin{align*}
        \mathcal{A}^k = \left\{ (e_i, {B}_{i}) \, :\, i\in I = \left[\bar{y} \frac{k}{\bar{y} -\frac{\underline{w}^{\mathcal{C}}(y')}{y'} \bar{y}} , \bar{y} \right] \right\},
    \end{align*}
    where, for all $i\in I$,
    \begin{align*}
        e_i = i - \frac{k \bar{y}}{\bar{y} -\frac{\underline{w}^{\mathcal{C}}(y')}{y'} \bar{y}} + k \left( \ln \left(\frac{k \bar{y}}{\bar{y} -\frac{\underline{w}^{\mathcal{C}}(y')}{y'} \bar{y}} \right) - \ln i \right).
    \end{align*}
    By definition of $e_i$, for all $w \in \mathcal{C}$,
    \begin{align*}
        \frac{\partial}{ \partial i} \left( \ell^w i - e_i \right) = \ell^w - \frac{i - k}{i} \geq 0 \Leftrightarrow \ell^w \geq \frac{i - k}{i},
    \end{align*}
    where $\ell^w = \frac{w(y')}{y'}$. So, $(e_i, B_{i})$ with $i<\bar{y}$ is implementable by $w \in \mathcal{C}$ if and only if
    \begin{align*}
        w(y') = \frac{i - k}{i} y',
    \end{align*}
    while action $i = \bar{y}$ is implementable by $w \in \mathcal{C}$ if and only if
    \begin{align*}
        w(y') \geq \frac{\bar{y} - k}{\bar{y}} y'.
    \end{align*}
    The contract $y \mapsto \frac{\bar{y}-k}{\bar{y}}y$ is \eqref{eq:FB_program}-optimal as it maximises total surplus and gives a profit of zero to the firm: $V(\mathcal{T}) = \alpha \left(\bar{y} - k - e_{\bar{y}}\right)$.
    
    Next, observe that, for all $i$, any contract that implements $(e_i, B_i)$ gives a profit of at most zero to the firm. So, the regulator's regret is
    \begin{align*}
        R\left(\mathcal{C,T}\right) = V(\mathcal{T}) = \alpha \left(\bar{y} - k - e_{\bar{y}}\right). 
    \end{align*}
    Plugging in the effort cost, the regret is equal to
    \begin{align*}
         \alpha \left(\bar{y} - k - e_{\bar{y}}\right) & =  \alpha \left(\bar{y} - k - \bar{y} + \frac{k \bar{y}}{y' - \underline{w}^{\mathcal{C}}(y')} - k \left( \ln \left(\frac{k \bar{y}}{\bar{y} -\frac{\underline{w}^{\mathcal{C}}(y')}{y'} \bar{y}} \right) - \ln \bar{y} \right) \right) \\
         & = \alpha k \left(\frac{\bar{y}}{\bar{y} -\frac{\underline{w}^{\mathcal{C}}(y')}{y'} \bar{y}} - \ln \left(\frac{k}{\bar{y} -\frac{\underline{w}^{\mathcal{C}}(y')}{y'} \bar{y}} \right) -1 \right).
    \end{align*}
    As in the proof of \cref{lemma:lowerbound_no_production_2}, taking 
    \begin{align*}
        k = \begin{cases}
            e^{\frac{ \frac{\underline{w}^{\mathcal{C}}(y')}{y'}}{1 - \frac{\underline{w}^{\mathcal{C}}(y')}{y'}} - 1} \left( 1 - \frac{\underline{w}^{\mathcal{C}}(y')}{y'} \right) \bar{y} \text{ if } \frac{\underline{w}^{\mathcal{C}}(y')}{y'} \leq \frac{1}{2}, \\
            \bar{y} -\frac{\underline{w}^{\mathcal{C}}(y')}{y'} \bar{y} \text{ if } \frac{\underline{w}^{\mathcal{C}}(y')}{y'} >\frac{1}{2},
        \end{cases}
    \end{align*}
    we get
    \begin{align*}
        WCR(\mathcal{C}) \geq  R\left(\mathcal{C},\left(k, \mathcal{A}^k \right) \right) =  \alpha e^{\frac{2\underline{w}^{\mathcal{C}}(y')-y'}{y' - \underline{w}^{\mathcal{C}}(y')}} \left(1-\frac{\underline{w}^{C}(y')}{y'}\right)\bar{y} \mathds{1}_{\left\{ \frac{\underline{w}^{\mathcal{C}}(y')}{y'} \leq \frac{1}{2}\right\}} + \alpha \frac{\underline{w}^{\mathcal{C}}(y')}{y'} \bar{y} \mathds{1}_{\left\{ \frac{\underline{w}^{\mathcal{C}}(y')}{y'} > \frac{1}{2}\right\}}.
    \end{align*}
    Since $\frac{\underline{w}^{\mathcal{C}}(y')}{y'} > \frac{\alpha -1}{2\alpha-1}$ by assumption, \cref{lemma:binding_cases} implies that the right-hand side is strictly greater than $\bar{R}$. 

    The proof for the case $\frac{\underline{w}^{\mathcal{C}}(y')}{y'} < \frac{\alpha -1}{2\alpha-1}$ for some $y'>\bar{y}$ similary adapts the proof of \cref{lemma:lowerbound_surplus_extraction_2}, and is thus omitted.
\end{proof}
This concludes the proof of \cref{theorem:optimalregulationnecessity} (i).
\end{proof}

\begin{proof}[Proof of point (ii)]
    We prove the contrapositive. Let $\mathcal{C} \subset\mathbb{C}_0$ and suppose that there exists $w \in \mathcal{C}$ such that $w(\bar{y})=\ell_\alpha^\star\cdot\bar{y}$ and $\underline{conv}(w)(\bar{y}) < \ell^{\star}_\alpha\cdot\bar{y}$. Observe that it implies that $\alpha>1$. We show that $\mathcal{C}$ is not optimal.
    
    First, note that, by point (i), we can suppose that $\underline{w}^\mathcal{C}(y)=\ell^\star_\alpha\cdot y$ for $y\geq \bar{y}$. The argument then mimics the proof of \cref{lemma:lowerbound_surplus_extraction_2}: we construct a similar technology while also exploiting the fact that the firm can offer the contract $w$ and gain some profit. We then show that this technology leads to a regret that exceeds $\bar{R}$ when the regulation is $\mathcal{C}$.
    
    Since $\underline{conv}(w)(\bar{y}) < \ell^{\star}_\alpha\cdot\bar{y}$, there exists $y_1< \bar{y} <y_2$ and $p \in (0,1)$ such that $p y_1 + (1-p) y_2 = \bar{y}$ and $p w(y_1) + (1-p) w(y_2) < \ell^{\star}_\alpha\cdot\bar{y}$. Let $w^F \in \mathcal{C}$ be the contract that minimises $p w(y_1) + (1-p)w(y_2)$, $\underline{\ell} =\displaystyle \frac{p w^F(y_1) + (1-p) w^F(y_2)}{\bar{y}}$, and observe that $\underline{\ell} < \ell^\star_\alpha$. Finally, let $\mu_F \in [0, \bar{y}]$, $\Pi = (1-\underline{\ell})\mu_F$, and define the action set 
    \begin{align*}
        \mathcal{A}^{\mu_F} & = \left\{\left( \mathbb{E}_{D_{\mu_F}} \left[w^F(y)\right], D_{\mu_F} = \left(1-\frac{\mu_F}{\bar{y}}\right)\delta_{\{0\}} + \frac{\mu_F}{\bar{y}} \left( p \delta_{\{y_1\}} + (1-p) \delta_{\{y_2\}}  \right) \right) \right\}  \\
        & \qquad \qquad \cup \left\{ (e_i, B_{i}) \, :\, i\in I = \left[\mu_F, \bar{y} \right] \right\},
    \end{align*}
    where, for all $i\in I$,
    \begin{align*}
        e_i =\begin{cases}
            \ell_\alpha^\star\cdot i & \text{if } i \in \left[\mu_F, \frac{1-\underline{\ell}}{1-\ell_\alpha^\star}\mu_F\right], \\
            i + \Pi \left( \ln \frac{1-\underline{\ell}}{1-\ell_\alpha^\star}\mu_F - \ln i \right) - (1- \underline{\ell}) \mu_F & \text{if } i \in \left(\frac{1-\underline{\ell}}{1-\ell_\alpha^\star}\mu_F, \bar{y}\right].
        \end{cases}
    \end{align*}
    We refer to the only action supported on $\{0, y_1, y_2\}$ as action $0$. For all $0\leq \mu_F \leq \bar{y}$, all outputs' distributions in $\mathcal{A}^{\mu_F}$ are supported on $\{0, y_1, \bar{y}, y_2\}$. Thus, $\mathcal{A}^{\mu_F}$ is compact. We consider the technology $\mathcal{T} = \left(0, \mathcal{A}^{\mu_F}\right)$, assuming that the firm has no production cost. We will maximise regret over $\mu_F$ subsequently.

    Following the same argument as in the proof of \cref{lemma:lowerbound_surplus_extraction_2}, the firm can incentivise action $i \in \left[\frac{1-\underline{\ell}}{1-\ell_\alpha^\star}\mu_F, \bar{y}\right)$ with the contract $w$ if and only 
    \begin{align*}
        \ell^w = \frac{i - \Pi}{i},
    \end{align*}
    where $\ell^w = \frac{w(\bar{y})}{\bar{y}}$. Action $i = \bar{y}$ is implementable by $w \in \mathcal{C}$ if and only if
    \begin{align*}
       \ell^w \geq \frac{\bar{y} - \Pi}{\bar{y}}.
    \end{align*}
    So, for each of these actions, the profit is at most $\Pi$. 
    
    Observe that if $i= \frac{1-\underline{\ell}}{1-\ell_\alpha^\star}\mu_F$, then $\frac{i - \Pi}{i}=\ell^\star_\alpha$. Furthermore, for any $w$ such that $\ell^w=\ell^\star_\alpha$ (that includes $w^F$), the worker is indifferent between any $i\in \left[\mu_F, \frac{1-\underline{\ell}}{1-\ell_\alpha^\star}\mu_F\right]$ and action $0$. In any case, his surplus is $0$. Among those actions, only action $0$ would then give a profit of $\Pi$ to the firm, by definition. This implies that, replacing $\Pi$ by $(1-\underline{\ell})\mu_F$ the regulator's regret is:
    \begin{align*}
        & \alpha \left(\bar{y}-\bar{y} - (1-\underline{\ell})\mu_F\ln{\frac{1-\underline{\ell}}{1-\ell^\star_\alpha}\frac{\mu_F}{\bar{y}}} + (1-\underline{\ell}\mu_F\right) - (1-\underline{\ell})\mu_F \\
        = & -\alpha(1-\underline{\ell})\mu_F \ln{\frac{1-\underline{\ell}}{1-\ell^\star_\alpha}\frac{\mu_F}{\bar{y}}} + (\alpha-1)(1-\underline{\ell})\mu_F.
    \end{align*}
    Maximising over $\mu_F$, we obtain a maximiser equal to $e^{-\frac{1}{\alpha}}\bar{y}\frac{1-\ell^\star_\alpha}{1-\underline{\ell}}$, giving a regret of
    \begin{align*}
        \alpha e^{-\frac{1}{\alpha}} (1-\ell_\alpha^\star) \bar{y} + e^{-\frac{1}{\alpha}}\ell_\alpha^\star\bar{y} = \bar{R} + e^{-\frac{1}{\alpha}}\ell_\alpha^\star\bar{y} > \bar{R},
    \end{align*}
    where the inequality is strict because $\alpha>1$.
\end{proof}

\begin{proof}[Proof of point (iii)]
    We prove the contrapositive. Let $\mathcal{C} \subset\mathbb{C}_0 $ and suppose that there exists $y \geq \bar{y}$ such that $[\ell^{\star}_{\alpha} \cdot y, (1-\rho^{*})y] \not\subset \mathrm{Im}(\mathcal{C})(y)$. We show that $\mathcal{C}$ is not optimal.

    By \cref{theorem:optimalregulationnecessity} (i), we can suppose that $\underline{w}^{\mathcal{C}}(y) = \ell^{\star}_{\alpha} \cdot y$, for otherwise $\mathcal{C}$ is not be optimal. Then, there exists $w_1, w_2 \in \left[\ell^{\star}_{\alpha} \cdot y, (1 - \rho^{*})y \right]$ such that $\ell^{\star}_{\alpha} \cdot y < w_1<w_2$ and $(w_1, w_2) \cap Im\left( \mathcal{C}\right) (y) = \emptyset$ (where we used that $\mathrm{Im}(\mathcal{C})(y)$ is closed).

    To show that $\mathcal{C}$ is not optimal, we exhibit a technology akin to the technology considered in \cref{lemma:lowerbound_no_production_2} such that the regulator regret under this technology and the regulation $\mathcal{C}$ exceeds $\bar{R}$. 
    
    We first slightly modify the technology obtained in the proof of \cref{lemma:lowerbound_no_production_2}: Define $\mathcal{T} = \left(k^*, \mathcal{A}^{k^*}\right)$, where\footnote{The second equality follows from the definitions of $\ell_\alpha^\star$ and $\rho^\star$.}
    \begin{align*}
        k^* = e^{\frac{2 \ell_\alpha^\star-1}{1 - \ell_\alpha^\star}}(1 - \ell^\star_\alpha)\bar{y}  = \rho^*\cdot\bar{y},
    \end{align*}
    and 
    \begin{align*}
        \mathcal{A}^{k^*} = \left\{ (e_i, {B}_{i}) \mid  i\in I = \left[\frac{k^*}{1 - \ell^\star_\alpha} \bar{y}, \bar{y} \right] \right\},
    \end{align*}
    where, for all $i\in I$, $B_i$ is supported on $\{0, y\}$ with mean $i$, and
    \begin{align*}
        e_i = i - \frac{k^*}{1- \ell^{\star}_{\alpha}} + k^* \left( \ln \left(\frac{k^*}{1-\ell^{\star}_{\alpha}} \right) - \ln i \right).
    \end{align*}
    The only change from the technology constructed in \cref{lemma:lowerbound_no_production_2} is the support of the binary distributions over output $B_i$: $\{0, y\}$ instead of $\{0, \bar{y}\}$. Under this technology, the same argument as in the proof of \cref{lemma:lowerbound_no_production_2} shows that the regret is 
    \begin{align*}
        R\left(\mathcal{C,T}\right) = V(\mathcal{T}) = \alpha \left( \bar{y} - e_{\bar{y}} - k^* \right) = \alpha e^{\frac{2 \ell_\alpha^\star-1}{1 - \ell_\alpha^\star}}(1 - \ell^\star_\alpha)\bar{y} = \bar{R}.
    \end{align*}
    
    We then modify the technology $\mathcal{T}$ and consider $\tilde{\mathcal{T}} = \left(k^*, \tilde{\mathcal{A}}^{k^*}\right)$, where
    \begin{align*}
        \tilde{\mathcal{A}}^{k^*} = \left\{ (\tilde{e}_i, {B}_{i}) \colon i\in I = \left[\frac{k^*}{1 - \ell^\star_\alpha} \bar{y}, \bar{y} \right] \right\},
    \end{align*}
    with
    \begin{align*}
        \tilde{e}_i=\begin{cases}
            e_i & \text{if } i \in \left[\frac{k^\star}{1-\ell^\star_\alpha} \bar{y}, i_1\right],\\
            e_{i_1} + (i-i_1)(\frac{y_1}{y}+\epsilon) & \text{if } i \in (i_1, i_2], \\
            e_i - (e_{i_2} - \tilde{e}_{i_2}) & \text{if } i\in (i_2, \bar{y}],
        \end{cases}
    \end{align*}
    where $\epsilon>0$ is small and $i_n$ is defined as the $i$ such that $ \frac{w_n}{y} = \frac{\partial e_i}{\partial i} \left(= \frac{i-k^{\star}}{i}\right) $, for $n=1,2$. The slope of the cost curve $e_i$ is increasing, continuous, and comprised between $\ell_\alpha^\star$ and $1-\rho^\star$, which implies that each $i_n$ is well defined.

    The logic behind this construction is similar to the proof of \cref{lemma:lowerbound_no_production_2}. However, we exploit in addition the fact that no contract $w$ with ``slope'' $\frac{w(y)}{y}$ in $\left[\frac{w_1}{y},\frac{w_2}{y}\right]$ can be offered. Consequently, it is not necessary for the cost curve's slope to be always equal to $\frac{i-k^\star}{k^\star}$ to eliminate profitable opportunities for the firm. In the interval $[i_1, i_2]$, we can simply fix the slope of the cost curve to be slightly above $\frac{w_1}{y}$. The $\epsilon$ ensures that the firm cannot incentivise any of these actions (except action $i_1$ with the contract $\frac{w_1}{y}$). Consequently, as in the proof of \cref{lemma:lowerbound_no_production_2}, the firm cannot make any positive profit. The regret is thus:
    \begin{align*}
        R\left(\mathcal{C}, \tilde{\mathcal{T}}\right) & = V(\tilde{\mathcal{T}}) = \alpha \left( y -\tilde{e}_y - k^*\right) = V(\mathcal{T}) + \alpha (e_{\bar{y}}-\tilde{e}_{\bar{y}}) \\
        & = V(\mathcal{T}) + \alpha \left(e_{i_2} - e_{i_1} - \left(\frac{y_1}{y}+\epsilon\right)(i_2 - i_1) \right) > \bar{R},
    \end{align*}
    where the inequality follows for $\epsilon$ small enough from the strict convexity of $i \mapsto e_i$ since its slope at $i_1$ equals $\frac{w_1}{y}$ and $i_2>i_1$. This concludes the proof \cref{theorem:optimalregulationnecessity} (iii).
\end{proof}

\appendix

\setcounter{section}{2}
\setcounter{lemma}{6}
\setcounter{equation}{26}

\section{Supplementary Lemmas}\label{app:supporting_results}
The section contains additional results that we use in other proofs, in particular the proofs of \cref{theorem:main,theorem:optimalregulationnecessity}. The first lemma simplifies the search for regret-maximising technologies for MPR regulations, in the scenario where the firm is able to obtain positive profit.
\begin{lemma}\label{lemma:support_simplification_production}
    For any MPR regulation $\mathcal{C}_\ell$ with a minimum contract $\ell$, and any technology $\mathcal{T}$ such that $\Pi\left(\mathcal{C,T}\right) >0$, there exists a technology $\hat{\mathcal{T}} = (\hat{k}, \hat{\mathcal{A}})$ that generates a greater regret, $R\left(\mathcal{C,T}\right) \leq R(\mathcal{C},\hat{\mathcal{T}})$, and that satisfies the following conditions: 
    \begin{enumerate}[(i)]
        \item $\hat{k}=0$,
        \item the firm optimally offers the contract $\ell$ and the worker's surplus $WS(\mathcal{C,\hat{T}}, \hat{\sigma}) =0$ for some regret-maximising equilibrium strategy $\hat{\sigma} \in \Sigma(\mathcal{C,\hat{T}})$, 
        \item $\hat{\mathcal{T}}$ is binary,
    \end{enumerate}
\end{lemma}
\begin{proof}
    Let $\mathcal{C}_\ell \subset \mathbb{C}_0 $ be an MPR regulation with the minimum contract $\ell$, and $\mathcal{T} = \left(k, \mathcal{A}\right)$ a technology such that $\Pi\left(\mathcal{C_\ell,T}\right) >0$. Denote as $\sigma = (w_\pi, (e_\pi, F_\pi)) \in \mathcal{C} \times \mathcal{A}$ the equilibrium strategy of the contracting game. We construct $\hat{\mathcal{T}}$ gradually.
    \paragraph{Step 1.} Define the technology
    \begin{align*}
        \mathcal{T}^1 = \left( k^1 = 0, \mathcal{A}^1 = \left\{\left(e + WS(\mathcal{C, T}, \sigma), F\right)\right\}_{(e,F) \in \mathcal{A}}  \right).
    \end{align*}
    With technology $\mathcal{T}^1$, the firm implements the action $(e^1_\pi, F^1_\pi) = (e_\pi + WS(\mathcal{C, T}, \sigma), F_\pi)$, by offering the same contract $w_\pi$, when solving the profit maximisation problem \eqref{eq:profit_program} under technology $\mathcal{T}^0$. This is because increasing uniformly the cost of every action does not change the incentive constraints \eqref{eq:IC_W}, and given that each action's cost has increased by $WS(\mathcal{C,T},\sigma)$, the participation constraint \eqref{eq:IR_W} is still satisfied under $\sigma^1=(w_\pi, (e^1_\pi, F^1_\pi))$. Furthermore, the firm cannot increase her profit, hence $ \Pi(\mathcal{C,T}^1) = \Pi(\mathcal{C,T}) + k$, and $WS(\mathcal{C,T}^1, \sigma^1) = \mathbb{E}_{F^1_\pi}\left[ w_\pi(y) \right] -e^1_\pi = WS(\mathcal{C, T}, \sigma) - WS(\mathcal{C, T}, \sigma) =0$. 

    So, $\mathcal{T}^1$ satisfies (i).
    
    \paragraph{Step 2.} Define the technology
    \begin{equation*}
        \mathcal{T}^2 = \left(k^2=0, \mathcal{A}^1 \cup \left\{\left( \ell \cdot \mu^\star, B_{\mu^\star}\right) \right\} \right),
    \end{equation*}
    where $B_{\mu^\star}$ is the binary distribution with mean $\mu^\star$ that solves
    \begin{align*}
        \Pi\left( \mathcal{C}, \mathcal{T}^1\right) = \mu^\star - \ell\cdot \mu^\star.
    \end{align*}
    Offering contract $\ell$ implements action $\left( \ell\cdot \mu^\star, B_{\mu^\star}\right)$, since
    \begin{align}
        \mathbb{E}_{B_{\mu^\star}} [\ell\cdot y] - \ell\cdot \mu^\star = 0 = \mathbb{E}_{F^1_\pi} [{w}_\pi(y)] -e^1_\pi \geq \mathbb{E}_{F} [{w}_\pi(y)] -e \geq \mathbb{E}_{F} [\ell\cdot y] -e \label{ineq:sup_lem_step2}
    \end{align}
    for all $(e,F) \in \mathcal{A}^1$. The second equality comes from $WS(\mathcal{C,T}^1, \sigma^1) = 0$, the first inequality follows from the worker incentive compatibility constraints \eqref{eq:IC_W} when the firm maximises profit with technology $\mathcal{T}^1$, and the second inequality follows from the definition of $\ell$. Therefore, offering contract $\ell$ implements action $\left( \ell \cdot \mu^\star, B_{\mu^\star}\right)$ and maximises the firm's profit when the technology is $\mathcal{T}^2$, generating $\Pi(\mathcal{C,T}^2) = \Pi(\mathcal{C,T}^1) = \Pi(\mathcal{C,T}) + k$.

    So, $\mathcal{T}^2$ satisfies (i) and (ii).

    \paragraph{Step 3.} For any action $(e,F) \in \mathcal{A}^2$, we define its average output $\mu_F$ and the associated binary distribution $B_{\mu_F} \in \Delta(\{0, \bar{y}\})$ that generates the same average output. We define the following technology:
    \begin{equation*}
        \hat{\mathcal{T}} = \left(\hat{k} = 0, \hat{\mathcal{A}} = \{(e, B_{\mu_F})\}_{(e,F) \in \mathcal{A}^2}\right).
    \end{equation*}
    From \eqref{ineq:sup_lem_step2}, contract $\ell$ incentivises the agent to choose action $(\ell\cdot \mu^\star, B_{\mu^\star})$. Furthermore, the firm cannot get a greater profit by offering another contract. To see this, note that because these are only binary actions, every contract can be equivalently seen as a linear contract in terms of the payoffs it generates. Thus if a contract $\ell'>\ell$ could implement a more profitable action in $\hat{\mathcal{A}}$, then the same contract could also implement such an action in $\hat{\mathcal{A}}^2$, a contradiction.

    Consequently, the profile $\hat{\sigma}= (\ell, (\ell\cdot \mu^\star, \mu^\star)) \in \Sigma(\mathcal{C, \hat{T}})$ maximises the firm's profit, which equals $\Pi(\mathcal{C,\hat{T}})=\Pi(\mathcal{C,T}) + k$, gives a surplus of $0$ to the worker, and maximises the regulators regret. 

    So, $\hat{\mathcal{T}}$ satisfies properties (i)-(iii). There remains to prove that $\hat{\mathcal{T}}$ generates a greater regret. Given that $\hat{k}=0$, the regulator offers the contract $w_R(y) = y$ in problem \eqref{eq:FB_program} under technology $\hat{\mathcal{T}}$, which induces the worker to take an action that maximises the total surplus. Denote this action $(\hat{e}_R, \hat{F}_R)$ and $\hat{\mu}_R$ its average output. Similarly, denote $(e_R, F_R)$ the full-information action implemented by the regulator under technology $\mathcal{T}$, and $\mu_R$ its average output. Since $(\hat{e}_R, \hat{F}_R)$ maximises the total surplus in $\hat{\mathcal{T}}$ and $(e_R + WS(\mathcal{C, T}, \sigma), B_{\mu_R}) \in \hat{\mathcal{A}}$,
    \begin{align*}
        \hat{\mu}_R -\hat{e}_R \geq \mu_R - (e_R + WS(\mathcal{C, T}, \sigma)),
    \end{align*} 
    which implies, using that $WS(\mathcal{C, \hat{T}}, \hat{\sigma}) =0$ by (ii), that
    \begin{align}
        R(\mathcal{C,\hat{T}}) & = \alpha\left( \hat{\mu}_R - \hat{e}_R \right) -  \Pi(\mathcal{C,\hat{T}})= \alpha\left( \hat{\mu}_R - \hat{e}_R \right) -  \Pi(\mathcal{C,T}) - k \notag\\
        & \geq \alpha\left( \mu_R -e_R \right) - \left( \Pi(\mathcal{C,T}) + \alpha WS(\mathcal{C, T}, \sigma)\right) - k \notag\\
        & = \alpha\left( \mu_R -e_R - k\right) - \left( \Pi(\mathcal{C,T}) + \alpha WS(\mathcal{C, T}, \sigma)\right) +(\alpha-1) k \notag\\
        & \geq V(\mathcal{T}) - \left( \Pi(\mathcal{C,T}) + \alpha WS(\mathcal{C, T}, \sigma)\right) +(\alpha-1) k \label{ineq:sup_lem_fullinfo}\\
        & \geq R(\mathcal{C,T}), \label{ineq:sup_lem_regret}
    \end{align}
    where \eqref{ineq:sup_lem_fullinfo} holds since $\alpha(\mu_R-e_R-k)$ is an upper bound on $V(\mathcal{T})$, and \eqref{ineq:sup_lem_regret} follows from $(\alpha-1)k \geq 0$. This concludes the proof.
\end{proof}

\begin{lemma}\label{lemma:binding_cases} \
    \begin{enumerate}
        \item For any $\alpha \geq 1$ and $\underline{w} \in [\frac{1}{2}\bar{y},\Bar{y}]$, $ \underline{w} > e^{-\frac{1}{\alpha}}(\bar{y}-\underline{w}).$
        \item For any $\underline{w} \in [0,\frac{1}{2}\Bar{y}]$, $e^{\frac{2\underline{w}-\Bar{y}}{\Bar{y}- \underline{w}}}(\Bar{y}-\underline{w}) \geq \underline{w}$.
    \end{enumerate}
\end{lemma}
\begin{proof}
    \noindent 1. For any $\alpha \geq 1$ and $\underline{w} \in [0,\Bar{y}]$, 
    \begin{align*}
        e^{-\frac{1}{\alpha}}(\bar{y}-\underline{w}) \geq \underline{w} \iff \underline{w} \leq \frac{e^{-\frac{1}{\alpha}}}{1 + e^{-\frac{1}{\alpha}}} \bar{y} < \frac{1}{2}\bar{y}.
    \end{align*} \\
    \noindent 2. Define $f: [0, \frac{1}{2}\bar{y}] \to \mathbb{R}$ by $f(\underline{w}) = e^{\frac{2\underline{w}-\Bar{y}}{\Bar{y}- \underline{w}}}(\Bar{y}-\underline{w})- \underline{w}$. Observe that
    \begin{align*}
        f\left(\frac{1}{2}\bar{y}\right) = e^{2 \frac{\bar{y}-\bar{y}}{\bar{y}}}(\bar{y}-\frac{1}{2}\bar{y}) - \frac{1}{2}\bar{y} = 0.
    \end{align*}
    Moreover, $f$ is nonincreasing on its domain, since
    \begin{align*}
        \frac{\partial f}{\partial \underline{w}} = \frac{\bar{y}}{\bar{y}- \underline{w}}e^{\frac{2\underline{w}-\Bar{y}}{\Bar{y}- \underline{w}}} - e^{\frac{2\underline{w}-\Bar{y}}{\Bar{y}- \underline{w}}} -1 = \frac{\bar{w}}{\bar{y}- \underline{w}}e^{\frac{2\underline{w}-\Bar{y}}{\Bar{y}- \underline{w}}} -1 \leq 0.
    \end{align*}
    Therefore, for all $\underline{w} \in [0, \frac{1}{2}\bar{y}]$,
    \begin{align*}
        f(\underline{w}) \geq f\left(\frac{1}{2}\bar{y}\right) =0 \,  \Leftrightarrow \, e^{\frac{2\underline{w}-\Bar{y}}{\Bar{y}- \underline{w}}}(\Bar{y}-\underline{w}) \geq \underline{w}.
    \end{align*}
\end{proof}

\begin{lemma}\label{lemma:claim2inequality}
    For any $\alpha \geq 1$ and $\underline{w} \in [0,\Bar{y}]$, $\alpha e^{-\frac{1}{\alpha}}(\bar{y}-\underline{w}) \geq (\alpha-1)(\bar{y}-\underline{w})$.
\end{lemma}

\begin{proof}
    If $\underline{w} = \Bar{y}$, this holds as an equality. For any $\underline{w} \in [0,\Bar{y})$,
    \begin{align*}
        \alpha e^{-\frac{1}{\alpha}}(\bar{y}-\underline{w}) \geq (\alpha-1)(\bar{y}-\underline{w}) & \iff \alpha e^{-\frac{1}{\alpha}} - (\alpha-1) \geq 0 \\
        & \iff e^{-\frac{1}{\alpha}} -1 + \frac{1}{\alpha} \geq 0
    \end{align*}
    Define the function $f(\alpha)= e^{-\frac{1}{\alpha}} -1 + \frac{1}{\alpha}$. Then, for $\alpha \neq 0$, 
    \begin{align*}
        f'(\alpha) &= \frac{1}{\alpha^2}\left(e^{-\frac{1}{\alpha}}-1\right) \leq 0 \\
        & \Leftrightarrow e^{-\frac{1}{\alpha}}-1 \leq 0 \\
        &  \Leftarrow \alpha \geq 0
    \end{align*}
    So, $f$ is decreasing for $\alpha \geq 1$. Furthermore, $f(1) = e^{-1}>0$ and $\lim_{\infty} f = 0$. Therefore, $f(\alpha) \geq 0 $ for any $\alpha \geq 1$.\\
\end{proof}

\section{Supporting Lemmas for Step 1 in \cref{subsec:proof_sketch}}\label{app:proof_step1}

The following lemma proves a slightly stronger result than \cref{ineq:MPR}, namely that for any linear regulation whose smallest contract is $\ell$, the MPR regulation with the minimum contract $\ell$ achieves a weakly lower worst-case regret with respect to binary technologies.

\begin{lemma}
    \label{lemma:ineq:MPR}
    For any linear regulation $\mathcal{C}$, whose smallest contract is denoted $\ell$, define the MPR regulation $\mathcal{C}_\ell = \{w \in \mathbb{C}_0  \mid w \geq \ell \}$. Then 
    \begin{equation*}
        \sup_{\text{\textbf{binary} } \mathcal{T}} R(\mathcal{C_\ell, T}) \leq \sup_{\text{\textbf{binary} } \mathcal{T}} R(\mathcal{C, T}).
    \end{equation*}
\end{lemma}
\begin{proof}
    We show that any candidate worst-case binary technology for $\mathcal{C}_\ell$ would generate the same regret under $\mathcal{C}$. We differentiate between two scenarios, depending of whether $\Pi(\mathcal{C,T})>0$ or $\Pi(\mathcal{C,T})=0$.

    First, given that $\mathcal{C} \subset \mathcal{C}_\ell$, any technology such that the firm makes zero profit under regulation $\mathcal{C}_\ell$ is also a technology for which the firm makes zero profit under regulation $\mathcal{C}$. Consequently, if the worst-case regret for binary technologies under $\mathcal{C}_\ell$ results from a technology where the firm makes zero profit, then the conclusion of the lemma follows.

    Second, suppose that the worst-case for binary technologies under $\mathcal{C}_\ell$ results from a technology $\mathcal{T}=(k, \mathcal{A})$ such that $\Pi(\mathcal{C_\ell, T})>0$. From \cref{lemma:support_simplification_production}, we can assume that under $\mathcal{T}$ and $\mathcal{C}_\ell$, the firm optimally offers the contract $\ell$ to the worker. Given that $\ell\in\mathcal{C}\subset\mathcal{C}_\ell$, this implies that under $\mathcal{C}$ the firm would also offer $\ell$, hence generating the same regret: $R(\mathcal{C_\ell, T'}) = R(\mathcal{C, T'})$.
\end{proof}

The next lemma establishes that we can restrict ourselves to binary technologies to maximise regret against an MPR regulation, obtaining \cref{ineq:all_tech}.

\begin{lemma}
    \label{lemma:ineq:all_tech}
    For any MPR regulation $\mathcal{C}_\ell$, and $\mathcal{T}$, there exists a binary technology $\hat{\mathcal{T}}$ such that:
    \begin{equation*}
        R(\mathcal{C_\ell,T}) \leq R(\mathcal{C_\ell,\hat{T}}).
    \end{equation*}
\end{lemma}
\begin{proof}
    Fix $\mathcal{T} = (k, \mathcal{A})$ an arbitrary technology and $\mathcal{C}_\ell$ the MPR regulation with minimum contract $\ell$. Similarly to the proof of \cref{lemma:ineq:MPR}, we consider two scenarios, depending of whether $\Pi(\mathcal{C_\ell,T})>0$ or $\Pi(\mathcal{C_\ell,T})=0$. If $\Pi(\mathcal{C_\ell,T})>0$, the conclusion directly follows from \cref{lemma:support_simplification_production}.

    Suppose that $\Pi(\mathcal{C_\ell,T})=0$. For any action $(e,F)\in \mathcal{A}$, we write $\mu_F$ for its average output. We also denote as $(e_R, F_R)$ the \eqref{eq:FB_program}-optimal action chosen by the regulator if they knew the technology $\mathcal{T}$, and as $\mu_R$ its average output. We aim to gradually construct a binary technology $\hat{\mathcal{T}}$ that generates a weakly greater regret.

    \paragraph{Step 1.} If $\frac{\mu_{R}-k}{\mu_{R}} > \ell$, then we go directly to step 2. Otherwise, if $\frac{\mu_{R}-k}{\mu_{R}} \leq \ell$, we consider the following binary technology
    \begin{equation*}
        \hat{\mathcal{T}} = (k, \{(e_R, B_{\mu_R})\}),
    \end{equation*}
    which only contains a single binary action $(e_R, B_{\mu_R})$ that has the same cost and generates the same average output as $(e_R, F_R)$. In this case, the firm cannot implement this action and make a strictly positive profit, because $\frac{\mu_{R}-k}{\mu_{R}} \leq \ell$, thus she does not hire the worker. The regulators would implement this action with the contract $y \mapsto \frac{\mu_R-k}{\mu_R}\cdot y$, giving the entire generated surplus to the worker. Therefore,
    \begin{align*}
        R(\mathcal{C_\ell,\hat{T}}) = \alpha (\mu_R-e_R-k) \geq V(\mathcal{T}) = R(\mathcal{C_\ell,T}),
    \end{align*}
    which concludes the proof.
    
    \paragraph{Step 2.} In the case where $\frac{\mu_{R}-k}{\mu_{R}} > \ell$, consider the following binary technology
    \begin{align*}
        \mathcal{T}^2 = (k, \mathcal{A}^2=\{(e, B_{\mu_F})\}_{(e,F) \in \mathcal{A}}),
    \end{align*}
    which takes the binary equivalent of every action in $\mathcal{A}$. Note that the firm still does not want to produce when facing $\mathcal{T}^2$ instead of $\mathcal{T}$, because the actions have the same expected revenue and cost, but as they are all binary, only linear contracts can be used to incentivise the worker, which weakly reduce the firm's profit. If the contract $y \mapsto \frac{\mu_R-k}{\mu_R}\cdot y$ incentivises action $(e_R, B_{\mu_R})$, then the same conclusion as in step 1 holds by setting $\hat{\mathcal{T}}=\mathcal{T}^2$. Otherwise, we turn to Step 3.

    \paragraph{Step 3.} If $y \mapsto \frac{\mu_R-k}{\mu_R}\cdot y$ does not incentivise action $(e_R, B_{\mu_R})$, then it must be that any action $(e, B_\mu)$ that the worker would prefer to choose when facing this contract is such that $\mu < \mu_R$. This is because otherwise, the contract $y \mapsto \frac{\mu_R-k}{\mu_R}\cdot y$, which is feasible in regulation $\mathcal{C}$ (remember that $\frac{\mu_{R}-k}{\mu_{R}} > \ell$), could implement a strictly profitable action for the firm. Let $S= \max_{(e, B_\mu) \in \mathcal{A}^2} \frac{\mu_R-k}{\mu_R}\cdot \mu - e$ be the maximal surplus the worker obtains from the contract $y \mapsto \frac{\mu_R-k}{\mu_R}\cdot y$. Define
    \begin{equation*}
        e^\star =\mu_R-k - S < e_R,
    \end{equation*}
    where the inequality follows from $(e_R,B_{\mu_R})$ not being the worker's best-reply for the contract $y \mapsto \frac{\mu_R-k}{\mu_R}\cdot y$. We now define the following binary technology
    \begin{equation*}
        \hat{\mathcal{T}} = (k, \hat{\mathcal{A}} = \mathcal{A}^2 \setminus\{(e_R, B_{\mu_R})\} \cup \{(e^\star, B_{\mu_R})\}),
    \end{equation*}
    where we replace, in $\mathcal{A}^2$, $(e_R, B_{\mu_R})$ by $(e^\star, B_{\mu_R})$. Note that in $\mathcal{\hat{T}}$, the contract $y \mapsto \frac{\mu_R-k}{\mu_R}\cdot y$ incentivises action $(e^\star, B_{\mu_R})$:
    \begin{equation*}
        \frac{\mu_R-k}{\mu_R}\cdot\mu_R -e^\star = \mu_R-k-(\mu_R-k)+S = S = \max_{(e, B_\mu) \in \mathcal{\hat{T}}} \frac{\mu_R-k}{\mu_R}\cdot \mu - e.
    \end{equation*}
    Furthermore, any contract with a strictly smaller slope would not incentivise action $(e^\star, B_{\mu_R})$, hence the firm cannot make any strictly positive profit when facing $\mathcal{\hat{T}}$ and $\mathcal{C_\ell}$. Therefore,
    \begin{align*}
        R(\mathcal{C_\ell,\hat{T}}) = \alpha(\mu_R-e^\star-k) > \alpha (\mu_R-e_R-k) \geq V(\mathcal{T}) = R(\mathcal{C_\ell,T}),
    \end{align*}
    which concludes the proof.
\end{proof}

\section{Omitted Proofs}\label{app:omitted_proofs}

\subsection{Proof of \cref{corollary:minimaloptimalregulation}}

    \cref{theorem:optimalregulationnecessity} implies that, if a minimum contract regulation $\mathcal{C}$ is optimal, then the minimum contract $\underline{w} \geq \ell^{\star}_\alpha$. We prove that if $\underline{w}(y) > \ell^{\star} y$ for some $y$, then $\mathcal{C}$ is not optimal. If $y \geq \bar{y}$, this follows from \cref{theorem:optimalregulationnecessity} (i). So, suppose that $y < \bar{y}$. Then the concave envelop of $\underline{w}$, denoted $conc(\underline{w})$, is such that $conc(\underline{w})(\bar{y}) >\ell^{\star}_{\alpha} \bar{y}$, and there exist $y_1< \bar{y} <y_2$ and $p\in(0,1)$ such that $p y_1 + (1-p) y_2 = \bar{y}$ and $p \underline{w}(y_1) + (1-p) \underline{w}(y_2) = conv(\underline{w})(\bar{y}) > \ell^{\star}_\alpha\cdot\bar{y}$.

    For all $k \in [0, \bar{y} -  conv(\underline{w})(\bar{y}))$, define $\underline{i}(k) = \frac{k\bar{y}}{ \bar{y} -  conv(\underline{w})(\bar{y})} < \bar{y}$ and the three point distribution $D_0(k) = p_0 \delta_{\{0\}} + (1-p_0) \left(p \delta_{\{y_1\}} + (1-p) \delta_{\{y_2\}}\right)$ such that
    \begin{align*}
       (1-p_0) conv(\underline{w})(\bar{y}) = \frac{conv(\underline{w})(\bar{y})}{\bar{y}} \underline{i}(k).
    \end{align*}
    Consider then the family of technologies indexed by $k$
    \begin{align*}
        \mathcal{T}^k = \left(k, \left\{ \left(e_0 = 0, D_0(k)  \right) \cup \mathcal{A}^k = \left\{ (e_i, {B}_{i}) \, :\, i\in I = \left[\underline{i}(k) , \ \bar{y} \right] \right\} \right\} \right),
    \end{align*}
    where, for all $i\in I$,
    \begin{align*}
        e_i = i - \underline{i}(k) + k \left( \ln \left( \underline{i}(k) \right) - \ln (i) \right).
    \end{align*}
    By definition of $e_i$, for all $w \in \mathcal{C}$,
    \begin{align*}
        \frac{\partial}{ \partial i} \left( \ell^w i - e_i \right) = \ell^w - \frac{i - k}{i} \geq 0 \Leftrightarrow \ell^w \geq \frac{i - k}{i},
    \end{align*}
    where $\ell^w = \frac{w(\bar{y})}{\bar{y}}$. So, if $w \in \mathcal{C}$ implements action $i$, then it must be that
    \begin{align*}
        w(\bar{y}) = \frac{i - k}{i} \bar{y}  \text{ for } \underline{i}(k) \leq i<\bar{y}, \ w(\bar{y}) \geq \frac{i - k}{i} \bar{y}  \text{ for } i=\bar{y}.
    \end{align*}
    Action $0$ is implementable by $\underline{w}$ since
    \begin{align*}
        \mathbb{E}_{D_0} \left[\underline{w}(y)\right] -e_0 = \frac{conv(\underline{w}^{\mathcal{C}})(\bar{y})}{\bar{y}} \underline{i}(k) > \ell_\alpha^\star \cdot \underline{i}(k),
    \end{align*}
    where the last inequality indicates that it gives a greater surplus than choosing action $\underline{i}(k)$, the most attractive among the binary actions in $\mathcal{A}^k$ under contract $\underline{w}$.
    
    Next, observe that any contract that could implement any action in $\mathcal{A}^k$ gives a profit of at most zero. Furthermore, any contract that implements $(e_{\bar{y}}, B_{\bar{y}})$ while giving zero profit to the firm is \eqref{eq:FB_program}-optimal, as it maximises the total surplus (since $\frac{\partial}{\partial i} e_i \leq 1$), and allocates all of it to the worker:
    \begin{align*}
        V(\mathcal{T}) = \alpha \left(\bar{y} - k - e_{\bar{y}}\right).
    \end{align*}
    Since no contract gives the firm a strictly positive payoff, in the worst-case scenario for the regulator, the firm chooses her outside option. So the above is also equal to the regulator's regret. Plugging in the effort cost, the regret rewrites
    \begin{align*}
         \alpha \left(\bar{y} - k - e_{\bar{y}}\right) & =  \alpha \left(\bar{y} - k -\bar{y} + \underline{i}(k) - k \left( \ln \left(\underline{i}(k) \right) - \ln (\bar{y}) \right) \right) \\
         & = \alpha k \left(\frac{\bar{y}}{ \bar{y} -  conv(\underline{w})(\bar{y})} - \ln \left(\frac{k}{ \bar{y} -  conv(\underline{w})(\bar{y})} \right) -1 \right).
    \end{align*}
    This is the same expression as \cref{eq:lemma:lowerbound_surplus_extration_2} in \cref{lemma:lowerbound_no_production_2}, but with $\underline{w}^{\mathcal{C}}$ replaced by $conv(\underline{w})(\bar{y})$. So, taking the supremum with respect to $k \in[0, \bar{y} - \underline{w}^{C}(\bar{y}))$, we obtain a lower bound on the regulator's worst-case regret:
    \begin{align*}
        WCR(\mathcal{C}) \geq \underset{k \in [0, y' - \underline{w}^{\mathcal{C}}(y'))}{\sup} \,  R\left(\mathcal{C},\left(k, \mathcal{A}^k \right) \right) = \alpha e^{\frac{ 2 conv(\underline{w})(\bar{y})- \bar{y}}{\bar{y} -  conv(\underline{w})(\bar{y})}}(\bar{y} -  conv(\underline{w})(\bar{y}))> \bar{R}.
    \end{align*}
    where the last inequality follows since $conv(\underline{w})(\bar{y}) >\ell^{\star}_{\alpha}\bar{y}$. \hfill \qed

\subsection{Proof of \cref{prop:optimalregulationMLRP}}\label{app:proofprop:optimalregulationMLRP}

    By \cref{theorem:main} (ii),
    \begin{align*}
        \bar{R} = WCR\left(\mathcal{C}^{\star}_{\alpha}\right) \geq  \underset{\mathcal{T}\in \mathbb{T}^{MLRP}}{\sup}\, R\left(\mathcal{C}^{\star}_{\alpha}, \mathcal{T}\right).
    \end{align*}
    So, it is enough to show that, for all $\mathcal{C} \subset \mathbb{C}_0 $,
    \begin{align}\label{eq:proofoptimalregulationMLRP}
        \underset{\mathcal{T}\in \mathbb{T}^{MLRP}}{\sup}\, R\left(\mathcal{C,T}\right) \geq \bar{R}.
    \end{align}
    to prove \cref{prop:optimalregulationMLRP}. But, \cref{eq:proofoptimalregulationMLRP} holds if the technologies identified in \cref{lemma:lowerbound_no_production_1,lemma:lowerbound_no_production_2,lemma:lowerbound_surplus_extraction_2} belong to $\mathbb{T}^{MLRP}$. The technologies considered in  \cref{lemma:lowerbound_no_production_1} are such that $\# \mathcal{A} =1$. So, they vacuously belong to $\mathbb{T}^{MLRP}$. There remains to show that the technologies considered in \cref{lemma:lowerbound_no_production_2,lemma:lowerbound_surplus_extraction_2} also belong to $\mathbb{T}^{MLRP}$. Both cases are similar, so the proof for the latter is omitted.

    Recall that the technology constructed in \cref{lemma:lowerbound_no_production_2} was $\mathcal{T} = \left(k^*, \mathcal{A}^{k^*}\right)$ where
    \begin{align*}
        k^* = e^{\frac{2 \ell^{\star}_\alpha\cdot\bar{y} - \bar{y})}{\bar{y} - \ell^{\star}_\alpha\cdot\bar{y}}} \left( \bar{y} - \ell^{\star}_\alpha\cdot\bar{y} \right).
    \end{align*}
    and 
    \begin{align*}
        \mathcal{A}^{k^*} = \left\{ (e_i, {B}_{i}) \colon i\in I = \left[\bar{y}\frac{k^*}{\bar{y} - \ell^{\star}_\alpha\cdot\bar{y}} , \bar{y} \right] \right\},
    \end{align*}
    where, for all $i\in I$,
    \begin{align*}
        e_i = i - \frac{k^* \bar{y}}{\bar{y} - \ell^{\star}_\alpha\cdot\bar{y}} + k^* \left( \ln \left(\frac{k^* \bar{y}}{\bar{y}-\ell^{\star}_\alpha\cdot\bar{y}} \right) - \ln i \right).
    \end{align*}
    First, note that $i\geq j \Rightarrow e_i \geq e_j$, so the effort cost is increasing in $i$. Second, we verify that $i\geq j \Rightarrow B_i \geq_{MLRP} B_j$ according to Defintion A.1 in \cite{athey2002monotone}; i.e.,
    \begin{align}\label{eq:proofoptimalregulationMLRP_2}
        \frac{dB_i(y')}{dC_{ij}(y')} \frac{dB_j(y)}{dC_{ij}(y)} \geq  \frac{dB_j(y')}{dC_{ij}(y')} \frac{dB_i(y)}{dC_{ij}(y)}
    \end{align}
    for all $i\geq j \in I$ and all $y'\geq y$ in the support of $\mathcal{C}_{ij}(\cdot) = \frac{1}{2} B_i(\cdot) + \frac{1}{2} B_j(\cdot)$. By direct calculations, \cref{eq:proofoptimalregulationMLRP_2} is equivalent to
    \begin{align*}
        \frac{i}{\bar{y}}\left(1-\frac{j}{\bar{y}}\right) \geq \frac{j}{\bar{y}}\left(1-\frac{i}{\bar{y}}\right),
    \end{align*}
    which holds since $i\geq j$. So, $i\geq j \Rightarrow B_i \geq_{MLRP} B_j$ and $\mathcal{T}\in \mathbb{T}^{MLRP}$. \hfill \qed

\subsection{Proof of \cref{prop:optimalregulationquantitativeknowledge}}\label{app:proofprop:optimalregulationquantitativeknowledge}

The proof of \Cref{prop:optimalregulationquantitativeknowledge} mimics that of \Cref{theorem:main}. \cref{lemma:lowerbound_no_production_2constrained} and \cref{lemma:lowerbound_surplus_extraction_2constrained} below show that the regulator's worst-case regret for any regulation $\mathcal{C} \subset \mathbb{C}_0$ is bounded below by
\begin{align}\label{eq:Regretlowerbound}
    \min_{0\leq \ell^{\mathcal{C}} \leq 1} \max  \bigg\{ &\underset{y \in \left[ \underline{y}, \frac{\bar{k}}{1-\ell^{\mathcal{C}}} \wedge \bar{y} \right]}{\sup } \alpha \ell^{\mathcal{C}} y, \   \underset{k \in [\underline{k}, \bar{k}] }{\sup}\alpha \left(  (1-\ell^{\mathcal{C}})\underline{y} -k  - k \left( \ln \left( \underline{y} \right) - \ln \left( \bar{y}\right)\right)\right) \mathds{1}_{\{ (1-\ell^{\mathcal{C}}) \underline{y} \leq \bar{k} \}} \notag \\
    & + \underset{k \in [\underline{k}, \bar{k}] }{\sup}\alpha \left(  (1-\ell^{\mathcal{C}})\underline{y} -k  - k \left( \ln \left( \underline{y} \right) - \ln \left( \bar{y}\right)\right)\right) \mathds{1}_{\{ (1-\ell^{\mathcal{C}}) \underline{y} > \bar{k} \}}. \notag \\
    & \underset{y \in \left[\frac{\underline{k}}{1-\ell} \vee \underline{y}, \bar{y}\right]}{\sup} \ \alpha \int_{y}^{\bar{y}} \frac{1-\ell^{\mathcal{C}}}{\mu}y d\mu + (\alpha -1) \left( \left(1- \ell^{\mathcal{C}}\right) y - \underline{k} \right) \bigg\}.
\end{align}
To prove \cref{prop:optimalregulationquantitativeknowledge}, there remains to show that a MPR regulation achieve the upper bound above. This follows from \cref{lemma:boundonregretnoproductionconstrained} and \cref{lemma:boundonregretproductionconstrained} below, which shows that the regulator's regret when offering the MPR regulation $\mathcal{C}_{\ell^{\star}}$ is at most \eqref{eq:Regretlowerbound} for $\ell^{\star}$ a solution of the minimisation problem \eqref{eq:Regretlowerbound}. \hfill \qed

\vspace{1em}

\cref{lemma:lowerbound_no_production_2constrained} and \cref{lemma:lowerbound_surplus_extraction_2constrained} adapt the proofs of \cref{lemma:lowerbound_no_production_2,lemma:lowerbound_no_production_1,lemma:lowerbound_surplus_extraction_2} to establish a lower bound on the regulator's worst-case regret for any regulation. 

\begin{lemmap}{\ref{lemma:lowerbound_no_production_2}'}\label{lemma:lowerbound_no_production_2constrained}
    For any regulation $\mathcal{C} \subset \mathbb{C}_0 $,
    \begin{align}\label{eq:lowerbound_no_productionconstrained}
        \underset{\mathcal{T} \in \mathbb{T}^c}{\sup} \, R\left(\mathcal{C}, \mathcal{T}\right) & \geq  \max \bigg\{  \underset{y' \in [\underline{y}, \bar{y}], y \in [\underline{y}, \frac{\bar{k}}{1-\ell^{y'}}\wedge y']} \alpha \frac{\underline{w}^{\mathcal{C}}(y')}{y'}y, \ \underset{y', k \ : \ \underline{k} \leq k \leq y' - \underline{w}^{C}(y') \wedge \bar{k}}{\sup } \   \alpha \bigg( \frac{k y'}{y' - \underline{w}^{\mathcal{C}}(y')} \vee \underline{y} - k  \notag \\ &-  k \left( \ln \left(\frac{k y'}{y'-\underline{w}^{\mathcal{C}}(y')} \vee \underline{y} \right) - \ln (y') \right) - \frac{k y'}{y' - \underline{w}^{\mathcal{C}}(y')} \vee \underline{y} \mathds{1}_{\{ \bar{k} < (y' - \underline{w}^{\mathcal{C}}(y')) \frac{\underline{y}}{y'}\}} \bigg) \bigg\}.
    \end{align}
\end{lemmap}

\begin{proof}
    Let $\mathcal{C} \subset \mathbb{C}_0$ be a regulation. To show that the right-hand side of \cref{eq:lowerbound_no_productionconstrained} is a lower bound on the regulator's worst-case regret, it is enough to exhibit a technology that yields a regret at least equal to it. So, in the remainder of the proof, we construct two such technology, one for each term in the maximand.

    (i) Let $\underline{y} \leq y' \leq \bar{y}$ and denote by $\ell^{y'} = \frac{\underline{w}^{\mathcal{C}}}{y'}$. Suppose that the production set $\mathcal{A}$ contains a single binary action supported on $\{0,y'\}$ with expected output level $\underline{y}\leq y \leq y'\leq \bar{y}$ and zero cost: $\mathcal{A} = \left\{\left(0, (1- \frac{y}{y'})\delta_{\{ 0 \}} \frac{y}{y'} \delta_{\{ y' \}}\right)\right\}$. Given the technology $\mathcal{T}=\left(k, \mathcal{A} \right)$ and the regulation $\mathcal{C}$, the maximal \emph{profit} the firm can obtain is
    \begin{align*}
        (1-\ell^{y'})y - k.
    \end{align*}
    If $k > (1-\ell^{y'})y$, no contract is signed between the firm and the worker, as the firm strictly prefers her outside option to the best available contract in $\mathcal{C}$. The regret for the regulator is then
    \begin{align*}
        R(\mathcal{C,T}) = \alpha(y -k),
    \end{align*}
    since the contract $y \mapsto (y - k)\mathds{1}_{\{y'=y\}}$ implements action $\left(0, \delta_{\{ y' \}}\right)$ and allocates the entire surplus to the worker. Taking the supremum over $y,k$ such that $k >(1-\ell^{y'})y$, the regulator's regret is bounded below by
    \begin{align*}
        \underset{y' \in [\underline{y}, \bar{y}], y \in [\underline{y}, \frac{\bar{k}}{1-\ell^{y'}}\wedge y']} \alpha \ell^{y'}y.
    \end{align*}
    Finally, we obtain the desired bound by taking the supremum over $\underline{y} \leq y' \leq \bar{y}$.

    (ii) We can assume that there exists $y'$ such that $y' - \underline{w}^{\mathcal{C}}(y') > \underline{k}$, for otherwise we are done. Let $k \in [\underline{k}, \bar{k}]$ be (strictly) smaller than $ y' -\underline{w}^{\mathcal{C}}(y')$. Denote by $B_i$ the binary output distribution with support $\left\{0, y'\right\}$ and mean $i$. Define $\underline{i}=\displaystyle y' \frac{k}{y' - \underline{w}^{\mathcal{C}}(y')} \vee \underline{y}$ and the compact action set $\mathcal{A}^k = \left\{ (e_i, {B}_{i}) \mid i\in I = \left[\underline{i} , \ y' \right] \right\}$, where, for all $i\in I$,
    \begin{align*}
        e_i = i - \frac{k y'}{y' - \underline{w}^{\mathcal{C}}(y')}\vee \underline{y} + k \left( \ln \left(\frac{k y'}{y'-\underline{w}^{\mathcal{C}}(y')} \vee \underline{y} \right) - \ln (i) \right) + e_0,
    \end{align*}
    where $e_0 =0$ if $(y' - \underline{w}^{\mathcal{C}}(y')) \frac{\underline{y}}{y'} \leq \bar{k}$ and $e_0 > \frac{\underline{w}^{\mathcal{C}}(y'))}{y'}\bar{y}$ otherwise (when $\frac{k y'}{y' - \underline{w}^{\mathcal{C}}(y')} \leq \underline{y}$ for all $k$). We consider the family of technologies $\mathcal{T}^k= (k, \mathcal{A}^k)$, $k < y' -\underline{w}^{\mathcal{C}}(y')$.

    By definition of $e_i$, for all $w \in \mathcal{C}$,
    \begin{align*}
        \frac{\partial}{ \partial i} \left( \ell^w \cdot i - e_i \right) = \ell^w - \frac{i - k}{i} \geq 0 \Leftrightarrow \ell^w \geq \frac{i - k}{i},
    \end{align*}
    where $\ell^w = \frac{w(y')}{y'}$. So, for $\underline{i}<i<y'$, $(e_i, B_{i})$ is implementable by $w \in \mathcal{C}$ if and only if
    \begin{align*}
        w(y') = \frac{i - k}{i} y',
    \end{align*}
    by a simple first-order condition. Direct computations then shows that action $\underline{i}$ is only implementable by $w$ if $w(y') = \underline{w}^{\mathcal{C}}(y')$, and action $i = y'$ is implementable by $w \in \mathcal{C}$ if and only if
    \begin{align*}
        w(y') \geq \frac{y' - k}{y'} y' = y'-k.
    \end{align*}
    Since $\frac{\partial}{\partial i} e_i \leq 1$, action $y'$ maximises total surplus. Moreover, any contract $w$ such that $w(y') = y'-k$ implements this action and gives the entire surplus to the worker and zero profit to the firm. Thus, the regulator offers such a contract to solve \eqref{eq:FB_program}, obtaining
    \begin{align*}
        V(\mathcal{T}^k) = \alpha \left(y' - k - e_{y'}\right).
    \end{align*}
    On the other hand, observe that, for all $i\in [\underline{i}, y']$, any contract that implements $(e_i, B_i)$ gives a profit of at most zero to the principal. So, in the worst-case scenario for the regulator, the firm chooses her outside option. The regulator's regret is thus
    \begin{align*}
        R\left(\mathcal{C,T}^k\right) = V(\mathcal{T}^k) = \alpha \left(y' - k - e_{y'}\right). 
    \end{align*}
    Plugging in the effort cost, the regret is equal to
    \begin{align*}
         \alpha \left(y' - k - e_{y'}\right) & =  \alpha \left(y' - k - y' + \frac{k y'}{y' - \underline{w}^{\mathcal{C}}(y')} \vee \underline{y} - k \left( \ln \left(\frac{k y'}{y'-\underline{w}^{\mathcal{C}}(y')} \vee \underline{y} \right) - \ln (y') \right) -e_0 \right).
    \end{align*}
    The worst-case regret is obtained by maximising over all possible technologies. So, taking the supremum of the above expression with respect to $k \in[\underline{k}, (y' - \underline{w}^{C}(y')) \wedge \bar{k})$ and $y' \in [\underline{y}, \bar{y}]$ such that $y' -\underline{w}^{\mathcal{C}}(y') > \underline{k}$ yields a lower bound on the worst-case regret, which coincides with the second term in the maximand of \cref{eq:lowerbound_no_productionconstrained}.
\end{proof}

\begin{lemmap}{\ref{lemma:lowerbound_surplus_extraction_2}'}\label{lemma:lowerbound_surplus_extraction_2constrained}
    For any regulation $\mathcal{C} \subset \mathbb{C}_0 $,
    \begin{align}\label{eq:lowerbound_surplus_extractionconstrained}
         \underset{\mathcal{T} \in \mathbb{T}^c}{\sup} & \, R\left(\mathcal{C}, \mathcal{T}\right) \geq \max\bigg\{ \left( \alpha -1\right) \left[(1-\ell)\bar{y} - \underline{k} \right], \notag  \\
        & \underset{y, y' \ : \ \frac{\underline{k}}{1- \frac{\underline{w}^{\mathcal{C}}(y')}{y'}} \vee \underline{y} \leq y \leq y' \leq \bar{y}}{\sup} \ \alpha \int_{y}^{y'} \frac{1- \frac{\underline{w}^{\mathcal{C}}(y')}{y'}}{\mu}y d\mu + (\alpha -1) \left( \left(1- \frac{\underline{w}^{\mathcal{C}}(y')}{y'}\right)y - \underline{k} \right)\bigg\}.
    \end{align}
\end{lemmap}

\begin{proof}
    Let $\mathcal{C} \subset \mathbb{C}_0 $ be a regulation. As in the proof of \cref{lemma:lowerbound_no_production_2constrained}, to show that the right-hand side of \cref{eq:lowerbound_surplus_extractionconstrained} is a lower bound on the regulator's worst-case regret, it is enough to exhibit a technology that yields a regret at least equal to it. So, in the remainder of the proof, we construct such a technology.

    Let $\mu_F, y'$ such that $\underline{y} \leq \mu_F \leq y' \leq \bar{y}$, and $\Pi^{\mu_F} = \left(1-\frac{\underline{w}^{\mathcal{C}}(y')}{y'}\right) \mu_F$. We denote by $B_i$ the binary output distribution with support $\left\{0, y'\right\}$ and mean $i$ and we define the compact action set $\mathcal{A}^{\mu_F} = \left\{ (e_i, {B}_{i}) \, :\, i\in I = \left[\mu_F, y' \right] \right\}$, where, for all $i\in I$,
    \begin{align*}
        e_i = i - \mu_F + \Pi^{\mu_F} \left( \ln \left(\mu_F \right) - \ln i \right) + \frac{\underline{w}^{\mathcal{C}}(y')}{y'} \mu_F.
    \end{align*}
    We consider the family of technologies $\mathcal{T}^{\mu_F} = \left(\underline{k}, \mathcal{A}^{\mu_F}\right)$, $\mu_F \leq y'$.

    By definition of $e_i$, for all $w \in \mathcal{C}$,
    \begin{align*}
        \frac{\partial}{ \partial i} \left( \ell^w i - e_i \right) = \ell^w - \frac{i - \Pi^{\mu_F}}{i} \geq 0 \Leftrightarrow \ell^w \geq \frac{i - \Pi^{\mu_F}}{i},
    \end{align*}
    where $\ell^w = \frac{w(y')}{y'}$. So, $(e_i, B_{i})$ with $i<y'$ is implementable by $w \in \mathcal{C}$ if and only if
    \begin{align*}
        w(y') = \frac{i - \Pi^{\mu_F}}{i} y',
    \end{align*}
    while action $i = y'$ is implementable by $w \in \mathcal{C}$ if and only if
    \begin{align*}
        w(y') \geq \frac{y' - \Pi^{\mu_F}}{y'} y'.
    \end{align*}
    
    The contract $w(y) =y$ implements the surplus maximising action $(e_{y'}, B_{y'})$ (since $\frac{\partial}{\partial i} e_i \leq 1$), and leaves no profit to the firm, hence allocating the entire surplus to the worker. So, it is the regulator's preferred contract and
    \begin{align*}
        V(\mathcal{T}^{\mu_F}) = \alpha \left(y' - e_{y'} - k\right).
    \end{align*}
    
    Next, observe that, for all $i$, any contract that implements $(e_i, B_i)$ gives a profit of at most $\Pi^{\mu_F}$ to the principal, since
    \begin{align*}
        \left(1- \frac{i-\Pi^{\mu_F}}{i} \right) \mathbb{E}_{B_i}\left[y\right] = \Pi^{\mu_F}.
    \end{align*}
    So, in the worst-case scenario, the firm implements the least productive action, generating the average output $\mu_F$. After simple calculations, the regulator's regret rewrites
    \begin{align*}
        R\left(\mathcal{C,T}^{\mu_F}\right) = \alpha \left(y' - e_{y'} - \left( \mu_F - e_{\mu_F}\right)\right) + (\alpha-1) (\Pi^{\mu_F} - k). 
    \end{align*}
    Plugging in the effort cost, the regret is
    \begin{align*}
        \alpha \left(y' - e_{y'} - \left( \mu_F - e_{\mu_F}\right)\right) & +(\alpha-1) (\Pi^{\mu_F} - k) \\
        & =  - \alpha \Pi^{\mu_F}\left( \ln \left(\mu_F \right) - \ln y' \right) + (\alpha-1) (\Pi^{\mu_F} - k) \\
        & = \alpha \int_{\mu_F}^{y'} \frac{1- \frac{\underline{w}^{\mathcal{C}}(y')}{y'}}{\mu}\mu_F d\mu + (\alpha -1) \left( \left(1- \frac{\underline{w}^{\mathcal{C}}(y')}{y'}\right) \mu_F - k \right).
    \end{align*}
    The worst-case regret is obtained by maximising over all possible technologies. So, we can obtain a lower bound by taking the supremum of the above expression with respect to $\mu_F \in \left[\frac{\underline{k}}{1- \frac{\underline{w}^{\mathcal{C}}(y')}{y'}} \vee \underline{y}, y'\right]$ and $y' \in [\underline{y}, \bar{y}]$. Therefore, the regulator's worst-case regret is bounded below by the right-hand side in \cref{eq:lowerbound_surplus_extractionconstrained}.
\end{proof}

\cref{lemma:boundonregretnoproductionconstrained} and \cref{lemma:boundonregretproductionconstrained} adapt the proofs of \cref{lemma:boundonregretproduction,lemma:boundonregretnoproduction} to establish an upper bound on the regulator's regret for any MPR regulation. 

\begin{lemmap}{\ref{lemma:boundonregretnoproduction}'}\label{lemma:boundonregretnoproductionconstrained}
    For all MLC regulation $\mathcal{C}_{\ell}$ with minimum contract $\ell$,
    \begin{align}\label{eq:regret_claim_1_constrained}
       \underset{\mathcal{T} \in \mathbb{T}^c \colon \Pi\left(\mathcal{C}_{\ell}, \mathcal{T}\right) =0 }{\sup} & \, R\left(\mathcal{C}_{\ell}, \mathcal{T}\right)  \leq  \max\bigg\{ \underset{k \in [\underline{k}, \bar{k}] }{\sup} \alpha \left(  \underline{y}\vee \frac{k}{1-\ell} -k  - k \left( \ln \left( \underline{y}\vee \frac{k}{1-\ell} \right) - \ln \left( \bar{y}\right)\right)\right) \mathds{1}_{\{ (1-\ell^{\mathcal{C}}) \underline{y} \leq \bar{k} \}} \notag \\
       & + \underset{k \in [\underline{k}, \bar{k}] }{\sup}\alpha \left(  (1-\ell^{\mathcal{C}})\underline{y} -k  - k \left( \ln \left( \underline{y} \right) - \ln \left( \bar{y}\right)\right)\right) \mathds{1}_{\{ (1-\ell^{\mathcal{C}}) \underline{y} > \bar{k} \}}, \ \underset{y \in \left[\underline{y}, \bar{y}\wedge \frac{k}{1-\ell}\right] }{\sup} \alpha  \ell y \bigg\}.
    \end{align}
\end{lemmap}

\begin{proof}
    The proof follows closely the proof of \cref{lemma:boundonregretnoproduction}. Let $\mathcal{C}_{\ell}$ be a MLC regulation with mimimum contract $\ell$ and $\mathcal{T} = \left(k , \mathcal{A} = \left\{ \left(e_i, F_i\right) \right\}_{i\in I} \right) \in \mathbb{T}^c$ be a technology such that $\Pi\left(\mathcal{C}_{\ell}, \mathcal{T}\right) =0$. For all $i\in I$, we write $\mu_i$ for $\mathbb{E}_{F_i}\left[y\right]$. Denote by $\left(e_R, F_R\right)$ the \eqref{eq:FB_program}-optimal action. Since $\Pi\left(\mathcal{C}_{\ell}, \mathcal{T}\right) =0$, we can assume that the firm takes her outside option, as it maximises the regret among all possible contracting equilibria. We distinguish three cases:
    \begin{enumerate}
        \item[a)] if $\mathbb{E}_{F_R} \left[ y \right] =\mu_R \leq \frac{k}{1-\ell}$, we show that
        \begin{align*}
            R\left(\mathcal{C}_{\ell}, \mathcal{T}\right) \leq \underset{y \in \left[\underline{y}, \bar{y}\wedge \frac{\bar{k}}{1-\ell}\right] }{\sup} \alpha  \ell y;
        \end{align*}
    
        \item[b)] if $ \mu_R > \frac{k}{1-\ell}$ and $(1-\ell) \underline{y} - \bar{k} \leq 0$, we show that
        \begin{align}
            R\left(\mathcal{C}_{\ell}, \mathcal{T}\right) \leq \underset{k \in [\underline{k}, \bar{k}] }{\sup} \alpha \left(  \underline{y}\vee \frac{k}{1-\ell} -k  - k \left( \ln \left( \underline{y}\vee \frac{k}{1-\ell} \right) - \ln \left( \bar{y}\right)\right)\right); \label{eq:noproductionupperboundb)}
        \end{align}

        \item[c)] if $ \mu_R > \frac{k}{1-\ell}$ and $(1-\ell) \underline{y} - \bar{k} > 0$, we show that
        \begin{align}
            R\left(\mathcal{C}_{\ell}, \mathcal{T}\right) \leq \underset{k \in [\underline{k}, \bar{k}] }{\sup}\alpha \left(  (1-\ell)\underline{y} -k  - k \left( \ln \left( \underline{y} \right) - \ln \left( \bar{y}\right)\right)\right). \label{eq:noproductionupperboundc)}
        \end{align}
    \end{enumerate}
    Put together, they establish \cref{eq:regret_claim_1_constrained}.

\noindent a) If $\mu_R \leq \frac{k}{1-\ell}$, then the regulator's regret is bounded above by
\begin{align*}
    R\left( \mathcal{C}_{\ell}, \mathcal{T} \right) & \leq \alpha \left( \mu_R -e_R -k\right) \\
    & \leq \alpha \left( \mu_R - e_R - (1-\ell)\mu_R \right) \\
    & \leq \underset{\mu_R \in \left[\underline{y}, \bar{y}\wedge \frac{k}{1-\ell}\right] }{\sup} \alpha  \ell \mu_R.
\end{align*}

The result follows since the regret never exceeds $\alpha \bar{y}$.

\noindent b) + c) If $\mu_R > \frac{k}{1-\ell}$, then $\frac{\mu_R - k}{\mu_R} > \ell$. So, the linear contract $y\to 1- \frac{ k + \epsilon_1}{\mu_R} \cdot y$ belongs to the regulation $\mathcal{C}_{\ell}$ for $\epsilon_1>0$ small. Proceeding inductively as in the proof of \cref{lemma:boundonregretnoproduction}, case b), we can construct as sequence of actions, $(i_j)_j$, with $\mu_{i_{j+1}} < \mu_{i_j}$ and
\begin{align*}
    (e_{i_j} -e_{i_{j+1}}) \geq  \left(1- \frac{k + 2^{-j}\epsilon_{1}}{\mu_{i_j}} \right) (\mu_{i_j} - \mu_{i_{j+1}}),
\end{align*}
such that the limit of the sequence of expected revenue, $\mu_{\infty}$, falls below $\underline{y}\vee \frac{k}{1-\ell}$. Therefore, still following the proof of \cref{lemma:boundonregretnoproduction}, we obtain
\begin{align*}
    e_R \geq e_{\infty} + \int_{\underline{y}\vee \frac{k}{1-\ell}}^{\mu_R} \left(1 - \frac{k}{\mu}\right) d\mu.
\end{align*}
So,
\begin{align*}
    R\left(\mathcal{C}_{\ell}, \mathcal{T}\right) & \leq \alpha \left( \mu_R -e_R - k \right) \\
    & \leq \alpha \left(  \underline{y}\vee \frac{k}{1-\ell} -k -e_{\infty} - k \left( \ln \left( \underline{y}\vee \frac{k}{1-\ell} \right) - \ln \left( \mu_R\right)\right)\right).
\end{align*}
To obtain the upper bound in \cref{eq:noproductionupperboundb)}, simply note that $e_{\infty} \geq 0$ and take the supremum over $k\in [\underline{k}, \bar{k}]$ and $\mu_R \in [\underline{y}, \bar{y}]$. To obtain the upper bound in \cref{eq:noproductionupperboundc)}, recall that $(1-\ell)\underline{y} - \bar{k} >0$ and $\Pi(\mathcal{C}_{\ell}, \mathcal{T}) =0$. So, offering contract $\ell$ cannot incentivise any action: $e_i > \ell \mu_i $ for all $i$. Hence, $e_{\infty} \geq \ell \underline{y}$. Taking the supremum over $k\in [\underline{k}, \bar{k}]$ and $\mu_R \in [\underline{y}, \bar{y}]$ then yields the upper bound in \cref{eq:noproductionupperboundc)}, since $\frac{\bar{k}}{1-\ell} \leq \underline{y}$.

Combining cases a), b) and c) gives \cref{eq:regret_claim_1_constrained}, which concludes the proof.
\end{proof}

\begin{lemmap}{\ref{lemma:boundonregretproduction}'}\label{lemma:boundonregretproductionconstrained}
    For all MLC regulation $\mathcal{C}_{\ell}$ with minimum contract $\ell$,
    \begin{align}\label{eq:regret_claim_2_constrained}
        \underset{\mathcal{T} \in \mathbb{T}^c \colon \Pi\left(\mathcal{C}_{\ell}, \mathcal{T}\right) >0 }{\sup} \, R\left(\mathcal{C}_{\ell}, \mathcal{T}\right) & \leq \max\bigg\{ \underset{y \in \left[\frac{\underline{k}}{1-\ell} \vee \underline{y}, \bar{y}\right]}{\sup} \ \alpha \int_{y}^{\bar{y}} \frac{1-\ell}{\mu} \mu_F d\mu + (\alpha -1) \left(1-\ell \right) y - (\alpha-1)\underline{k},  \notag \\
        & \underset{k \in \left[\underline{k}, (1-\ell)\underline{y} \wedge \bar{k} \right]}{\sup } \ \alpha \left(  (1-\ell)\underline{y} -k  - k \left( \ln \left( \underline{y} \right) - \ln \left( \bar{y}\right)\right)\right), \ \left( \alpha -1\right) \left[(1-\ell)\bar{y} - \underline{k} \right] \bigg\}.
    \end{align}
\end{lemmap}

\begin{proof}
    The proof follows closely the proof of \cref{lemma:boundonregretproduction}. Let $\mathcal{C}_{\ell}$ be a MLC regulation with mimimum contract $\ell$ and $\mathcal{T} = \left(k , \mathcal{A} = \left\{ \left(e_i, F_i\right) \right\}_{i\in I} \right) \in \mathbb{T}^c$ be a technology such that $\Pi\left(\mathcal{C}_{\ell}, \mathcal{T}\right) >0$. For all $i\in I$, we write $\mu_i$ for $\mathbb{E}_{F_i}\left[y\right]$. Denote by $\left(e_R, F_R\right)$ the \eqref{eq:FB_program}-optimal action, and by $\left(e_F, F_F\right)$ the action optimally implemented by the firm in the regret-maximising equilibrium. As we can always uniformly increase the effort costs incurred by the worker, we can assume without loss that the worker's surplus is zero, $\mathbb{E}_{F_F}\left[ w^{\star}(y) \right] - e_F =0$, in the regret maximizing contracting equilibrium. We distinguish three cases:
    \begin{enumerate}
        \item[a)] if $\mu_R -e_R \leq \mu_F -e_F$, we show that
        \begin{align}
            R\left(\mathcal{C}_{\ell}, \mathcal{T}\right) \leq \left( \alpha -1\right) \left[(1-\ell)\bar{y} - \underline{k} \right] ; \label{eq:productionupperbounda)}
        \end{align}
    
        \item[b)] if $\mu_R -e_R > \mu_F -e_F$ and $\mu_R \geq \mu_F$, we show that
        \begin{align}\label{eq:productionupperboundb)}
            R\left(\mathcal{C}_{\ell}, \mathcal{T}\right) \leq \underset{y \in \left[\frac{\underline{k}}{1-\ell} \vee \underline{y}, \bar{y}\right]}{\sup} \ \alpha \int_{y}^{\bar{y}} \frac{1-\ell}{\mu} y d\mu + (\alpha -1) \left(1-\ell \right) y - (\alpha-1)\underline{k}; 
        \end{align}

        \item[c)] if $ \mu_R > \frac{k}{1-\ell}$ and $\mu_R < \mu_F$, we show that
        \begin{align}
            R\left(\mathcal{C}_{\ell}, \mathcal{T}\right) \leq \max\bigg\{ \underset{y \in \left[\frac{\underline{k}}{1-\ell} \vee \underline{y}, \bar{y}\right]}{\sup} & \ \alpha \int_{y}^{\bar{y}} \frac{1-\ell}{\mu} y d\mu + (\alpha -1) \left(1-\ell \right) y - (\alpha-1)\underline{k},  \notag \\
            & \underset{k \in \left[\underline{k}, (1-\ell)\underline{y}\right]}{\sup } \alpha \left(  (1-\ell)\underline{y} -k  - k \left( \ln \left( \underline{y} \right) - \ln \left( \bar{y}\right)\right)\right) \bigg\}. \label{eq:productionupperboundc)}
        \end{align}
    \end{enumerate}
    Put together, they establish \cref{eq:regret_claim_2_constrained}.

\noindent a) Suppose first that $\mu_R -e_R \leq \mu_F -e_F$. The regulator's regret is then bounded above by
\begin{align*}
    R\left(\mathcal{C}_{\ell}, \mathcal{T}\right) & = \alpha \left(\mu_R - e_R - (\mu_F - e_F) \right) + (\alpha-1) \left[\Pi(\mathcal{C}_{\ell}, \mathcal{T}) - k\right] \\
    & \leq (\alpha-1) \left[\Pi(\mathcal{C}_{\ell}, \mathcal{T}) - k\right] \\
    & \leq (\alpha-1) \left[(1-\ell)\mu_F - k\right].
\end{align*}
Taking the supremum over $k\in [\underline{k}, \bar{k}]$ and $\mu_F \in [\underline{y}, \bar{y}]$ yields the upper bound in \cref{eq:productionupperbounda)}.

\noindent b) Suppose that $\mu_R -e_R > \mu_F -e_F$ and $\mu_F < \mu_R$. The regulator's regret is bounded above by
\begin{align}\label{eq:proofclaim2_regretconstrained}
    R\left(\mathcal{C}_{\ell}, \mathcal{T}\right) \leq \alpha \left( \mu_R - e_R \right) - \Pi(\mathcal{C}_{\ell}, \mathcal{T}) - (\alpha-1) k.
\end{align}
Following the proof of the Claim in case b) of the proof of \cref{lemma:boundonregretproduction}, we obtain 
\begin{align*}
    e_F - e_R \leq - \int_{\mu_R}^{\mu_F} \left(1 - \frac{1-\ell}{\mu}\mu_F \right) d\mu.
\end{align*}
As a result, we can bound the regret in \cref{eq:proofclaim2_regretconstrained} from above:
\begin{align*}
    R\left(\mathcal{C}_{\ell}, \mathcal{T}\right) & \leq \alpha \left( \mu_R - \mu_F - \int_{\mu_F}^{\mu_{R}} \left(1- \frac{1-\ell}{\mu} \mu_F\right) d\mu\right) + (\alpha -1) \Pi(\mathcal{C}_{\ell}, \mathcal{T}) - (\alpha-1)k \\
    & \leq \alpha \int_{\mu_F}^{\mu_{R}} \frac{1-\ell}{\mu} \mu_F d\mu + (\alpha -1) \left(1-\ell \right) \mu_F - (\alpha-1)k.
\end{align*}
where we used that $\Pi(\mathcal{C}_{\ell}, \mathcal{T}) = \mu_F -e_F \leq (1-\ell)\mu_F$.
Maximising the right-hand side with respect to $\mu_R \in [\underline{y}, \bar{y}]$ and $k \in [\underline{k}, \bar{k}]$ and then taking the supremum over $\mu_F \in \left[\frac{k}{1-\ell} \vee \underline{y}, \bar{y}\right]$ we obtain the upper bound in \cref{eq:productionupperboundb)}.

\noindent c) Suppose that $\mu_R -e_R > \mu_F -e_F$ and $\mu_F > \mu_R$. To prove \cref{eq:productionupperboundc)}, we distinguish two subcases.
\begin{itemize}
    \item  If $\Pi(\mathcal{C}_{\ell}, \mathcal{T}) \geq (1-\ell) \underline{y}$, we show that there is another technology $\mathcal{T}' \in \mathbb{T}^b$ such that (i) $\mu_F \leq \mu_R$ and (ii) $R\left(\mathcal{C}_{\ell}, \mathcal{T}\right) \leq R\left(\mathcal{C}_{\ell}, \mathcal{T}'\right)$. To see this, start by noting that $\mu = \frac{\Pi(\mathcal{C}_{\ell}, \mathcal{T})}{1-\ell} \geq \underline{y}$. Define then $\mathcal{T}' = \left(k, \mathcal{A}' = \mathcal{A} \cup \left\{(\ell \mu, B_{\mu}), (\tilde{e}_R, B_{\mu_R}) \right\} \right)$ where 
    \begin{align*}
        \tilde{e}_R = \sup\left\{ e \in [0, e_R] \ : \ \mu_R - k -e_R \geq \frac{\mu_R - k}{\mu_R} \mu_F - e_F, \ \forall (e,F) \in \mathcal{A} \text{ such that } \mu_F \leq \mu_R   \right\}.
    \end{align*} 
    Observe first that $\mu_R -e_R > \mu_F -e_F = \Pi(\mathcal{C}_{\ell}, \mathcal{T})$ implies $\mu_F < \mu_R$, since $\ell \mu_R -e_R \leq \mathbb{E}_{F_R}\left[w^{\star}(y)\right] -e_R \leq 0 \Rightarrow e_R \geq \ell \cdot \mu_R$. So, the new technology $\mathcal{T}'$ satisfies (i). Moreover, under technology $\mathcal{T}'$, the firm optimally offers contract $\ell$, which implements $\left\{(\ell \mu, B_{\mu}) \right\}$. Her profit and the worker's surplus remain the same. Therefore, $\mathcal{T}'$ also satisfies (ii) if $V(\mathcal{T}') \geq V(\mathcal{T})$. That is, we need to show that adding actions $(\ell \mu, B_{\mu})$ and $(\tilde{e}_R, B_{\mu_R})$ does not hurt the fully-informed regulator. To see that it does not, note that the regulator can offer the linear contract $\ell = \frac{\mu_R - k}{\mu_R}$, which implements $(\tilde{e}_R, B_{\mu_R})$ or a more productive action, by definition of $\tilde{e}$, and, hence, increases the total surplus and leaves profit at least $k$ to the firm. So, $V(\mathcal{T}') \geq V(\mathcal{T})$. 

    Therefore,
    \begin{align*}
        R(\mathcal{C}_{\ell}, \mathcal{T}) \leq R\left(\mathcal{C}_{\ell}, \mathcal{T}\right) \leq \underset{y \in \left[\frac{\underline{k}}{1-\ell} \vee \underline{y}, \bar{y}\right]}{\sup} \ \alpha \int_{y}^{\bar{y}} \frac{1-\ell}{\mu} \mu_F d\mu + (\alpha -1) \left(1-\ell \right) y - (\alpha-1)\underline{k}.
    \end{align*}

    \item If, instead, $\Pi(\mathcal{C}_{\ell}, \mathcal{T}) < (1-\ell) \underline{y}$, then, for all $i$, $e_i > \ell \mu_i$. In particular, $\ell \mu_R - e_R <0$. Note that the regret is bounded above by
    \begin{align*}
        R(\mathcal{C}_{\ell}, \mathcal{T}) \leq \alpha \left(\mu_R - e_R - k\right) - (\underbrace{\mu_F - e_F}_{=\Pi} - k).
    \end{align*}
    Since the firm chooses to implement action $(e_F, F_F)$ instead of $(e_R,F_R)$, the linear contract with slope $\frac{\mu_R - \Pi}{\mu_R}$ cannot implement $(e_R, F_R)$ or any other action whose expected output is greater than $\mu_R$. Therefore there must exists $i$ with $\mu_i <\mu_R$ such that
    \begin{align*}
        \frac{\mu_R - \Pi}{\mu_R} \mu_i - e_i \geq  \frac{\mu_R - \Pi}{\mu_R} \mu_R - e_R \Leftrightarrow e_R -e_i \geq  \frac{\mu_R - \Pi}{\mu_R} (\mu_R - \mu_i).
    \end{align*}
    Following the proof of \cref{lemma:boundonregretnoproduction}, proceeding inductively as in Step b), we obtain a lower bound on the effort cost $e_R$:
    \begin{align*}
        e_R - e_{\infty} \geq \int_{\underline{y}}^{\mu_R}\left(1 - \frac{\Pi}{\mu}\right) d\mu.
    \end{align*}
    But, for all $i$, $e_i > \ell \mu_i$. So, $e_{\infty} \geq \ell \underline{y}$. Therefore, the regulator's regret is bounded above by:
    \begin{align*}
        R(\mathcal{C}_{\ell}, \mathcal{T}) \leq \alpha \left(\underline{y} - \ell \underline{y} + \Pi \int_{\underline{y}}^{\bar{y}} \frac{1}{\mu} d\mu - k\right) - (\Pi - k).
    \end{align*}
    Since the firm produces, $\Pi > k$ and, by assumption $\Pi < (1-\ell) \underline{y}$. Note then that the supremum of the right-hand side with respect to $\Pi \in [k, (1-\ell)\underline{y}]$ is obtained for $\Pi = k$ or $\Pi = (1-\ell)\underline{y}$, which establishes the upper bound in \cref{eq:productionupperboundc)}.
\end{itemize}

Combining cases a), b) and c) gives \cref{eq:regret_claim_2_constrained}, which concludes the proof.
\end{proof}

\section{Existence of an optimal contract}\label{app:existence}

    \begin{lemma}\label{lemma:existenceoptimalcontract}
        Suppose that either
        \begin{enumerate}
            \item[(i)] the regulation $\mathcal{C} \subset \mathbb{C}_0 $ is (sequentially) compact in the topology associated with the product topology on $\mathbb{R}^{\mathbb{R}_+}$,
            \item[(ii)] the regulation $\mathcal{C}$ is convex, 
        \end{enumerate}
        Then the regulator's problem under complete information \eqref{eq:FB_program} and the firm's problem \eqref{eq:profit_program} admit a solution. 
    \end{lemma}

    \begin{proof}
        We prove the existence of a solution to the firm's problem \eqref{eq:profit_program}. The proof of existence for the regulator's problem \eqref{eq:FB_program} follows the same steps. Hence it is omitted. 

        Let $\mathcal{T} = \left(k, \mathcal{A}\right)$ be a technology and $\mathcal{C} \subset \mathbb{C}_0 $ be a regulation that satisfies condition (i), (ii), or (iii). It is enough to show the existence of a solution to
        \begin{align}
            P(\mathcal{C,T}) & =  \sup_{w \in \mathcal{C}, (e,F) \in \mathcal{A}} \mathbb{E}_F[y-w(y)]-k \tag{$P$} \label{eq:profit_programP}\\
            \text{s.t.} \quad & \mathbb{E}_F[w(y)] - e \geq 0, \tag{$IR_W$}  \\
            & \mathbb{E}_F[w(y)] - e \geq \mathbb{E}_{\Tilde{F}}[w(y)] - \Tilde{e}, \text{ for all } (\tilde{e},\tilde{F}) \in \mathcal{A}. \tag{$IC_W$}
        \end{align}
        Consider a maximising sequence $\left(w_n, \left(e_n, F_n\right) \right)_{n \in \mathbb{N}}$. The value of the above program, $P(\mathcal{C,T})$, is finite since $\sup_{(e,F) \in \mathcal{A}} \mathbb{E}_F[y] \leq \bar{y}$. So, there exists $N \in \mathbb{N}$ such that, for all $n\geq N$,
        \begin{align*}
            \int w_n(y) dF_n(y) < \underset{(e,F) \in \mathcal{A}}{\sup} \, \mathbb{E}_F\left[y\right] - \mathcal{P}\left(\mathcal{C,T}\right) + 1.
        \end{align*}
        Relabelling if necessary, we can assume that $N=1$. Moreover, since $w_n$ implements $(e_n, F_n)$, the incentive compatibility constraint implies that, for all $(e, F) \in \mathcal{A}$, 
        \begin{align*}
            \int w_n(y) d {F}(y) \leq \int w_n(y) dF_n(y) + {e} -e_n \leq  \underset{(e,F) \in \mathcal{A}}{\sup} \,\left\{  \mathbb{E}_F\left[y\right] \right\} - \mathcal{P}\left(\mathcal{C,T}\right) + 1 + \underset{(e,F) \in \mathcal{A}}{\sup} \, \left\{ e - e_n \right\}.
        \end{align*}
        So, there exists a constant $K >0$ such that
        \begin{align}\label{eq:uniformboundedness}
            \underset{n \in \mathbb{N}}{\sup} \, \underset{({e}, {F})}{\sup } \, \int w_n(y) d{F}(y) < K < \infty.
        \end{align}
        Moreover, since $\mathcal{A}$ is compact in the topology associated with the total variation norm by assumption, there exists a finite dominating measure $\nu$ on $\mathbb{R}_+$ such that, for all $n \in \mathbb{N}$, $F_n << \nu$ by Theorem 4.6.3 in \cite{bogachev2007measure}. The compactness of $\mathcal{A}$ also implies that, passing to a subsequence if necessary, $(e_n, F_n) \to (e^*, F^*) \in \mathcal{A}$, and, thus, that $\frac{dF_n}{d\nu} \to \frac{dF^*}{d\nu}$ in $L^1(\nu)$. Passing again to a subsequence if necessary, we can finally assume that $\frac{dF_n}{d\nu} \to \frac{dF^*}{d\nu}$ $\nu$-almost surely.
        
        If $\mathcal{C}$ is compact, then there exists a subsequence of $\left( w_n \right)_{n \in \mathbb{N}}$ that converges to some $w^{\star} \in \mathcal{C}$. Otherwise, by the Dunford-Pettis Theorem (Theorem 4.7.18 in \cite{bogachev2007measure}) and \cref{eq:uniformboundedness}, there exists a subsequence of $\left( w_n \right)_{n \in \mathbb{N}}$ that converges weakly in $L^1(F^*)$ to some $w^{\star}$. Moreover, \cref{eq:uniformboundedness} implies that we can invoke Koml\'{o}s' Subsequence Theorem (Theorem 4.7.24 in \cite{bogachev2007measure}). Therefore, there exists a further subsequence such that every subsubsequences C\'{e}saro converges to $w^{\star}$, ${F}^{\star}$-almost surely. Since $\mathcal{A}$, is compact, by a diagonal argument, we can extract a further subsequence such that every subsubsequences C\'{e}saro converges $\nu$-almost surely to $w^{\star}$ invoking again Koml\'{o}s' Theorem. Finally, the richness condition (ii) guarantees that $w^* \in \mathcal{C}$ since $\mathcal{C}$ is convex.

        Summarising, we can choose the maximising sequence $\left(w_n, (e_n, F_n)\right)$ such that it converges to some $(w^{\star}, (e^*, F^*)) \in \mathcal{C} \times \mathcal{A}$, where the $w_n$'s convergence is understood to be C\'{e}saro $\nu$-almost surely and the $F_n$'s convergence is understood to be the $\nu$-almost sure convergence of their Radon-Nykodym derivatives.  

        We are now ready to prove that $\left(w^{\star}, (e^*, F^*)\right)$ maximises \eqref{eq:profit_programP}. We first show that $w^{\star}$ implements $(e^*, F^*)$. Observe that
        \begin{align*}
            \left( \frac{\sum_{i=1}^n w_{i}(y)}{n} \right) \frac{dF_n(y)}{d\nu}  \to w^{\star}(y) \frac{dF^*(y)}{d\nu}, \quad \nu\text{-almost surely.}
        \end{align*}
        \cref{eq:uniformboundedness} and Vitali's convergence Theorem (Theorem 4.5.4 in \cite{bogachev2007measure}) then implies that, for all $({e},{F})$,
        \begin{align*}
            \int \frac{\sum_{i=1}^n w_{i}(y)}{n} d{F}(y) \to \int w^{\star} (y) d{F}(y) \text{ and } \int \left( \frac{\sum_{i=1}^n w_{i}(y)}{n} \right) d{F_n}(y) \to \int w^{\star} (y) d{F}^{\star}(y).
        \end{align*}
        It follows that
        \begin{align*}
            \int w^{\star} (y) d{F}^{\star}(y) -e^* \geq \max\left\{ \int w^{\star} (y) d{F}(y) - e, 0\right\} \text{ for all } (e, F) \in \mathcal{A}.
        \end{align*}
        Therefore $w^{\star}$ implements $(e^*,F^*)$. To conclude, there remains to show that
        \begin{align*}
            \int \left(y  - \frac{\sum_{i=1}^n w_{i}(y)}{n} \right) d F_n(y) \to  \int \left( y - w^{\star} (y)\right) d F^*(y).
        \end{align*}
        This again follows from Vitali convergence theorem since the sequence 
        \begin{align*}
            \left( \left(y  - \frac{\sum_{i=1}^n w_{i}(y)}{n} \right) \frac{dF_n(y)}{d\nu} \right)_{n\in \mathbb{N}}
        \end{align*}
        is uniformly bounded by \eqref{eq:uniformboundedness} and converges $\nu$-almost surely to $\left(y - w^{\star}(y)\right) \frac{dF^*(y)}{d\nu}$. 
    \end{proof}

\end{document}